\documentclass{article}

\usepackage{ifpdf}
\usepackage{natbib}
\usepackage{amsmath}
\usepackage{amssymb}
\usepackage{epubtk}
\usepackage{graphicx}
\usepackage{textcomp}
\usepackage{booktabs}
\usepackage{array}

\newcommand{\aap}{{Astron. Astrophys.}}
\newcommand{\aj}{{Astronomical J.}}
\newcommand{\apj}{{Astrophys. J.}}
\newcommand{\apjl}{{Astrophys. J. Lett.}}
\newcommand{\apss}{{Astrophys. Space Sci.}}
\newcommand{\grl}{{Geophys. Res. Lett.}}
\newcommand{\jgr}{{J. Geophys. Res.}}
\newcommand{\mnras}{{Mon. Not. Roy. Astron. Soc.}}
\newcommand{\nat}{{Nature}}
\newcommand{\pasp}{{Pub. Astron. Soc. Pacific}}

\newcommand{\solphys}{{Solar Phys.}}
\newcommand{\zap}{{Z. Astrophys.}}
\newcommand{\ssr}{{Space Sci. Rev.}}

\begin{document}

\title{The Solar Cycle}

\author{\epubtkAuthorData{David H.\ Hathaway}{%
    Mail Stop: 258-5, \\
    NASA Ames Research Center, \\
    Moffett Field, CA 94035, U.S.A.}{%
    david.hathaway@nasa.gov}{%
    http://solarscience.msfc.nasa.gov/}
}

\date{}
\maketitle

\begin{abstract}
The Solar Cycle is reviewed. The 11-year cycle of solar activity is 
characterized by the rise and fall in the numbers and surface area of 
sunspots. A number of other solar activity indicators  also vary in
association with the sunspots including; 
the 10.7~cm radio flux, the total solar irradiance, the magnetic field, 
flares and coronal mass ejections, geomagnetic activity, galactic cosmic ray 
fluxes, and radioisotopes in tree rings and ice cores.
Individual solar cycles are characterized
by their maxima and minima, cycle periods and 
amplitudes, cycle shape, the equatorward drift of the active latitudes,
hemispheric asymmetries, and active longitudes.
Cycle-to-cycle variability includes the Maunder 
Minimum, the Gleissberg Cycle, and the Gnevyshev--Ohl (even-odd) Rule.
Short-term variability includes the 154-day periodicity, quasi-biennial variations, and 
double-peaked maxima. We conclude with an examination of prediction 
techniques for the solar cycle and a closer look at cycles 23 and 24.
\end{abstract}

\epubtkKeywords{Solar activity, Solar cycle, Solar cycle prediction,
  Sunspots}

\newpage 

\section{Introduction}
\label{sec:intro}

Solar activity rises and falls with an 11-year cycle that affects modern life in many 
ways. Increased solar activity includes increases in extreme ultraviolet and 
X-ray emissions from the Sun that produce dramatic effects in Earth's 
upper atmosphere. The associated atmospheric heating increases both the 
temperature and density of the atmosphere at many spacecraft altitudes. The 
increase in atmospheric drag on satellites in low Earth orbit can 
dramatically shorten the orbital lifetime of these valuable assets
\citep[for a review see][]{Pulkkinen:2007}.

Increases in the number of solar flares and coronal mass ejections (CMEs) 
raise the likelihood that sensitive instruments in space will be damaged by 
energetic particles accelerated in these events. These solar energetic 
particles (SEPs) can also threaten the health of both astronauts in space 
and airline travelers in high-altitude, polar routes.

Solar activity apparently affects terrestrial climate as well. Although the 
change in the total solar irradiance seems too small to produce significant 
climatic effects, there is good evidence that, to some extent, the Earth's 
climate heats and cools as solar activity rises and falls
\citep[for a review see][]{Haigh:2007}.

There is little doubt that the solar cycle is magnetic in nature and 
is produced by dynamo processes within the Sun
\citep[for a review of the solar dynamo see][]{Charbonneau:2010}.
Although the details concerning, how, when, and where the dynamo processes
operate are still uncertain, several basic features of the dynamo
are fairly well accepted and provide a framework for understanding
the solar cycle.

Within the Sun's interior magnetic fields and the ionized plasma move
together.
(Any motion of the plasma relative to the magnetic field or vice versa
will set up currents that counter those relative displacements.)
Furthermore, throughout most of the Sun's interior the plasma pressure
exceeds the magnetic pressure and the plasma kinetic energy exceeds the
magnetic energy so that the motion of the plasma controls
the magnetic field---the magnetic field is transported and transformed by the
plasma flows.
(A notable exception is in sunspots where the magnetic field is strong enough
to choke off the convective heat flow---leaving sunspots cooler and darker than
their surroundings.)

Two basic processes are involved in most dynamo models -- shearing motions
that strengthen the magnetic field and align it with the flow (the Omega-effect)
and helical motions that lift and twist the magnetic field into a different plane
(the alpha-effect).
\cite{Babcock:1961} described a phenomenological dynamo model in which the
shearing motions are those of the Sun's differential rotation (which he
assumed was just a latitudinal shear).
His model starts with a global dipole field (a poloidal field) closely aligned with
the rotation axis at solar cycle minimum.
He assumed that this field threaded through a shallow surface layer and
connected to the opposite pole along meridional lines.
The observed latitudinal differential rotation should take this weak poloidal
field and shear it out to produce a much stronger toroidal field wrapped
around the Sun nearly parallel to lines of latitude.

Babcock noted that this toroidal field becomes strongest at latitudes near
30\textdegree where the shear is strongest (and where sunspots first
appear at the start of each cycle).
He suggested that sunspot groups form once this toroidal field becomes strong enough
to make the magnetized plasma buoyant.
As the cycle progresses and the shearing continues, the latitudes at which
the toroidal field becomes
buoyant should spread to both higher and lower latitudes.

In Babcock's model the toroidal field is not directed purely east-west along
lines of latitude, but retains a small north-south component from the
original poloidal field.
This gives a slight tilt to the emerging active regions (the alpha-effect)
with the following (relative to the direction of rotation) polarity sunspots in
a group at slightly higher latitudes.

At the time that Babcock presented his model little was known about the
Sun's meridional circulation other than the fact that it was much weaker
than the differential rotation.
There were, however, reports of sunspot groups moving equatorward at low
latitudes and poleward at high latitudes \citep{Dyson:1913, Tuominen:1942}.
Those observations (possibly coupled with considerations of the effects
of the Coriolis force on the differential rotation) led Babcock to suggest
the presence of a meridional flow that was equatorward at low latitudes
and poleward at high latitudes.

In his model, this meridional flow pattern transports the low latitude
(predominantly leading-polarity) magnetic field toward the equator,
where it cancels with the opposite polarity fields in the other hemisphere.
Meanwhile, the high latitude, following-polarity, magnetic field 
is transported to the poles.
This new cycle flux cancels with the opposite
polarity polar field that were there at the start of the cycle and then
builds-up new polar fields with reversed polarity---thus completing
the magnetic cycle. 

While Babcock's model does help to explain many characteristics of the solar cycle,
it fails in other areas.
It does not explain why the sunspot zones drift toward the equator.
It assumes a highly simplified initial state.
It incorporates a meridional flow that does not agree with modern measurements.
It neglects the diffusive effects of the convective motions on the magnetic field
(convective motions that were unrecognized at the time).
Later dynamo models have gone on to include processes that help to explain
these other features but, almost without exception, these later models
have also faced observations that conflict with the models themselves.
The solar cycle remains one of the oldest and biggest unsolved problem in solar physics.

Here, we examine the nature of the solar cycle and the characteristics
that must be explained by any viable dynamo model.

\newpage 

\section{The Solar Cycle Discovered}
\label{sec:discovery}

Sunspots (dark patches on the Sun where intense magnetic fields loop up 
through the surface from the deep interior) were almost certainly seen by 
prehistoric humans viewing the Sun through hazy skies. The earliest actual 
recordings of sunspot observations were from China over 2000 years ago 
\citep{Clark:1978, Wittmann:1987}. Yet, the existence of 
spots on the Sun came as a surprise to westerners when telescopes were first 
used to observe the Sun in the early 17th century. This is usually 
attributed to western philosophy in which the heavens and the Sun were 
thought to be perfect and unblemished \citep[see][]{Bray:1965, Noyes:1982}.

The first mention of possible periodic behavior in sunspots came from 
Christian Horrebow, who wrote in his 1776 diary:

\begin{quote}
    Even though our observations conclude that changes of
    sunspots must be periodic, a precise order of regulation and
    appearance cannot be found in the years in which it was
    observed. That is because astronomers have not been making the
    effort to make observations of the subject of sunspots on a
    regular basis. Without a doubt, they believed that these
    observations were not of interest for either astronomy or
    physics. One can only hope that, with frequent observations of
    periodic motion of space objects, that time will show how to
    examine in which way astronomical bodies that are driven and lit
    up by the Sun are influenced by sunspots. \citep[][translation
    by Elke Willenberg]{Wolf:1877}
\end{quote}

\subsection{Schwabe's discovery}

Although Christian Horrebow mentions this possible periodic variation in 
1776 the solar (sunspot) cycle was not truly discovered until 1844. In that 
year Heinrich Schwabe reported in \textit{Astronomische Nachrichten} \citep{Schwabe:1844} 
that his observations of the numbers of sunspot groups and spotless days 
over the previous 18 years indicated the presence of a cycle of activity 
with a period of about 10 years. Figure~\ref{fig:Schwabe} shows his data for the number of 
sunspot groups observed yearly from 1826 to 1843. 

\epubtkImage{SchwabeData.png}{%
\begin{figure}[htbp]
\centerline{\includegraphics[width=0.8\textwidth]{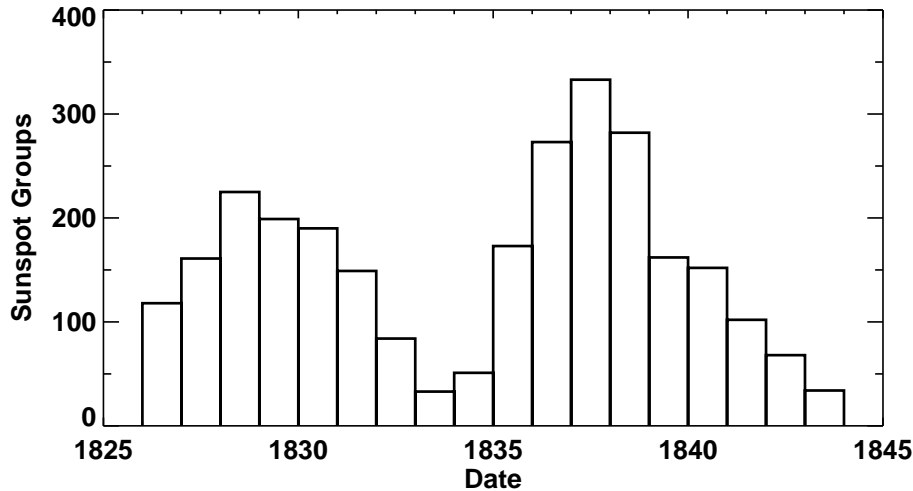}}
\caption{Sunspot groups observed each year from 1826 to 1843 by Heinrich 
Schwabe (\citeyear{Schwabe:1844}). These data led Schwabe to his discovery of the sunspot 
cycle.}
\label{fig:Schwabe}
\end{figure}}

\subsection{Wolf's relative sunspot number}

Schwabe's discovery was probably instrumental in initiating the work of 
Johann Rudolf Wolf (first at the Bern Observatory and later at Z\"{u}rich) toward 
acquiring daily observations of the Sun and extending the records to 
previous years \citep{Wolf:1861}. Wolf recognized that it was far easier to 
identify sunspot groups than to identify each individual sunspot. His 
``relative'' sunspot number, $R$, thus emphasized sunspot groups with
\begin{equation}
\label{eq1}
R = k\,(10\,g+n)
\end{equation}
where $k$ is a correction factor for the observer, $g$ is the number
of identified sunspot groups, and $n$ is the number of individual
sunspots. These Wolf, Z\"{u}rich, or International Sunspot Numbers
have been obtained daily since 1849.

Wolf instituted a system based on the use of a primary observer.
The sunspot number for the day was that found by the primary observer.
If the primary observer was unable to make a count then the count from a designated
secondary or tertiary observer was used instead.
Wolf himself was the primary
observer from 1849 to 1893 and had a personal correction factor,
$k=1.0$. 
He was followed by Alfred Wolfer from 1894 to 1926, William Otto Brunner from
1926 to 1944, and Max Waldmeier from 1945 to 1979.
Both Wolf and Wolfer observed the Sun in parallel over a 16-year period.
Wolfer counted more spots (different instruments were used and Wolf had a more
restrictive definition of what constituted a spot).
Thus, the $k$-factor for Wolfer (and subsequent primary observers) was set at $k=0.60$ by
comparing the sunspot numbers calculated by Wolfer to those calculated
by Wolf over the same days.

Beginning in 1981, and continuing through the present, the
International Sunspot Number has been provided by the Royal
Observatory of Belgium with S. Cortesi as the primary observer.
The process was changed from using the numbers from a single
primary/secondary/tertiary observer to using a weighted average of
many observers but with their $k$-factors tied to the primary observer.

\subsection{Wolf's reconstruction of earlier data}

Wolf himself extended the record back another 100 years using as primary
observers Staudacher from 1749 to 1787, Flaugergues from 1788 to 1825, 
and Schwabe from 1826 to 1847.
Although Wolf included many secondary observers, much of that earlier data is incomplete.
Wolf often filled in gaps in the sunspot observations using geomagnetic activity 
measurements as proxies for the sunspot number. 
The sunspot numbers are quite reliable since Wolf's time but those earlier 
numbers are far less reliable. The monthly averages of the daily numbers are 
shown in Figure~\ref{fig:MonthlySSN}.

\epubtkImage{MonthlySSN.png}{%
\begin{figure}[htbp]
\centerline{\includegraphics[width=\textwidth]{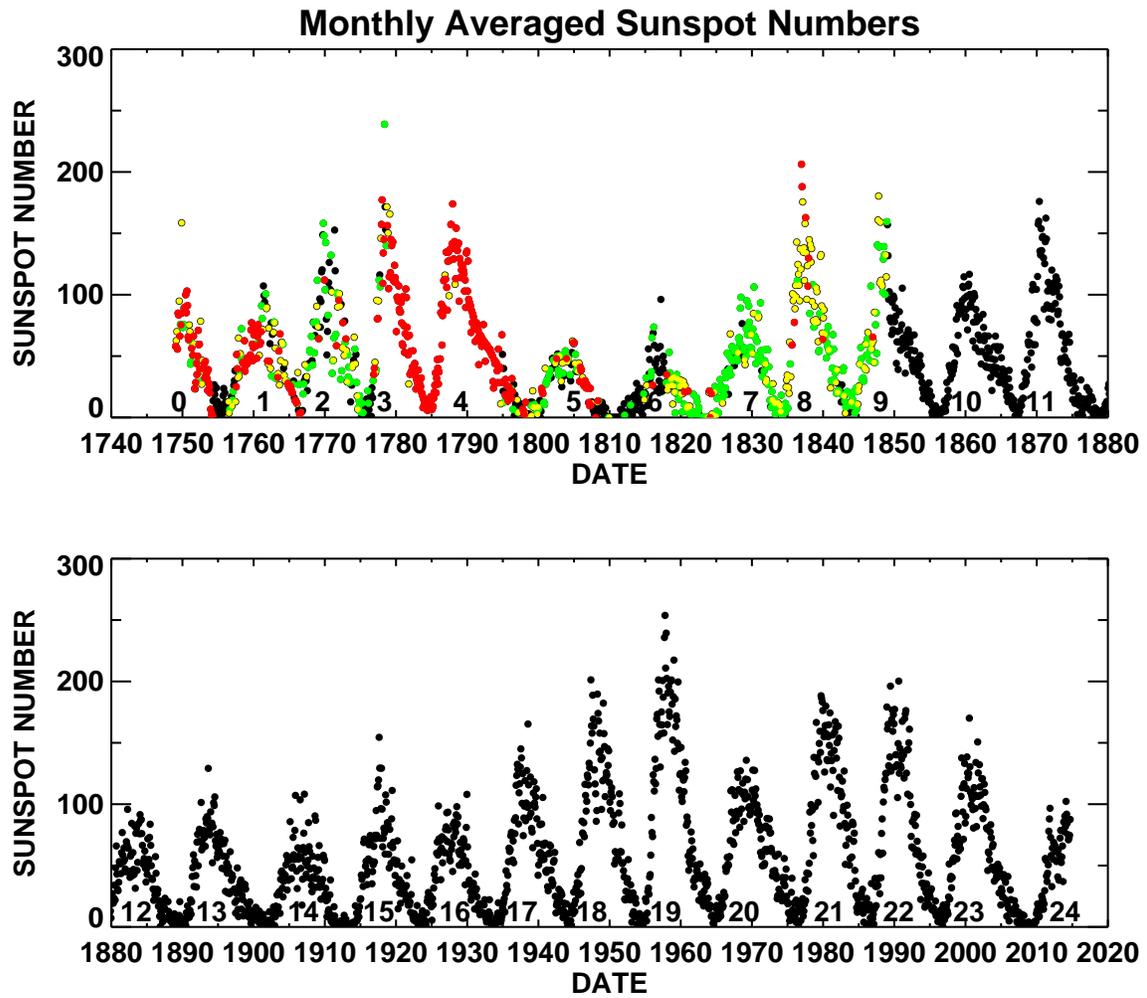}}
\caption{Monthly averages of the daily International Sunspot Number. This 
illustrates the solar cycle and shows that it varies in amplitude, shape, 
and length. Months with observations from every day are shown in black. 
Months with 1\,--\,10 days of observation missing are shown in green. Months with 
11\,--\,20 days of observation missing are shown in yellow. Months with more than 
20 days of observation missing are shown in red. [Missing days from 1818 to 
the present were obtained from the International daily sunspot numbers. 
Missing days from 1750 to 1818 were obtained from the Group Sunspot Numbers 
and probably represent an over estimate.]}
\label{fig:MonthlySSN}
\end{figure}}

\newpage 

\section{Solar Activity Data}
\label{sec:data}

\subsection{Sunspot numbers}

The International Sunspot Number, $R_I$, is the key indicator of solar activity. 
This is not because everyone agrees that it is the best indicator but rather 
because of the length of the available record. Traditionally, sunspot 
numbers are given as daily numbers, monthly averages, yearly averages, and 
smoothed numbers. The standard smoothing is a 13-month running mean centered 
on the month in question and using half-weights for the months at the start 
and end. Solar cycle maxima and minima are usually given in terms of these 
smoothed numbers.

Additional sunspot numbers do exist. The Boulder Sunspot Number is derived 
from the daily Solar Region Summaries \citep{NOAA:SRS} produced by the US Air Force and 
National Oceanic and Atmospheric Administration (USAF/NOAA) from sunspot 
drawings obtained from the Solar Optical Observing Network (SOON) sites 
since 1977. These summaries identify each sunspot group and list the number 
of spots in each group. The Boulder Sunspot Number is then obtained using 
Equation~(\ref{eq1}) with $k=1.0$. This Boulder Sunspot Number is typically about 55\% 
larger than the International Sunspot Number (corresponding to a correction 
factor $k=0.65$) but is available promptly on a daily basis, while the 
International Sunspot Number is posted monthly. The relationship between the 
smoothed Boulder and International Sunspot Number is shown in Figure~\ref{fig:BoulderVsInternational}. 

\epubtkImage{BoulderVsInternational.png}{%
\begin{figure}[htbp]
\centerline{\includegraphics[width=0.8\textwidth]{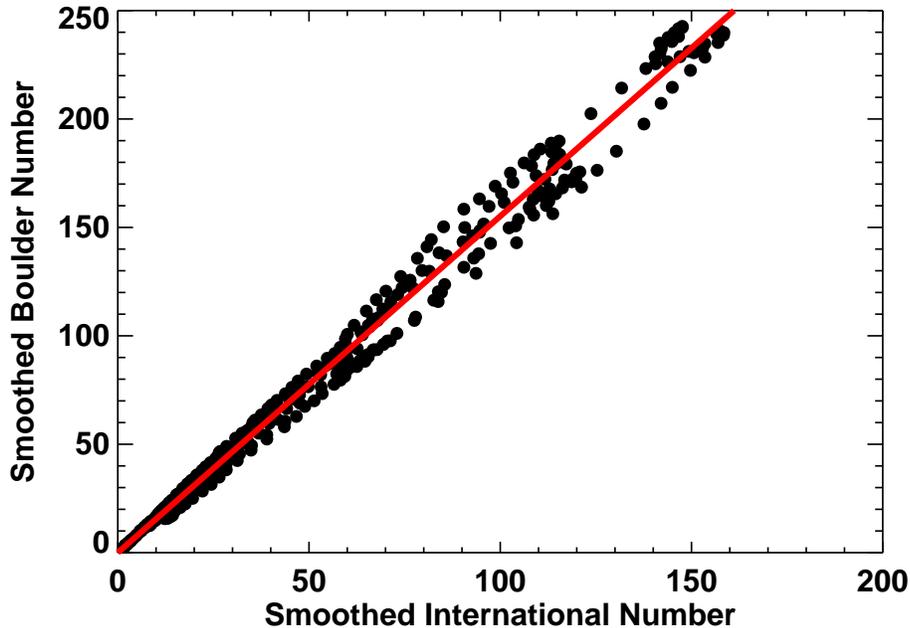}}
\caption{Boulder Sunspot Number vs.\ the International Sunspot Number at 
monthly intervals from 1981 to 2014. The average ratio of the two is 1.55 
and is represented by the solid line through the data points. The Boulder 
Sunspot Numbers can be brought into line with the International Sunspot 
Numbers by using a correction factor $k=0.65$ for Boulder.}
\label{fig:BoulderVsInternational}
\end{figure}}

A third sunspot number estimate is provided by the American Association of 
Variable Star Observers (AAVSO) and is usually referred to as the American 
Sunspot Number. These numbers are available from 1944 to the present. While 
the American Number occasionally deviates systematically from the 
International Number for years at a time, it is usually kept closer to the 
International Number than the Boulder Number through its use of correction 
factors. (The American Number is typically about 3\% lower than the 
International Number.) The relationship between the American and 
International Sunspot number is shown in Figure~\ref{fig:AAVSOvsInternational}.

\epubtkImage{AAVSOvsInternational.png}{%
\begin{figure}[htbp]
\centerline{\includegraphics[width=0.8\textwidth]{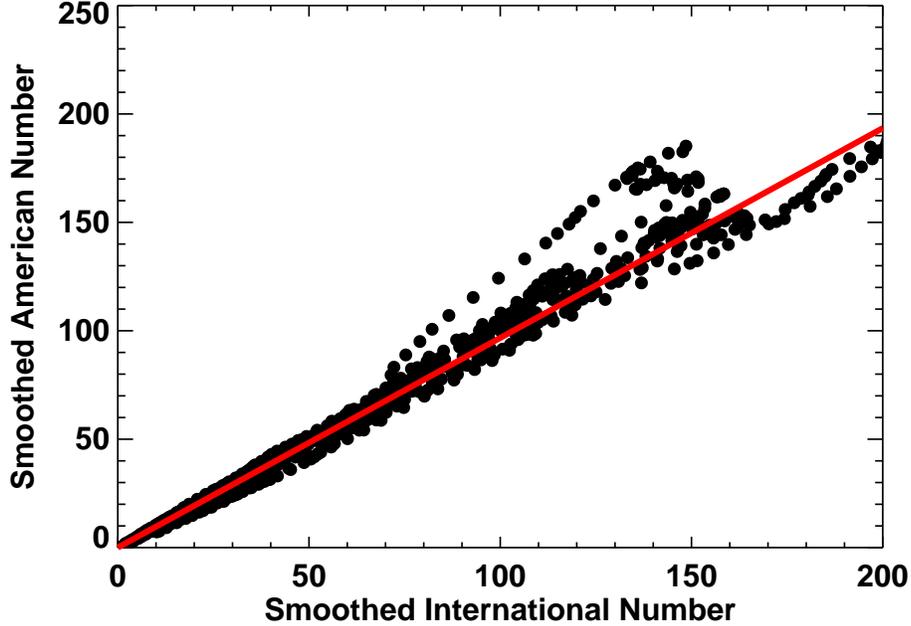}}
\caption{American Sunspot Number vs.\ the International Sunspot Number at 
monthly intervals from 1944 to 2014. The average ratio of the two is 0.97 
and is represented by the solid line through the data points.}
\label{fig:AAVSOvsInternational}
\end{figure}}

A fourth sunspot number is the Group Sunspot Number, $R_{G}$, devised
by \cite{Hoyt:1998}. This index counts only the number of sunspot
groups, averages together the observations from multiple observers
(rather than using the primary/secondary/tertiary observer system), and
normalizes the numbers to the International Sunspot Numbers using
\begin{equation}
\label{eq2}
R_G =\frac{12.08}{N}\sum\limits_{i=1}^N {k_i G_i } 
\end{equation}
where $N$ is the number of observers, $k_{i}$ is the $i$-th observer's correction 
factor, $G_{i}$ is the number of sunspot groups observed by observer $i$, and 
12.08 normalizes the number to the International Sunspot Number.
\cite{Hathaway:2002} found that the Group Sunspot Number follows the
International Number fairly closely but not to the extent that it
should supplant the International Number. In fact, the Group Sunspot
Numbers are not readily available after 1995. The primary utility of
the Group Sunspot number is in extending the sunspot number
observations back to the earliest telescopic observations in 1610. The
relationship between the Group and International Sunspot number is
shown in Figure~\ref{fig:GroupVsInternational} for the period 1874 to 1995. For this
period the numbers agree quite well, with the Group Number being about
1\% higher than the International Number. For earlier dates the Group
Number is a significant 24\% lower than the International Number.

\epubtkImage{GroupVsInternational.png}{%
\begin{figure}[htbp]
\centerline{\includegraphics[width=0.8\textwidth]{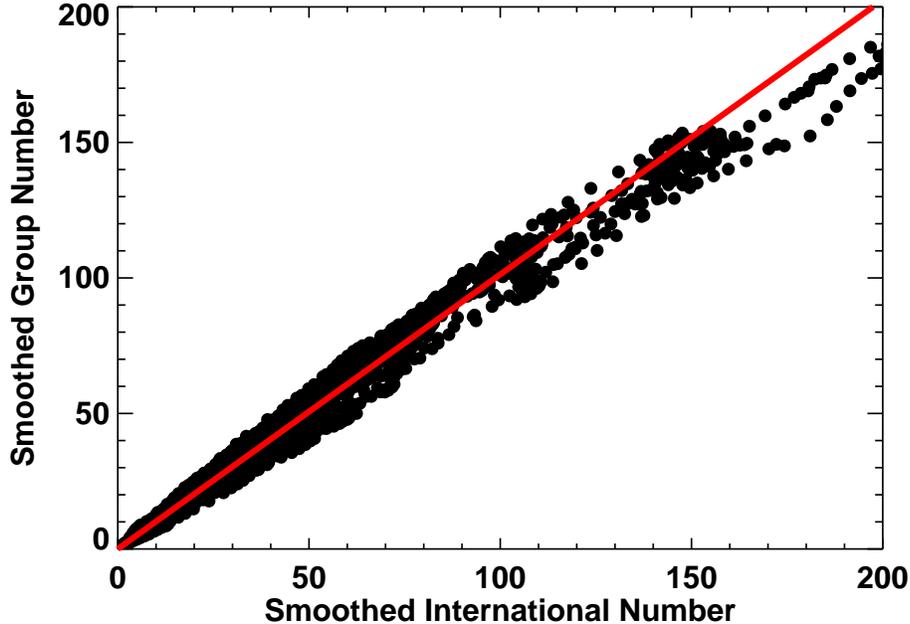}}
\caption{Group Sunspot Number vs.\ the International Sunspot Number at 
monthly intervals from 1874 to 1995. The average ratio of the two is 1.01 
and is represented by the solid line through the data points.}
\label{fig:GroupVsInternational}
\end{figure}}

These sunspot numbers are available from \cite{NOAA}. The
International Number can be obtained monthly directly from
\cite{SILSO}.

\subsection{Revised sunspot numbers}

As noted earlier, Wolf revised his sunspot numbers based on new information.
The previous sections also show that the sunspot number for a given day, month, or year,
can vary substantially depending on the source.
There is now good evidence that even the modern record (1849 to the present)
may need substantial revision.
\citet{Svalgaard:2013} has noted that when Waldmeier became the primary observer in 1946
he changed the way the sunspot number was calculated \citep{Waldmeier:1968}.
Instead of counting each spot within a group once, he gave greater counts (2, 3, or 5) to larger spots.
While this change went largely unnoticed by the community, the practice has continued
up to the present with the numbers provided by SILSO.
By comparing sunspot number counts with and without this weighting, \citet{Svalgaard:2013}
estimates that the modern sunspot numbers since 1946 have been inflated by about 20\%.

The earlier sunspot numbers, of course, have always been considered much less reliable.
There are many days (or even months) without any reported observations prior to 1849.
Even when observations are reported it can be difficult to determine a sunspot number from the reports.
This has led to divergent sunspot numbers for earlier times, as can be seen in Figure~\ref{fig:SSNyearlyRevised}.

\epubtkImage{SSNyearlyRevised.png}{%
\begin{figure}[htbp]
\centerline{\includegraphics[width=0.8\textwidth]{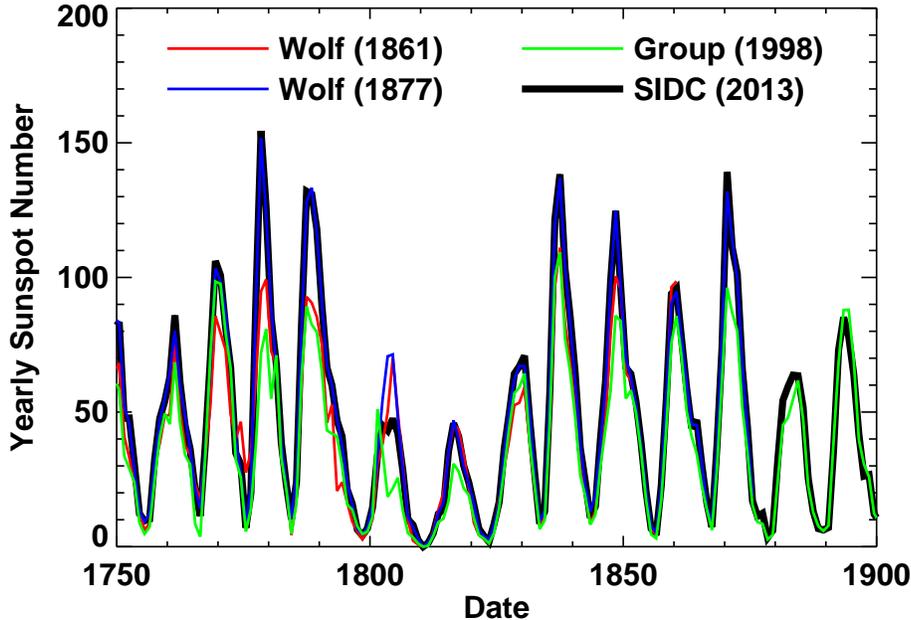}}
\caption{Sunspot number revisions.
Yearly sunspot numbers as reported by \cite{Wolf:1861} (red line), \cite{Wolf:1877b} (blue),
\citet{Hoyt:1998} (green), and by SILSO in 2013 (black).
These sunspot numbers have disagreements as late as 1900.
}
\label{fig:SSNyearlyRevised}
\end{figure}}

Since the Group Sunspot Number work of \cite{Hoyt:1998}, new observations have come to light.
For example, \cite{Vaquero:2007B, Vaquero:2007A} and \cite{Arlt:2008} have uncovered previously unknown 18th century observations that indicate possible changes to the Group Sunspot Number in the late 18th century, shortly after the Maunder Minimum.
\cite{Vaquero:2011} also found observations by G. Marcgraf for the critical years of 1636-1642. 
These observations suggest that the amplitude of the sunspot cycle just prior to the onset of the Maunder Minimum was much smaller than that given by the Group Sunspot Number (20 vs. 60-70 sunspots at maximum).

New analysis methods \citep[e.g.][]{Pop:2012, Arlt:2013, Leussu:2013} have also been developed and these continue to reveal errors and inconsistencies in the various sunspot number records.
As this is being written, there is a significant effort within the solar physics community to reconcile the differences in the sunspot numbers and to provide a more reliable sunspot record (with error estimates) from 1610 to the present.
Any revisions can have far-reaching impact on other areas.
Sunspot numbers are used to estimate the Sun's contribution to climate change \citep[e.g.][]{Lean:2008} and to the modulation of galactic cosmic rays and the radioisotopes they produce in Earth's atmosphere \citep[e.g.][]{Usoskin:2013}.

\subsection{Sunspot areas}

Sunspot areas are thought to be more physical measures of solar activity. 
Sunspot areas and positions were diligently recorded by the Royal 
Observatory, Greenwich (RGO) from May 1874 to the end of 1976 using 
measurements from photographic plates obtained from RGO itself and its 
sister observatories in Cape Town, South Africa, Kodaikanal, India, and 
Mauritius. Both umbral areas and whole spot areas were measured and 
corrected for foreshortening on the visible disc. Sunspot areas were given 
in units of millionths of a solar hemisphere ({\textmu}Hem). Comparing the 
corrected whole spot areas to the International Sunspot Number (Figure~\ref{fig:RGOareaVsInternational}) 
shows that the two quantities are indeed highly correlated ($r=0.994$, 
$r^{2}=0.988$). Furthermore, there is no evidence for any lead or lag 
between the two quantities over each solar cycle. Both measures could almost 
be used interchangeably except for one aspect---the zero point. Since a 
single, solitary sunspot gives a sunspot number of 11 (6.6 for a correction 
factor $k=0.6$) the zero point for the sunspot number is shifted slightly from 
zero. The best fit to the data shown in Figure~\ref{fig:RGOareaVsInternational} gives an offset of about 4 
and a slope of 16.7.

\epubtkImage{RGOareaVsInternational.png}{%
\begin{figure}[htbp]
\centerline{\includegraphics[width=0.8\textwidth]{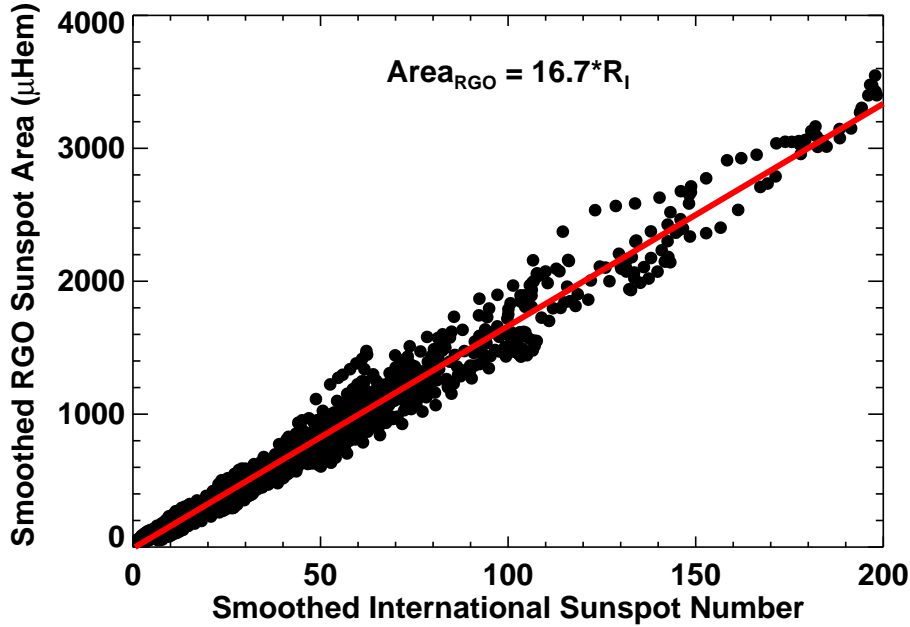}}
\caption{Smoothed RGO Sunspot Area vs.\ the International Sunspot Number at monthly 
intervals from May 1874 to December 1976. The two quantities are correlated at 
the 99.4\% level with a proportionality constant of about 16.7.}
\label{fig:RGOareaVsInternational}
\end{figure}}

In 1977, NOAA began reporting much of the same sunspot area and
position information in its Solar Region Summary reports. These
reports are derived from measurements taken from sunspot drawings done
at the USAF SOON sites. The sunspot areas were initially estimated by
overlaying a grid and counting the number of cells that a sunspot
covered. In late 1981, this procedure was changed to employ an overlay
with a number of circles and ellipses with different areas. The
sunspot areas reported by USAF/NOAA are significantly smaller than
those from RGO \citep{Fligge:1997, Baranyi:2001, Hathaway:2002,
Balmaceda:2009}.
Figure~\ref{fig:USAFareaVsInternational} shows the relationship between
the USAF/NOAA sunspot areas and the International Sunspot Number. The
slope in the straight line fit through the data is 11.22,
significantly less than that found for the RGO sunspot areas. This
indicates that these later sunspot area measurements should be
multiplied by 1.49 to be consistent with the earlier RGO sunspot
areas. The combined RGO USAF/NOAA datasets are available online \citep{RGO}.

The source of this substantial (40-50\%) difference in reported sunspot areas
is still uncertain. Sunspot area measurements using the SOHO/MDI intensity images
confirm that the error lies with the USAF/NOAA data.
While the measurements methods are clearly different (counting squares or pixels vs. selecting the appropriate ellipse) and the images are clearly different (photographic
plates or CCD images vs. drawings) it is not clear that this would give an underestimate
with the USAF/NOAA method.
\cite{Foukal:2013} has suggested that the source of the error is in the small spots
that appear as single dots with a pencil on the USAF drawings.
He argues that these are more accurately recorded by RGO and others using photographs
or CCD image and that the large number of such spots can account for the size of the
underestimation by USAF/NOAA.

\epubtkImage{USAFareaVsInternational.png}{%
\begin{figure}[htbp]
\centerline{\includegraphics[width=0.8\textwidth]{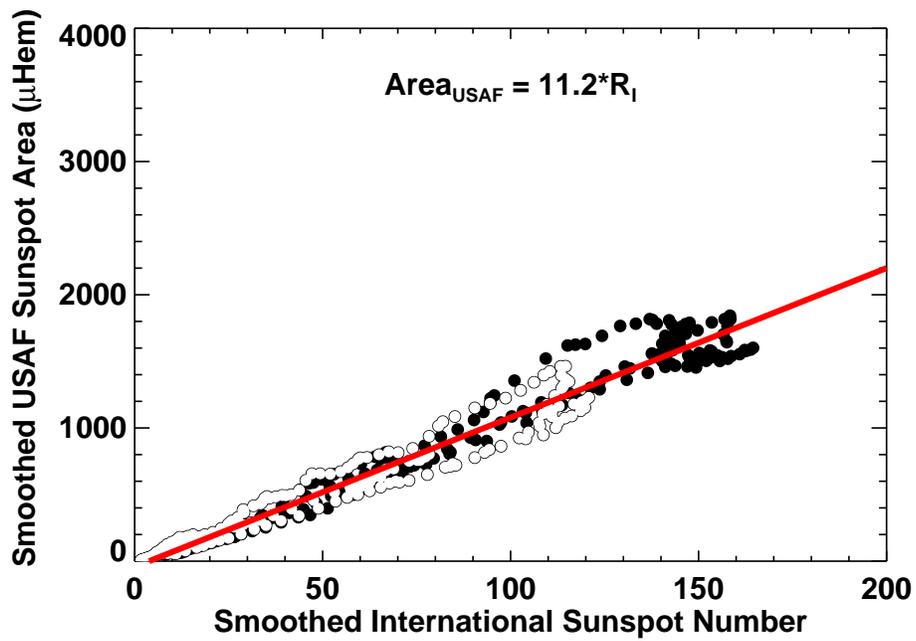}}
\caption{Smoothed USAF/NOAA Sunspot Area vs.\ the International Sunspot Number at 
monthly intervals from January 1977 to August 2014. The two quantities are correlated at 
the 99.1\% level with a proportionality constant of about 11.2. These 
sunspot areas have to be multiplied by a factor of 1.49 to bring them into line 
with the RGO sunspot areas.
Data obtained prior to cycle~23 are shown with filled dots, while data obtained
after 1997 are shown with open circles.}
\label{fig:USAFareaVsInternational}
\end{figure}}

Sunspot areas are also available from a number of other solar observatories, 
with links to much of that data available at \cite{NOAA:NGDC}.
While individual observatories have data gaps, their data are 
very useful for helping to maintain consistency over the full interval from 
1874 to the present.
Many of these observatories (notably \cite{Debrecen}) provide images in white light, Calcium K, and/or magnetic field as well.

These datasets have additional information that is not reflected in sunspot 
numbers---positional information---both latitude and longitude. The 
distribution of sunspot area with latitude (Figure~\ref{fig:ButterflyDiagram})
shows that sunspots appear in two bands on either side of the Sun's equator.
At the start of each cycle spots appear at latitudes above about 20\,--\,25\textdegree.
As the cycle progresses the range of latitudes with sunspots broadens and the central 
latitude slowly drifts toward the equator, but with a zone of avoidance near 
the equator. This behavior is referred to as ``Sp\"{o}rer's Law of Zones'' 
by \cite{Maunder:1903} and was famously illustrated by his ``Butterfly
Diagram'' \citep{Maunder:1904}.

\epubtkImage{ButterflyDiagram.png}{%
\begin{figure}[htbp]
\centerline{\includegraphics[angle=90, width=\textwidth]{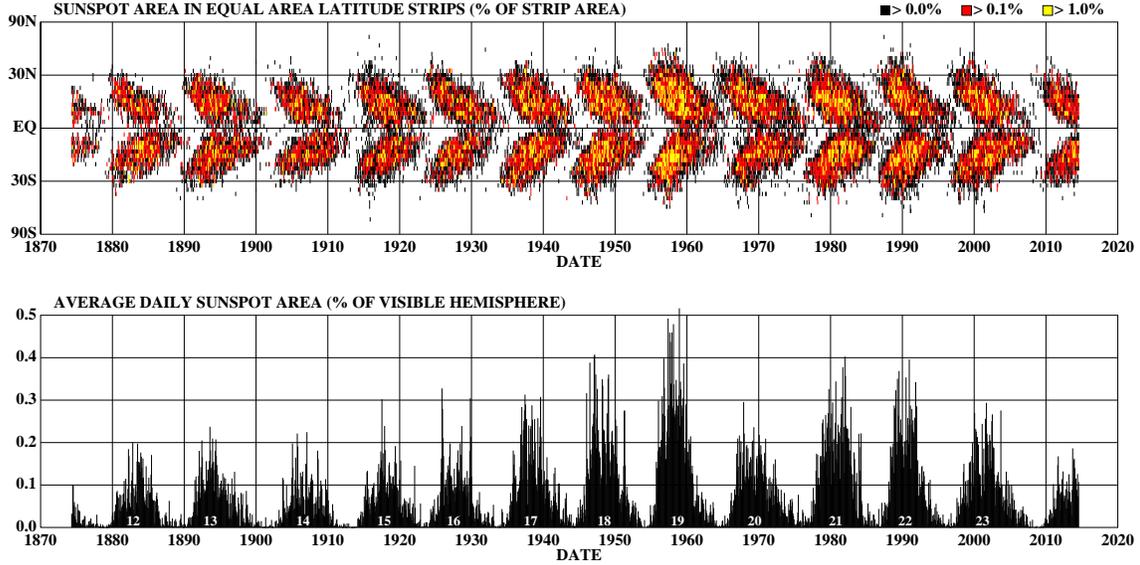}}
\caption{Sunspot area as a function of latitude and time. The average daily 
sunspot area for each solar rotation since May 1874 is plotted as a function 
of time in the lower panel. The relative area in equal area latitude strips 
is illustrated with a color code in the upper panel. Sunspots form in two 
bands, one in each hemisphere, which start at about 25\textdegree\  from the equator 
at the start of a cycle and migrate toward the equator as the cycle 
progresses.}
\label{fig:ButterflyDiagram}
\end{figure}}

\subsection{10.7~cm solar flux}

The 10.7~cm Solar Flux is the disc-integrated emission from the Sun at the 
radio wavelength of 10.7~cm (2800~MHz) \citep[see][]{Tapping:1994}. 
This measure of solar activity has advantages over sunspot numbers and areas 
in that it is completely objective and can be made under virtually all 
weather conditions. Measurements of this flux have been taken daily by the 
Canadian Solar Radio Monitoring Programme since 1946. Several measurements 
are taken each day and care is taken to avoid reporting values influenced by 
flaring activity. Observations were made in the Ottawa area from 1946 to 
1990. In 1990, a new flux monitor was installed at Penticton, British 
Columbia and run in parallel with the Ottawa monitor for six months before 
moving the Ottawa monitor itself to Penticton as a back-up. Measurements are 
provided daily \citep{DRAO} and the full dataset is archived \citep{Penticton}.

The relationship between the 10.7~cm radio flux and the International 
Sunspot Number is somewhat more complicated than that for sunspot area. 
First of all, the 10.7~cm radio flux has a base level of about 67 solar flux 
units. Secondly, the slope of the relationship changes as the sunspot number 
increases up to about 30. This is captured in a formula given by
\cite{Holland:1984} as:
\begin{equation}
\label{eq3}
F_{10.7} = 67+0.97\,R_I +17.6\left( {e^{-0.035\,R_I }-1} \right)
\end{equation}
In addition to this slightly nonlinear relationship there is evidence that 
the 10.7~cm radio flux lags behind the sunspot number by about one month 
 \citep{Bachmann:1994}.

\epubtkImage{F107VsInternational.png}{%
\begin{figure}[htbp]
\centerline{\includegraphics[width=0.8\textwidth]{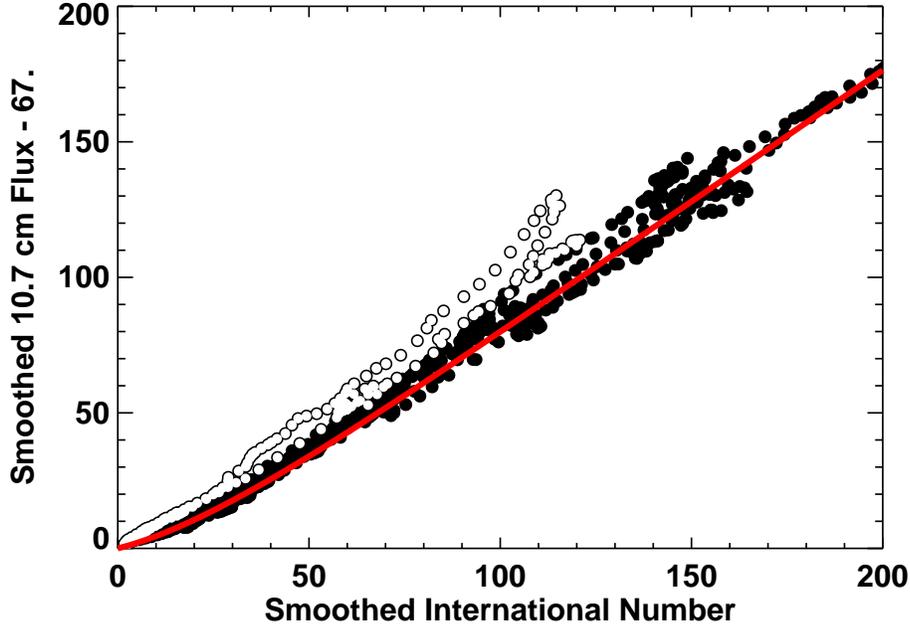}}
\caption{10.7cm radio flux vs.\ International Sunspot Number for the period 
of August 1947 to January 2014. Data obtained prior to cycle~23 are shown 
with filled dots while data obtained after 1997 are shown with open 
circles. The Holland and Vaughn formula relating the radio flux to the 
sunspot number is shown with the solid line. These two quantities are 
correlated at the 99.5\% level.}
\label{fig:F107VsInternational}
\end{figure}}

Figure~\ref{fig:F107VsInternational} shows the relationship between
the 10.7~cm radio flux and the International Sunspot Number.
The two measures are highly correlated ($r=0.995$, $r^{2}=0.990$).
The Holland and Vaughn formula fits the early data quite well.
However, the data after 1997 lies systematically higher than the levels
given by the Holland and Vaughn formula.
Speculation concerning the cause of this change is discussed in
Section~\ref{sec:Cycle23/24}.

\subsection{Total irradiance}

The Total Solar Irradiance (TSI) is the radiant energy emitted by the Sun at 
all wavelengths crossing a square meter each second outside Earth's 
atmosphere. Although ground-based measurements of this ``solar constant'' 
and its variability were made decades ago \citep{Abbot:1913}, 
accurate measurements of the Sun's total irradiance have only become 
available since our access to space. Several satellites have carried instruments 
designed to make these measurements: Nimbus-7 from November 1978 to 
December 1993; the Solar Maximum Mission (SMM) ACRIM-I from February 1980 
to June 1989; the Earth Radiation Budget Satellite (ERBS) from October 
1984 to December 1995; NOAA-9 from January 1985 to December 1989; NOAA-10 
from October 1986 to April 1987; Upper Atmosphere Research Satellite (UARS) 
ACRIM-II from October 1991 to November 2001; ACRIMSAT ACRIM-III from 
December 1999 to the present; SOHO/VIRGO from January 1996 to the present; 
and SORCE/TIM from January 2003 to the present.

\epubtkImage{TSI_Data.png}{%
\begin{figure}[htbp]
\centerline{\includegraphics[width=0.8\textwidth]{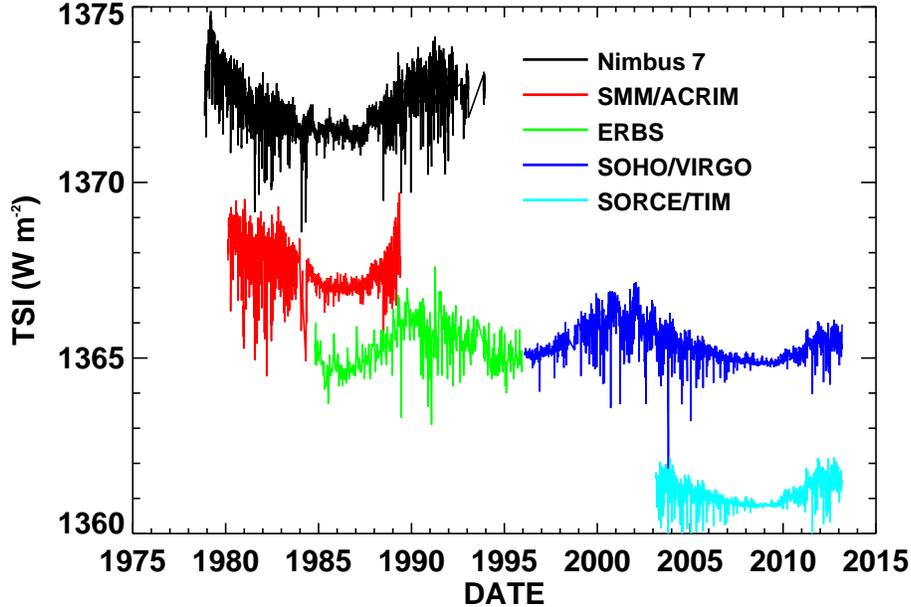}}
\caption{Daily measurements of the Total Solar Irradiance (TSI) from 
instruments on different satellites. The systematic offsets among 
measurements taken with different instruments complicate determinations of 
the long-term behavior.}
\label{fig:TSI_Data}
\end{figure}}

While each of these instruments is extremely precise in its measurements, 
their absolute accuracies vary in ways that make some important
aspects of the TSI subjects of controversy. Figure~\ref{fig:TSI_Data} shows
daily measurements of TSI from some of these instruments. Each
instrument measures the drops in TSI due to the formation and disc
passages of large sunspot groups as well as the general rise and fall
of TSI with the sunspot cycle \citep{Willson:1988}. However, there are
significant offsets among the absolute measured
values. Intercomparisons of the data have lead to different
conclusions. \cite{Willson:1997} combined the SMM/ACRIM-I data with
the later UARS/ACRIM-II data by using intercomparisons with 
Nimbus-7 and ERBS and concluded that the Sun was brighter by about
0.04\% during the cycle~22 minimum than it was during the cycle~21
minimum.  \cite{Frohlich:1998} constructed a composite (the PMOD
composite) that includes Nimbus-7, ERBS, SMM/ACRIM-I, UARS/ACRIM-II, and SOHO/VIRGO,
which does not show this increase.

This situation has not improved with the addition of data from the decline of
cycle~23 and the extraordinary cycle 23/24 minimum.
\cite{Frohlich:2013} found that the PMOD composite irradiance dropped well
below the lowest values seen at the previous two minima.
\cite{Scafetta:2014} found that the ACRIM composite irradiance at cycle 23/24
minimum was intermediate between the values seen at the previous two minima.

Comparing the PMOD composite to the sunspot number (Figure~\ref{fig:TSI_PMODvsInternational}) shows a strong correlation between the two quantities but with different behavior during 
cycle~23 (the VIRGO era). At its peak, cycle~23 had sunspot numbers about 20\% smaller 
than cycle~21 or 22. However, the cycle~23 peak PMOD composite TSI was 
similar to that of cycles~21 and 22. This behavior is similar to that seen 
in the 10.7~cm flux in Figure~\ref{fig:F107VsInternational},
but is complicated by the fact that the cycle~23 PMOD composite falls well below that for cycle~21 and 22 during the decline 
of cycle~23 toward minimum, while the 10.7~cm flux remained above the 
corresponding levels for cycles~21 and 22.

\epubtkImage{TSI_PMODvsInternational.png}{%
\begin{figure}[htbp]
\centerline{\includegraphics[width=0.8\textwidth]{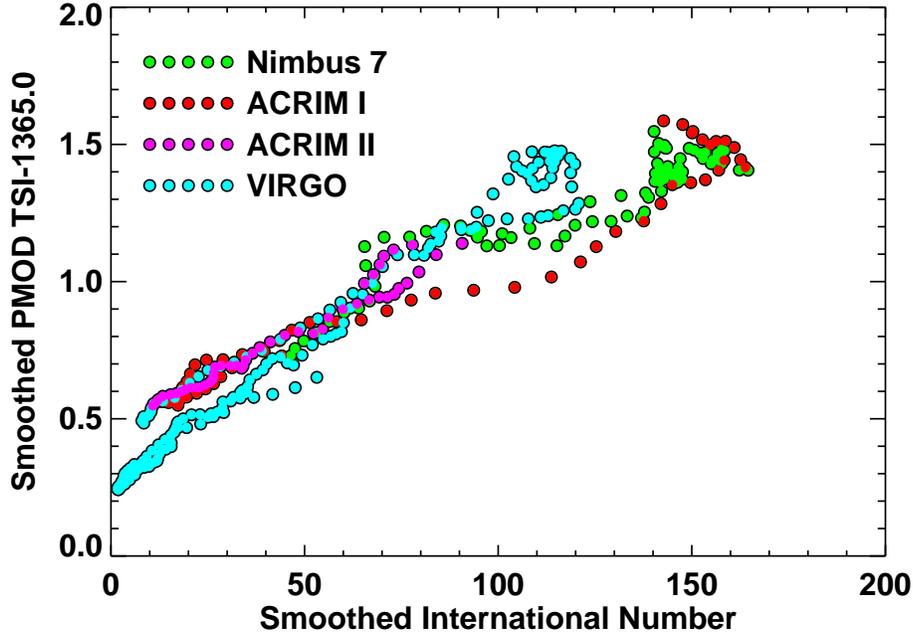}}
\caption{The PMOD (version d41-62-1204) composite TSI vs.\ International Sunspot Number.
The filled circles represent smoothed monthly averages, with different colors representing
data from the different instruments.}
\label{fig:TSI_PMODvsInternational}
\end{figure}}

Comparing the ACRIM composite to the sunspot number (Figure~\ref{fig:TSI_ACRIMvsInternational}) shows a much weaker correlation between the two quantities.
While the tendency for TSI to increase with sunspot number is evident in some intervals
(during an individual cycle's rising and fall phases), any simple proportionality
to sunspot number appears less likely.
The unresolved differences in the TSI measurements make further conclusions difficult.
Both composites indicate that either further adjustments need to be made to
the measurements or the Sun's irradiance is not tied solely to magnetic features.

\epubtkImage{TSI_ACRIMvsInternational.png}{%
\begin{figure}[htbp]
\centerline{\includegraphics[width=0.8\textwidth]{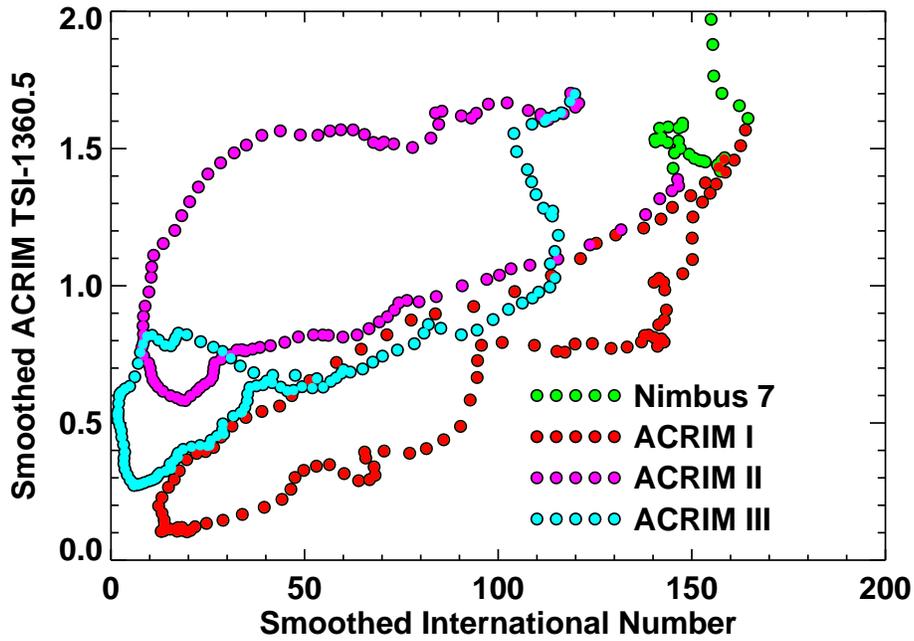}}
\caption{The ACRIM composite TSI vs.\ International Sunspot Number.
The filled circles represent smoothed monthly averages with different colors representing
data from the different instruments.}
\label{fig:TSI_ACRIMvsInternational}
\end{figure}}

\subsection{Magnetic field}
\label{sec:MagneticField}

Magnetic fields on the Sun were first measured in sunspots by
\cite{Hale:1908}. The magnetic nature of the solar cycle became
apparent once these observations extended over more than a single
cycle \citep{Hale:1919}.
While it is now well recognized that the solar cycle is best represented in
terms of the magnetic field itself, systematic daily observations are only available
starting in the 1970s and thus only characterize the last three-and-a-half
solar cycles. Nonetheless, a number of key characteristics were clear from
even the first observations.

\cite{Hale:1919} noted ``Hale's Polarity Laws'' for sunspots (illustrated in Figure~\ref{fig:HalesLaw}):

\begin{quote}
    \ldots the preceding and following spots of binary
    groups, with few exceptions, are of opposite polarity, and that
    the corresponding spots of such groups in the Northern and
    Southern hemispheres are also of opposite sign. Furthermore, the
    spots of the present cycle are opposite in polarity to those of
    the last cycle.
\end{quote}

\epubtkImage{HalesLaw.png}{%
\begin{figure}[htbp]
\centerline{\includegraphics[width=0.8\textwidth]{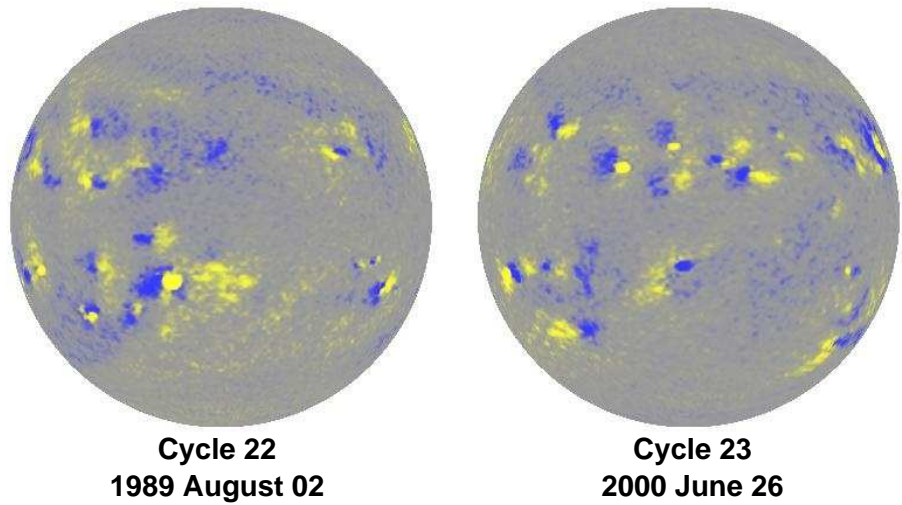}}
\caption{Hale's Polarity Laws. A magnetogram from sunspot cycle~22 (1989 
August 2) is shown on the left, with yellow denoting positive polarity and 
blue denoting negative polarity. A corresponding magnetogram from sunspot 
cycle~23 (2000 June 26) is shown on the right. Leading spots in one 
hemisphere have opposite magnetic polarity to those in the other hemisphere 
and the polarities flip from one cycle to the next.}
\label{fig:HalesLaw}
\end{figure}}

In addition to Hale's Polarity Laws for the changing polarity of sunspots,
it was found that the Sun's polar fields changed polarity as well.
\cite{Babcock:1958} noted that the Sun's south 
polar field reversed in mid-1957.
A year later, \cite{Babcock:1959} reported that the north polar field had reversed in 
late-1958 and suggested that these field reversals occur systematically at about the
time of cycle maximum (the maximum for cycle~19 occurred in late-1957). The polar fields 
are thus out of phase with the sunspot cycle---polar fields are at their 
peak near sunspot cycle minima.

The polar fields have been measured almost daily from the Wilcox Solar Observatory
at Stanford University since the mid 1970s \citep{Scherrer:1977}.
While the measurements have a very coarse spatial resolution,
great care has been taken to account for scattered light and other instrumental effects.
Their smoothed polar field strengths are shown in Figure~\ref{fig:PolarFields}, along
with the sunspot number for reference.
The polar fields reach their peaks late in each cycle at about the time of cycle minimum,
and the fields reverse polarity at about the time of cycle maximum.
It is also clear that the polar fields vary in strength from cycle to cycle.

\epubtkImage{PolarFields.png}{%
\begin{figure}[htbp]
\centerline{\includegraphics[width=0.8\textwidth]{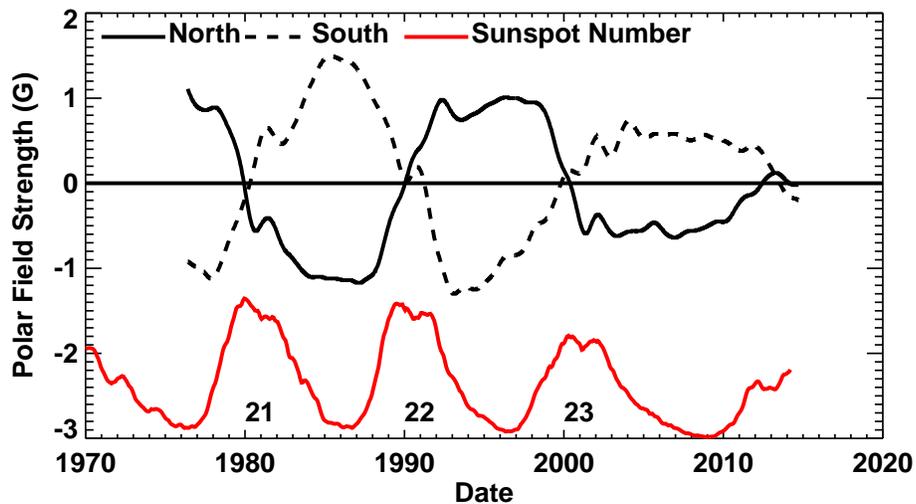}}
\caption{The Sun's polar fields as reported by the Wilcox Solar Observatory.
The smoothed field strength in their northernmost pixel is shown with the solid black line.
The smoothed field strength in their southernmost pixel is shown with the dashed line.
The smoothed sunspot number (scaled to fit on the figure) is shown with the red line.}
\label{fig:PolarFields}
\end{figure}}

Systematic, high-resolution, daily observations of the Sun's magnetic field over the visible 
solar disc were initiated at the Kitt Peak National Observatory in the early 
1970s. Synoptic maps from these measurements are nearly continuous from 
early-1975 through mid-2003. Shortly thereafter, similar (and even higher 
resolution) data became available from the National Solar Observatory (NSO) 
Synoptic Optical Long-term Investigations of the Sun (SOLIS) facility 
\citep{Keller:1998}. Gaps between these two datasets and within the SOLIS 
dataset can be filled with data from the Michelson Doppler Imager (MDI) on 
the Solar and Heliospheric Observatory (SOHO) mission
\citep{Scherrer:1995}. These synoptic maps are presented in an
animation here (Figure~\ref{mov1}).

\epubtkMovie{MagnetogramMovie2009.avi}{MagnetogramMovie.png}{%
\begin{figure}[htbp]
\centerline{\includegraphics[width=0.6\textwidth]{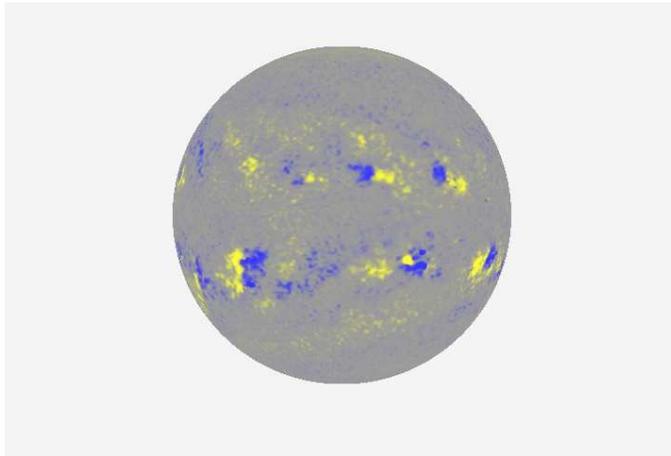}}
\caption{A full-disc magnetogram from NSO/Kitt Peak used in constructing
  magnetic synoptic maps over the last two sunspot cycles. Yellow
  represents magnetic field directed outward. Blue represents magnetic
  field directed inward.}
\label{mov1}
\end{figure}}

The radial magnetic fields from these synoptic maps are averaged over longitude for
each solar rotation to produce a ``Magnetic Butterfly Diagram,'' as
shown in Figure~\ref{fig:MagBfly}. 
In addition to showing the 11-year cycle, the equatorward drift of the sunspot zones,
and the overlapping cycles at minimum, this Magnetic Butterfly Diagram also exhibits
Hale's Polarity Laws, the polar field reversals, and ``Joy's Law''
\citep{Hale:1919}:

\begin{quote}
 The following spot of the pair tends to appear farther from the
 equator than the preceding spot, and the higher the latitude, the
 greater is the inclination of the axis to the equator.
\end{quote}
Joy's Law and Hale's Polarity Laws are apparent in
the ``butterfly wings.'' The equatorial sides of these wings are
dominated by the lower latitude, preceding-spot polarities, while the
poleward sides are dominated by the higher latitude,
following-spot polarities. These polarities are opposite in
opposing hemispheres and from one cycle to the next (Hale's Law). This
figure also shows that the higher latitude fields are transported
toward the poles where they eventually reverse the polar field at
about the time of sunspot cycle maximum.
The number of key characteristics of the solar cycle that are evident within
Figure~\ref{fig:MagBfly} make it a litmus test for dynamo theories.

\epubtkImage{MagBfly.jpg}{%
\begin{figure}[htbp]
\centerline{\includegraphics[width=\textwidth]{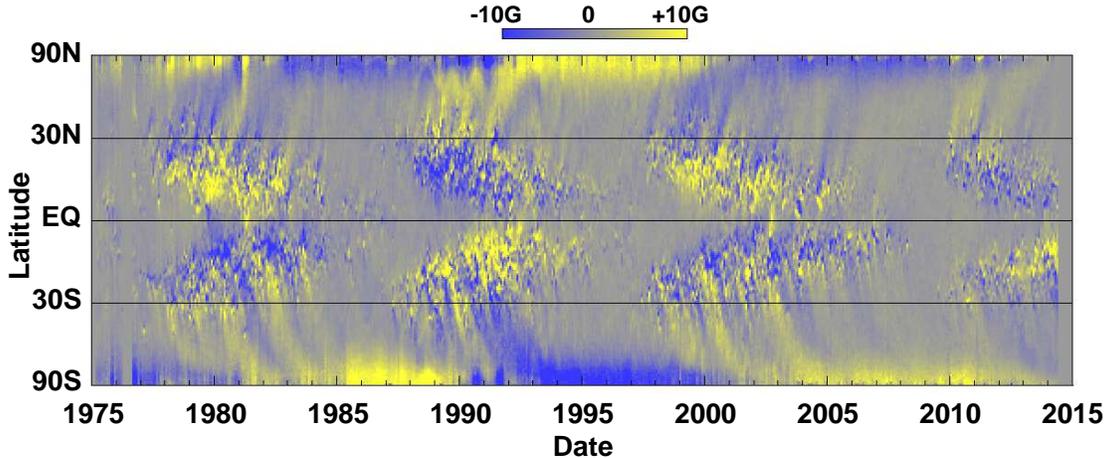}}
\caption{A Magnetic Butterfly Diagram constructed from the longitudinally 
averaged radial magnetic field obtained from instruments on Kitt Peak and 
SOHO. This illustrates Hale's Polarity Laws, Joy's Law, polar field 
reversals, and the transport of higher latitude magnetic field elements 
toward the poles.}
\label{fig:MagBfly}
\end{figure}}

\subsection{Ephemeral Regions}
\label{sec:EMR}

Ephemeral Regions are small ($<30$Mm across) bipolar magnetic regions that are typically observed
for only 1-2 days and usually do not produce sunspots.
They appear to extend the spectrum of the bipolar active regions that typically \textit{do} produce sunspots \citep[see][]{Parnell:2009} to smaller sizes.
They were first mentioned by \cite{Dodson:1953} in reference to a small, short-lived, high-latitude sunspot.

\cite{Harvey:1973} examined magnetograms and H$\alpha$ images acquired over three 4-day intervals in 1970 and 1971 (after the peak of cycle 20) and concluded that as many as 100
ephemeral regions may erupt per day with as much total magnetic flux as erupts in the larger
active regions.
They also noted that the distribution in latitude was broader than that of the
active regions and suggested that the occurrence of ephemeral regions did not
vary with the sunspot cycle.
Later, however, in a larger study extending from 1970 to 1973, \cite{Harvey:1975}
\textit{did} find a direct solar cycle dependence.
They also found that while the spatial orientation was almost random, there was a small excess of new cycle orientations at the high latitudes in 1973.

A solar cycle dependence for the number of ephemeral regions was also found by
\cite{Martin:1979} but with a slight shift in phase due to the early appearance
of new cycle ephemeral regions.
Their observations also led to the conclusion that there was more overlap between
solar cycles than is seen in sunspots alone (see Section~\ref{sec:ExtendedCycle} on
the extended solar cycle).

The small sizes and short lifetimes of ephemeral regions made these early observations
with ground-based magnetographs quite difficult.
This situation was greatly improved with the advent of space-based magnetographs.
\cite{Hagenaar:2001} studied the properties of ephemeral regions using the
SOHO/MDI instrument and found far more (smaller) ephemeral regions with a
rate of emergence sufficient to replace the quiet Sun magnetic field in just 14 hours.

\cite{Hagenaar:2003} extended these observations to include the rise from cycle
minimum in 1996 to maximum in 2001 and found that the number of the small
ephemeral regions varied in {\em anti}-phase with the sunspot cycle.
Later studies \citep{Abramenko:2006,Hagenaar:2008} found that fewer
ephemeral regions emerge in unipolar regions (coronal holes).
This might explain some of the cycle dependence since more unipolar regions
are found at cycle maximum in the studied area (within 60\textdegree of disc center).

\subsection{Flares and coronal mass ejections}

\cite{Carrington:1859} and \cite{Hodgson:1859} reported the first
observations of a solar flare from white-light observations on
September 1, 1859. While observing the Sun projected onto a viewing
screen, Carrington noticed a brightening that lasted for about 5
minutes. Hodgson also noted a nearly simultaneous geomagnetic
disturbance. Since that time, flares have been observed in H-alpha from
many ground-based observatories and characterizations of flares from
these observations have been made \citep[e.g.][]{Benz:2008}.

X-rays from the Sun were measured by instruments on early rocket flights and 
their association with solar flares was recognized immediately. NOAA has 
flown solar X-ray monitors on its Geostationary Operational Environmental 
Satellites (GOES) since 1975 as part of its Space Environment Monitor. The 
solar X-ray flux has been measured in two bandpasses by these instruments: 
0.5 to 4.0~{\AA}\ and 1.0 to 8.0~{\AA}. The X-ray flux is given on a 
logarithmic scale with A and B levels as typical background levels (depending 
upon the phase of the cycle), and C, M, and X levels indicating increasing 
levels of flaring activity. The number of M-class and X-class flares seen in 
the 1.0\,--\,8.0~{\AA}\ band tends to follow the sunspot number, as
shown in Figure~\ref{fig:FlaresVsSSN}. The two measures are well correlated
($r=0.95$, $r^{2}=0.90$) but there is a tendency to have more flares
on the declining phase of a sunspot cycle (the correlation is
maximized for a 2-month lag). In spite of this correlation,
significant flares can, and have, occurred at all phases of the
sunspot cycle. X-class flares have occurred during the few months
surrounding the sunspot cycle minima for three of the last four cycles
(Figure~\ref{fig:FlaresAndSSN}).

\epubtkImage{FlaresVsSSN.png}{%
\begin{figure}[htbp]
\centerline{\includegraphics[width=0.8\textwidth]{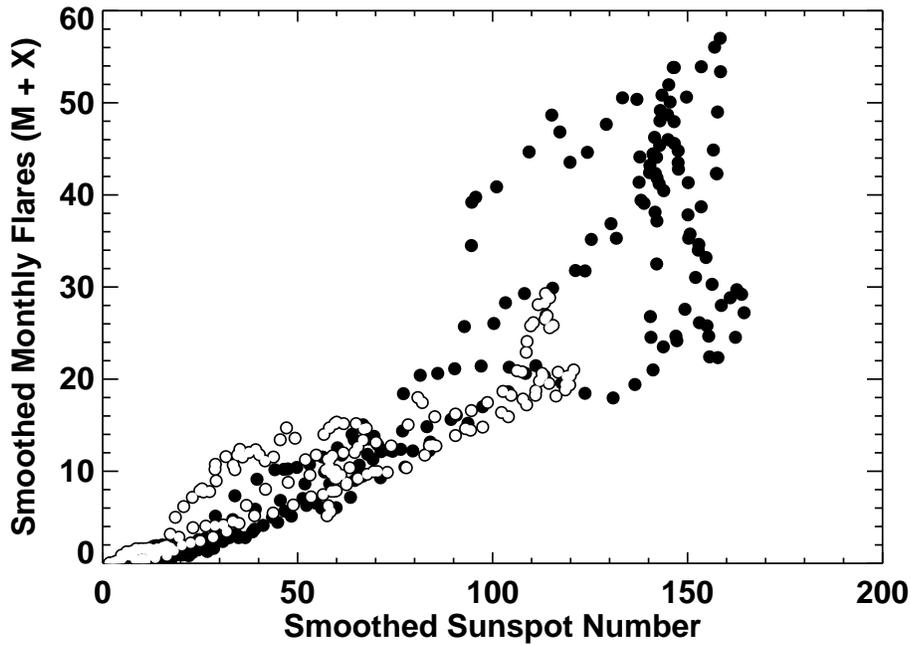}}
\caption{Monthly M- and X-class flares vs.\ International Sunspot Number 
for the period of March 1976 to December 2013. These two quantities are 
correlated at the 95\% level but show significant scatter when the 
sunspot number is high (greater than $\sim 100$).
Data obtained prior to cycle~23 are shown with filled dots, while data obtained
after 1997 are shown with open circles.}
\label{fig:FlaresVsSSN}
\end{figure}}

\epubtkImage{FlaresAndSSN.png}{%
\begin{figure}[htbp]
\centerline{\includegraphics[width=0.8\textwidth]{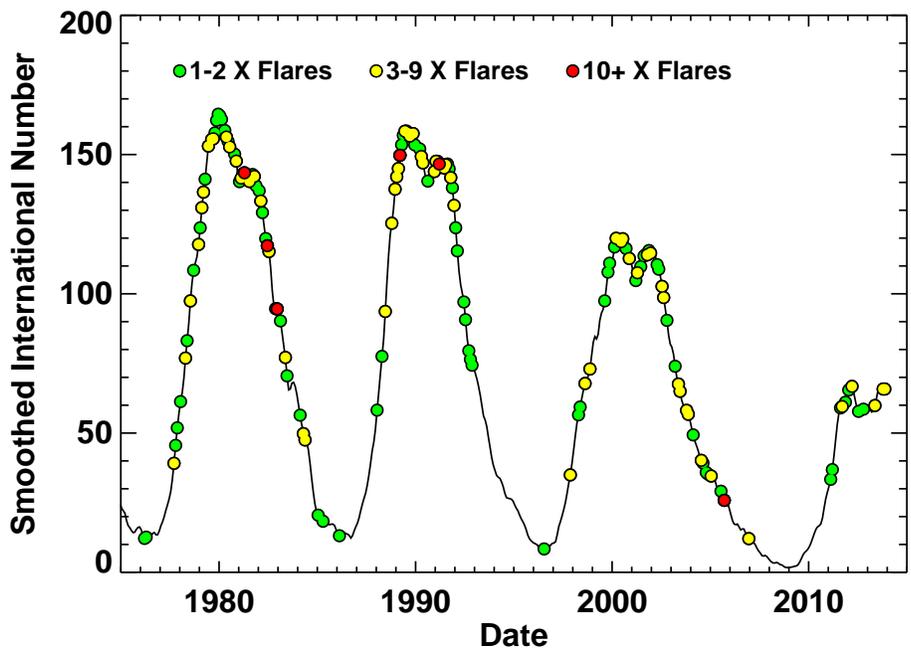}}
\caption{Monthly X-class flares and International Sunspot Number. X-class 
flares can occur at any phase of the sunspot cycle---including cycle 
minimum.}
\label{fig:FlaresAndSSN}
\end{figure}}

Coronal mass ejections (CMEs) are often associated with flares but can also 
occur in the absence of a flare. CMEs were discovered in the early 1970s 
via spacecraft observations from OSO~7 \citep{Tousey:1973} and from Skylab 
\citep{MacQueen:1974}. Routine~CME observations began with the Solar 
Maximum Mission and continue with SOHO. The frequency of occurrence of~CMEs 
is also correlated with sunspot number \citep{Webb:1994} but with differences
depending on the data used, the definition of what constitutes a CME, and
the method used for finding them \citep[e.g.][]{Webb:2012}.

\subsection{Geomagnetic activity}
\label{sec:geomagnetic-activity}

Geomagnetic activity also shows a solar cycle dependence but one that is 
more complex than seen in sunspot area, radio flux, or flares and~CMEs. 
There are a number of indices of geomagnetic activity; most measure rapid 
(hour-to-hour) changes in the strength and/or direction of Earth's
magnetic field from small networks of ground-based observatories. The
\textit{ap} index is a measure of the range of variability in the
geomagnetic field (in 2~nT units), measured in three-hour intervals
from a network of about 13 high-latitude stations. The average of the
eight daily \textit{ap} values is given as the equivalent daily
amplitude \textit{Ap}. These indices extend from 1932 to the
present. The \textit{aa} index extends back further \citep[to 1868;
  see][]{Mayaud:1972}, and is similarly derived from three-hour
intervals but from two antipodal stations located at latitudes of
about 50\textdegree. The locations of these two stations have changed
from time to time and there is evidence \citep{Svalgaard:2004} that
these changes are reflected in the data itself. Another frequently
used index is Dst, disturbance storm time, derived from measurements
obtained at four equatorial stations, since 1957.

\epubtkImage{aaVsTime.png}{%
\begin{figure}[htbp]
\centerline{\includegraphics[width=0.8\textwidth]{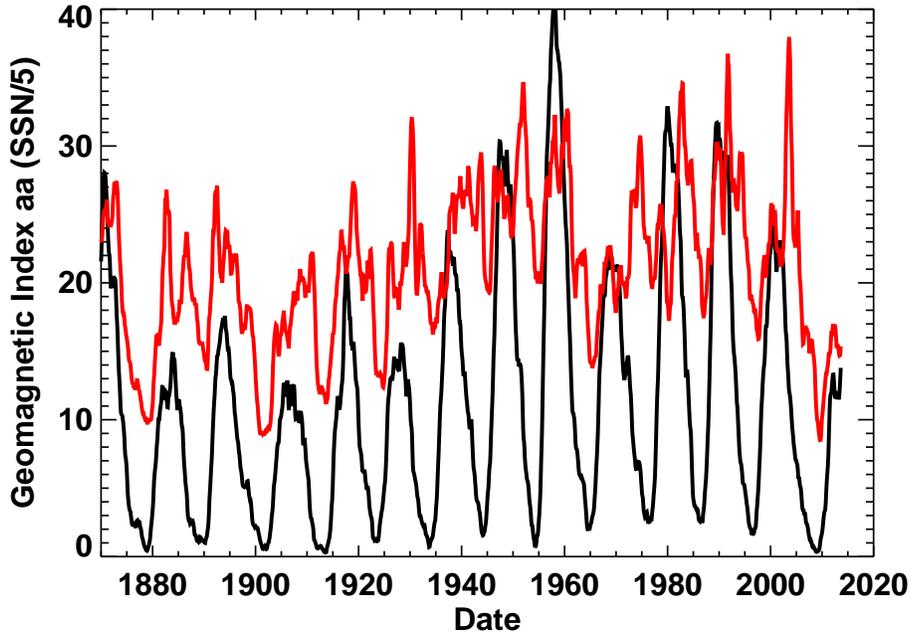}}
\caption{Geomagnetic activity and the sunspot cycle. The geomagnetic 
activity index \textit{aa} is plotted in red. The sunspot number (divided by five) is 
plotted in black.}
\label{fig:aaVsTime}
\end{figure}}

Figure~\ref{fig:aaVsTime} shows the smoothed monthly geomagnetic index
\textit{aa} as a function of time along with the sunspot number, for
comparison. The minima in geomagnetic activity tend to occur just
after those for the sunspot number and the geomagnetic activity tends
to remain high during the declining phase of each cycle. This late-cycle
geomagnetic activity is attributed to the effects of high-speed
solar wind streams from low-latitude coronal holes
\citep[e.g.,][]{Legrand:1985}. Figure~\ref{fig:aaVsTime} also shows the
presence of multi-cycle trends in geomagnetic activity that may be
related to changes in the Sun's magnetic field \citep{Lockwood:1999}.

\cite{Feynman:1982} decomposed geomagnetic variability into two components---one 
proportional to and in phase with the sunspot cycle (the R, or Relative 
sunspot number component) and another out of phase with the sunspot cycle 
(the I, or Interplanetary component).
Figure~\ref{fig:aaVsInternational} shows the relationship 
between geomagnetic activity and sunspot number. As the sunspot number 
increases there is a baseline level of geomagnetic activity that increases as well.
However, uniformly high levels of geomagnetic activity are found even when
the sunspot number is quite low.

\epubtkImage{aaVsInternational.png}{%
\begin{figure}[htbp]
\centerline{\includegraphics[width=0.8\textwidth]{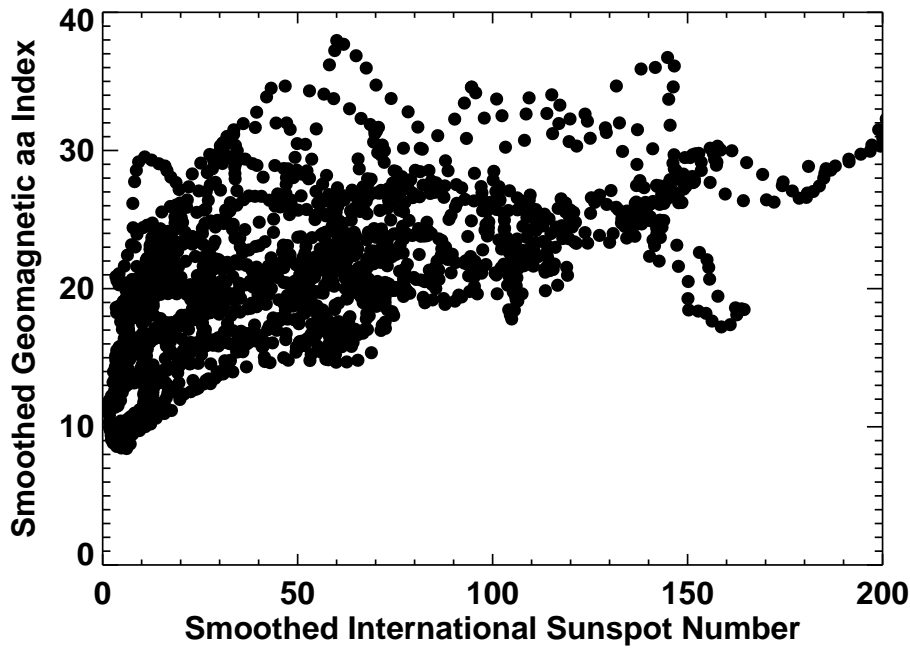}}
\caption{Geomagnetic activity index \textit{aa} vs.\ Sunspot Number. As Sunspot Number 
increases the baseline level of geomagnetic activity increases as
well.}
\label{fig:aaVsInternational}
\end{figure}}

\subsection{Cosmic rays}
\label{sec:cosmic-rays}

The flux of galactic cosmic rays at 1~AU is modulated by the solar cycle. 
Galactic cosmic rays consist of electrons and bare nuclei that are accelerated to GeV 
energies and higher at shocks produced by supernovae. The positively charged 
nuclei produce cascading showers of particles in Earth's upper 
atmosphere that can be measured by neutron monitors at high-altitude 
observing sites. The oldest continuously operating neutron monitor is 
located in Climax, Colorado, USA. Daily observations extend from 1951 to 
2006. Monthly averages of the neutron counts are shown as a function of time 
in Figure~\ref{fig:ClimaxCosmicRays}, along with the sunspot number.
As the sunspot numbers rise the 
neutron counts fall. This anti-correlation is attributed to scattering of 
the cosmic rays by tangled magnetic field within the heliosphere
\citep{Parker:1965}. At times of high solar activity, magnetic
structures are carried outward on the solar wind. These structures
scatter incoming cosmic rays and reduce their flux in the inner solar system.

The reduction in cosmic ray flux tends to lag behind solar activity by 6 to 
12 months \citep{Forbush:1954} but with significant differences between the even 
numbered and odd numbered cycles. In the even numbered cycles (cycles~20 and 
22) the cosmic ray variations seen by neutron monitors lag sunspot number 
variations by only about 2 months. In the odd numbered cycles (cycles~19, 
21, and 23) the lag is from 10 to 14 months. Figure~\ref{fig:ClimaxCosmicRays} also shows that the 
shapes of the cosmic ray maxima at sunspot cycle minima are different for 
the even and odd numbered cycles. The cosmic ray maxima (as measured by the 
neutron monitors) are sharply peaked at the sunspot cycle minima leading up 
to even numbered cycles and broadly peaked prior to odd numbered sunspot 
cycles. This behavior is accounted for in the transport models for galactic 
cosmic rays in the heliosphere \citep[e.g.,][]{Ferreira:2004}. The 
positively charged cosmic rays drift in from the heliospheric polar regions 
when the Sun's north polar field is directed outward (positive). When the 
Sun's north polar field is directed inward (negative) the positively charged 
cosmic rays drift inward along the heliospheric current sheet where they are 
scattered by corrugations in the current sheet and by magnetic clouds from 
CMEs. The negatively charged cosmic rays (electrons) drift inward from 
directions (polar or equatorial) opposite to the positively charged cosmic 
rays that are detected by neutron monitors.

\epubtkImage{ClimaxCosmicRays.png}{%
\begin{figure}[htbp]
\centerline{\includegraphics[width=0.8\textwidth]{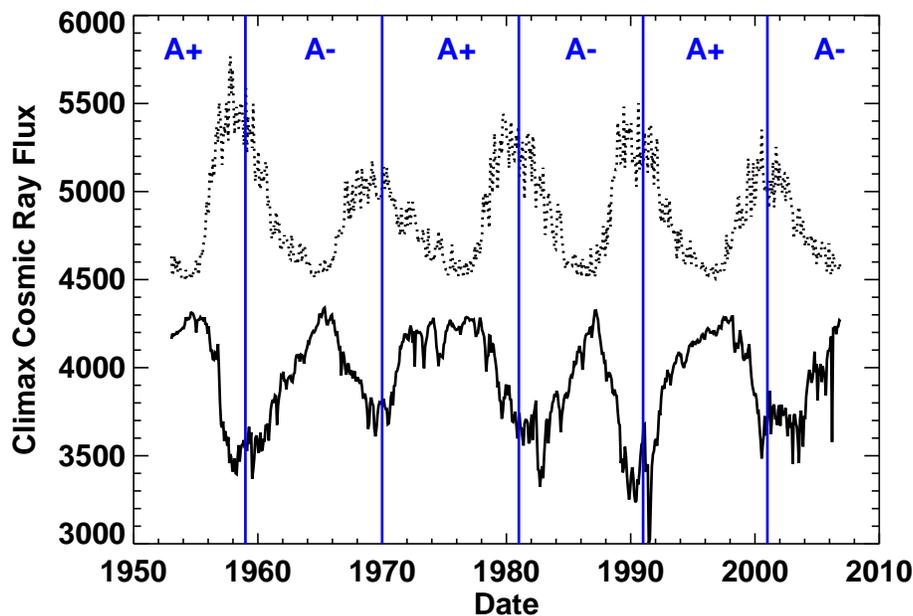}}
\caption{Cosmic Ray flux from the Climax Neutron Monitor and rescaled 
Sunspot Number. The monthly averaged neutron counts from the Climax Neutron 
Monitor are shown by the solid line. The monthly averaged sunspot numbers 
(multiplied by five and offset by 4500) are shown by the dotted line. Cosmic 
ray variations are anti-correlated with solar activity but with differences 
depending upon the Sun's global magnetic field polarity (A+ indicates 
periods with positive polarity north pole, while A-- indicates periods with 
negative polarity).}
\label{fig:ClimaxCosmicRays}
\end{figure}}

\subsection{Radioisotopes in tree rings and ice cores}
\label{sec:trees-and-ice}

The radioisotopes \super{14}C and \super{10}Be are produced in Earth's 
stratosphere by the impact of galactic cosmic rays on \super{14}N and \super{16}O. 
The \super{14}C gets oxidized to form CO\sub{2}, which is taken up by plants in 
general and trees in particular, where it becomes fixed in annual growth 
rings. The \super{10}Be gets oxidized and becomes attached to aerosols that can 
precipitate in snow, where it then becomes fixed in annual layers of ice. The 
solar cycle modulation of the cosmic ray flux can then lead to solar cycle 
related variations in the atmospheric abundances of \super{14}C
\citep{Stuiver:1980} and \super{10}Be \citep{Beer:1990}. While the
production rates of these two radioisotopes in the stratosphere should
be anti-correlated with the sunspot cycle, the time scales involved in
the transport and ultimate deposition in tree rings and ice tends to
reduce and delay the solar cycle variations
\citep[see][]{Masarik:1999}. The production rates in the
stratospheric are functions of magnetic latitude, which changes as Earth's
magnetic dipole wanders and varies in strength.
Furthermore, the latency in the stratosphere/troposphere
is a function of the changing reservoirs for these chemical
species. This rather complicated
production/transport/storage/deposition process makes direct
comparisons between $\Delta ^{14}$C (the difference between
measured \super{14}C abundance and that expected from its 5730-year
half-life) and sunspot number difficult.
For more details on the influence of solar activity on radioisotopes
and on what is learned about solar activity from radioisotopes, see the
review by \citet{Usoskin:2013}.

\newpage 

\section{Individual Cycle Characteristics}
\label{sec:charact}

Each sunspot cycle has its own characteristics. Many of these 
characteristics are shared by other cycles and these  
provide important information for models of the solar activity cycle. A 
paradigm shift in sunspot cycle studies came about when \cite{Waldmeier:1935} 
suggested that each cycle should be treated as an individual
\textit{outburst} with its own characteristics. Prior to that time,
the fashion was to consider solar activity as a superposition of
Fourier components. This superposition idea probably had its roots in
the work of \cite{Wolf:1859}, who suggested a formula based on the
orbits of Venus, Earth, Jupiter, and Saturn to fit Schwabe's data for
the years 1826 to 1848.

Determining characteristics such as period and amplitude would seem simple 
and straightforward but the published studies show that this is not true. A 
prime example concerns determinations of the dates (year and month) of cycle 
minima. A frequently used method is to take monthly averages of the daily 
International Sunspot Number and to smooth these with the 13-month running 
mean. Unfortunately, this leaves several uncertain dates. With this method, 
the minimum that occurred in 1810 prior to cycle 6 could be taken as any 
month from April to December---all nine months had smoothed sunspot numbers 
of 0.0!

\subsection{Minima and maxima}

The dates and values for the cycle minima and maxima are the primary data 
for many studies of the solar cycle. These data are sensitive to the methods 
and input data used to find them. Solar activity is inherently noisy and it 
is evident that there are significant variations in solar activity on time 
scales shorter than 11 years (see Section~\ref{sec:stvar}).
\cite{Waldmeier:1961} published tables of
sunspot numbers along with dates and values of minima and maxima for
cycles~1 to 19. \cite{McKinnon:1987} extended the data to include
cycles~20 and 21. The values they give for sunspot number maxima and
minima are those found using the 13-month running mean. However, the
dates given for maxima and minima may vary after considering
additional indicators. According to McKinnon:

\begin{quote}
 {\ldots} maximum is based in part on an average of the times
 extremes are reached in the monthly mean sunspot number, the smoothed
 monthly mean sunspot number, and in the monthly mean number of spot
 groups alone.
\end{quote}
These dates and the values for sunspot cycle maxima are given in
Table~\ref{tab1} (the number of groups is multiplied by 12.08 to
produce group sunspot numbers that are comparable to the relative
sunspot numbers). It is clear from this table that considerably more
weight is given to the date provided by the 13-month running mean. The
dates provided by Waldmeier and McKinnon are far closer to those given
by the 13-month running mean than they are to the average date of the
three indicators. (One exception is the date they give for the maximum
of cycle~14, which should be a half-year earlier by almost any
averaging scheme.) The monthly numbers of sunspots and spot groups
vary widely and, in fact, should be less reliable indicators and given
lesser weight in determining maximum. 

\begin{table}[htbp]
\caption{Dates and values for sunspot cycle maxima.}
\label{tab1}
\centering
\begin{tabular}{c| lr| lr| lr| lr}
\toprule
Cycle &
\multicolumn{2}{>{\centering\arraybackslash}p{1in}|}{Waldmeier/ \ifpdf\par\else\fi McKinnon} & 
\multicolumn{2}{>{\centering\arraybackslash}p{1in}|}{13-month Mean \ifpdf\par\else\fi Maximum} & 
\multicolumn{2}{>{\centering\arraybackslash}p{1in}|}{Monthly Mean \ifpdf\par\else\fi Maximum} & 
\multicolumn{2}{>{\centering\arraybackslash}p{1in}}{Monthly Group \ifpdf\par\else\fi Maximum} \\
~ & Date & Value & Date & Value & Date & Value & Date & Value \\
\midrule
1  & 1761.5  &  86.5 & 1761/06 &  86.5 & 1761/05 & 107.2 & 1761/05 & 109.4 \\
2  & 1769.7  & 115.8 & 1769/09 & 115.8 & 1769/10 & 158.2 & 1771/05 & 162.5 \\
3  & 1778.4  & 158.5 & 1778/05 & 158.5 & 1778/05 & 238.9 & 1778/01 & 144.0 \\
4  & 1788.1  & 141.2 & 1788/02 & 141.2 & 1787/12 & 174.0 & 1787/12 & 169.0 \\
5  & 1805.2  &  49.2 & 1805/02 &  49.2 & 1804/10 &  62.3 & 1805/11 &  67.0 \\
\midrule
6  & 1816.4  &  48.7 & 1816/05 &  48.7 & 1817/03 &  96.2 & 1817/03 &  57.0 \\
7  & 1829.9  &  71.7 & 1829/11 &  71.5 & 1830/04 & 106.3 & 1830/04 & 101.5 \\
8  & 1837.2  & 146.9 & 1837/03 & 146.9 & 1836/12 & 206.2 & 1837/01 & 160.7 \\
9  & 1848.1  & 131.6 & 1848/02 & 131.9 & 1847/10 & 180.4 & 1849/01 & 130.9 \\
10 & 1860.1  &  97.9 & 1860/02 &  98.0 & 1860/07 & 116.7 & 1860/07 & 103.4 \\
\midrule
11 & 1870.6  & 140.5 & 1870/08 & 140.3 & 1870/05 & 176.0 & 1870/05 & 122.3 \\
12 & 1883.9  &  74.6 & 1883/12 &  74.6 & 1882/04 &  95.8 & 1884/01 &  86.0 \\
13 & 1894.1  &  87.9 & 1894/01 &  87.9 & 1893/08 & 129.2 & 1893/08 & 126.7 \\
14 & 1907.0? &  64.2 & 1906/02 &  64.2 & 1907/02 & 108.2 & 1906/07 & 111.6 \\
15 & 1917.6  & 105.4 & 1917/08 & 105.4 & 1917/08 & 154.5 & 1917/08 & 157.0 \\
\midrule
16 & 1928.4  &  78.1 & 1928/04 &  78.1 & 1929/12 & 108.0 & 1929/12 & 121.8 \\
17 & 1937.4  & 119.2 & 1937/04 & 119.2 & 1938/07 & 165.3 & 1937/02 & 154.5 \\
18 & 1947.5  & 151.8 & 1947/05 & 151.8 & 1947/05 & 201.3 & 1947/07 & 149.3 \\
19 & 1957.9  & 201.3 & 1958/03 & 201.3 & 1957/10 & 253.8 & 1957/10 & 222.2 \\
20 & 1968.9  & 110.6 & 1968/11 & 110.6 & 1969/03 & 135.8 & 1968/05 & 132.3 \\
\midrule
21 & 1979.9  & 164.5 & 1979/12 & 164.5 & 1979/09 & 188.4 & 1979/01 & 179.4 \\
22 &      ~  &     ~ & 1989/07 & 158.5 & 1990/08 & 200.3 & 1990/08 & 195.9 \\
23 &      ~  &     ~ & 2000/04 & 120.7 & 2000/07 & 169.1 & 2000/07 & 153.9 \\
24 &      ~  &     ~ & 2014/04 & 81.9 & 2014/02 & 102.3 & 2011/11 & 101.1 \\
\bottomrule
\end{tabular}
\end{table}

The minima in these three indicators have been used along with additional 
sunspot indicators to determine the dates of minima. The number of spotless 
days in a month tends to maximize at the time of minimum and the number of 
new cycle sunspot groups begins to exceed the number of old cycle sunspot 
groups at the time of minimum. Both Waldmeier and McKinnon suggest using 
these indicators as well when setting the dates for minima. These dates are 
given in Table~\ref{tab2} where both the spotless days per month and the number of 
old-cycle and new-cycle groups per month are smoothed with the same 13-month 
mean filter. The average date given in the last column is the average of: the 
13-month mean minimum date; the 13-month mean spotless days per month 
maximum date; and the date when the 13-month mean of the number of new-cycle 
groups exceeds the 13-month mean of the number of old-cycle groups. For the 
early cycles, where spotless days and old-cycle and new-cycle groups are not 
available, the 13-month mean minimum date is used for those dates in forming 
the average.

\begin{table}[htbp]
\caption{Dates and values for sunspot cycle minima. The value is
  always the value of the 13-month mean of the International Sunspot
  Number. The dates differ according to the indicator used.}
\label{tab2}
\centering
\begin{tabular}{c| lr| 
    >{\centering\arraybackslash}p{0.8in}|
    >{\centering\arraybackslash}p{1in}|
    >{\centering\arraybackslash}p{0.8in}|
    >{\centering\arraybackslash}p{0.8in}
}
\toprule
Cycle & 
\multicolumn{2}{>{\centering\arraybackslash}p{1in}|}
{13-month Mean \ifpdf\par\else\fi Minimum} & 
Waldmeier/ \ifpdf\par\else\fi McKinnon & 
Spotless Days \ifpdf\par\else\fi Maximum & 
New $>$ Old & 
Average \\
~ & Date & Value & Date & Date & Date & Date \\
\midrule
 1 & 1755/02 &  8.4 & 1755.2 &       ~ &       ~ & 1755/02 \\
 2 & 1766/06 & 11.2 & 1766.5 &       ~ &       ~ & 1766/06 \\
 3 & 1775/06 &  7.2 & 1775.5 &       ~ &       ~ & 1775/06 \\
 4 & 1784/09 &  9.5 & 1784.7 &       ~ &       ~ & 1784/09 \\
 5 & 1798/04 &  3.2 & 1798.3 &       ~ &       ~ & 1798/04 \\
\midrule
 6 & 1810/08 &  0.0 & 1810.6 &       ~ &       ~ & 1810/08 \\
 7 & 1823/05 &  0.1 & 1823.3 & 1823/02 &       ~ & 1823/04 \\
 8 & 1833/11 &  7.3 & 1833.9 & 1833/11 &       ~ & 1833/11 \\
 9 & 1843/07 & 10.6 & 1843.5 & 1843/07 &       ~ & 1843/07 \\
10 & 1855/12 &  3.2 & 1856.0 & 1855/12 &       ~ & 1855/12 \\
\midrule
11 & 1867/03 &  5.2 & 1867.2 & 1867/05 &       ~ & 1867/04 \\
12 & 1878/12 &  2.2 & 1878.9 & 1878/10 & 1879/01 & 1878/12 \\
13 & 1890/03 &  5.0 & 1889.6 & 1890/02 & 1889/09 & 1890/01 \\
14 & 1902/01 &  2.7 & 1901.7 & 1902/01 & 1901/11 & 1901/12 \\
15 & 1913/07 &  1.5 & 1913.6 & 1913/08 & 1913/04 & 1913/06 \\
\midrule
16 & 1923/08 &  5.6 & 1923.6 & 1923/10 & 1923/09 & 1923/09 \\
17 & 1933/09 &  3.5 & 1933.8 & 1933/09 & 1933/11 & 1933/10 \\
18 & 1944/02 &  7.7 & 1944.2 & 1944/02 & 1944/03 & 1944/02 \\
19 & 1954/04 &  3.4 & 1954.3 & 1954/04 & 1954/04 & 1954/04 \\
20 & 1964/10 &  9.6 & 1964.9 & 1964/11 & 1964/08 & 1964/10 \\
\midrule
21 & 1976/03 & 12.2 & 1976.5 & 1975/09 & 1976/08 & 1976/03 \\
22 & 1986/09 & 12.3 &      ~ & 1986/03 & 1986/10 & 1986/07 \\
23 & 1996/05 &  8.0 &      ~ & 1996/07 & 1996/12 & 1996/08 \\
24 & 2008/12 &  1.7 &      ~ & 2008/12 & 2008/09 & 2008/11 \\
\bottomrule
\end{tabular}
\end{table}

When available, all three indicators tend to give dates that are fairly 
close to each other and the average of the three is usually close to the 
dates provided by Waldmeier and McKinnon. There are, however, two notable 
exceptions. The dates given by Waldmeier for the minima preceding cycles~13 
and 14 are both significantly earlier than the dates given by all
three indicators. The cycle~13 minimum date of 1889.6 was adopted from
\cite{Wolf:1892} while the cycle~14 minimum date of 1901.7 was adopted
from \cite{Wolfer:1903}.

Since many researchers simply adopt the date given by the minimum in the 
13-month running mean, the date for the minimum preceding cycle~23 is also 
problematic. The minimum in the smoothed sunspot number came in May of 1996. The 
maximum in the smoothed number of spotless days per month came in July of 
1996. However, the cross-over in the smoothed number of groups from 
old-cycle groups to new-cycle groups occurred in December of 1996. \cite{Harvey:1999} 
provide a good discussion of the problems in determining cycle minimum and 
have argued that the minimum for cycle~23 should be taken as September 1996 
(based on their determination that new-cycle groups exceed old-cycle groups 
in January of 1997). The average of the three indicators gives August  
1996.

Additional problems in assigning dates and values to maxima and minima can 
be seen when using data other than sunspot numbers. Table~\ref{tab3} lists the dates 
and values for cycle maxima using the 13-month running mean on sunspot 
numbers, sunspot areas, and 10.7~cm radio flux. The sunspot areas have been 
converted to sunspot number equivalents using the relationship shown in 
Figure~\ref{fig:RGOareaVsInternational}, and the 10.7~cm radio flux has been converted into sunspot number 
equivalents using Equation~(\ref{eq3}). Very significant differences can be seen in the 
dates. Over the last five cycles the ranges in dates given by the different 
indices have been: 4, 27, 25, 1, and 22 months.

\begin{table}[htbp]
\caption{Dates and values of maxima using the 13-month running mean
  with sunspot number data, sunspot area data, and 10.7~cm radio flux
  data.}
\label{tab3}
\centering
\begin{tabular}{c| lr| lr| lr}
\toprule
Cycle & 
\multicolumn{2}{>{\centering\arraybackslash}p{1in}|}{13-month Mean \ifpdf\par\else\fi Maximum} & 
\multicolumn{2}{>{\centering\arraybackslash}p{1in}|}{13-month Mean \ifpdf\par\else\fi Sunspot Area} & 
\multicolumn{2}{>{\centering\arraybackslash}p{1in}}{13-month Mean \ifpdf\par\else\fi 10.7~cm Flux} \\
 ~ & Date    & Value & Date    & R-Value & Date  & R-Value \\
\midrule
 1 & 1761/06 &  86.5 &       ~ &     ~ &       ~ & ~ \\
 2 & 1769/09 & 115.8 &       ~ &     ~ &       ~ & ~ \\
 3 & 1778/05 & 158.5 &       ~ &     ~ &       ~ & ~ \\
 4 & 1788/02 & 141.2 &       ~ &     ~ &       ~ & ~ \\
 5 & 1805/02 &  49.2 &       ~ &     ~ &       ~ & ~ \\
\midrule
 6 & 1816/05 &  48.7 &       ~ &     ~ &       ~ & ~ \\
 7 & 1829/11 &  71.5 &       ~ &     ~ &       ~ & ~ \\
 8 & 1837/03 & 146.9 &       ~ &     ~ &       ~ & ~ \\
 9 & 1848/02 & 131.9 &       ~ &     ~ &       ~ & ~ \\
10 & 1860/02 &  98.0 &       ~ &     ~ &       ~ & ~ \\
\midrule
11 & 1870/08 & 140.3 &       ~ &     ~ &       ~ & ~ \\
12 & 1883/12 &  74.6 & 1883/11 &  88.3 &       ~ & ~ \\
13 & 1894/01 &  87.9 & 1894/01 & 100.4 &       ~ & ~ \\
14 & 1906/02 &  64.2 & 1905/06 &  75.4 &       ~ & ~ \\
15 & 1917/08 & 105.4 & 1917/08 &  93.0 &       ~ & ~ \\
\midrule
16 & 1928/04 &  78.1 & 1926/04 &  92.3 &       ~ & ~ \\
17 & 1937/04 & 119.2 & 1937/05 & 133.3 &       ~ & ~ \\
18 & 1947/05 & 151.8 & 1947/05 & 166.5 &       ~ & ~ \\
19 & 1958/03 & 201.3 & 1957/11 & 216.5 & 1958/03 & 201.2 \\
20 & 1968/11 & 110.6 & 1968/04 & 100.9 & 1970/07 & 109.6 \\
\midrule
21 & 1979/12 & 164.5 & 1982/01 & 156.0 & 1981/05 & 159.4 \\
22 & 1989/07 & 158.5 & 1989/06 & 158.5 & 1989/06 & 168.0 \\
23 & 2000/04 & 120.7 & 2002/02 & 126.7 & 2002/02 & 152.3 \\
\bottomrule
\end{tabular}
\end{table}

These tables illustrate the problems in determining dates and values for 
cycle minima and maxima. The crux of the problem is in the short-term 
variability of solar activity. One solution is to use a different smoothing method.

\subsection{Smoothing}

The monthly averages of the daily International Sunspot Number are noisy and 
must be smoothed in some manner in order to determine appropriate values for 
parameters such as minima, maxima, and their dates of occurrence. The daily 
values themselves are highly variable. They depend upon the number and 
the quality of observations as well as the time of day when they are taken 
(the sunspot number changes over the course of the day as spots form and 
fade away). The monthly averages of these daily values are also problematic. 
The Sun rotates once in about 27 days but the months vary in length from 28 
to 31 days. If the Sun is particularly active at one set of longitudes then 
some monthly averages will include one appearance of these active longitudes 
while other months will include two. This aspect is particularly important 
for investigations of short-term (months) variability (see
Section~\ref{sec:predict}). For long-term (years) variability this can
be treated as noise and filtered out.

The traditional 13-month running mean (centered on a given month with equal 
weights for months --5 to +5 and half-weight for months --6 and +6) is both 
simple and widely used, but does a poor job of filtering out high-frequency 
variations (although it is better than the simple 12-month average). 
Gaussian-shaped filters are preferable because they have Gaussian shapes in 
the frequency domain and effectively remove high-frequency variations 
\citep{Hathaway:1999}. A tapered (to make the filter 
weights and their first derivatives vanish at the end points) Gaussian 
filter is given by
\begin{equation}
\label{eq4}
W(t)= e^{-t^{2}/2a^{2}} - e^{-2} \left(3-t^{2}/2a^{2}\right)
\end{equation}
with
\begin{equation}
\label{eq5}
-2a+1\le t\le +2a-1
\end{equation}
where $t$ is the time in months and $2a$ is the Full Width at Half Maximum (FWHM)
of the filter
(note that this formula is slightly different than that given in
\cite{Hathaway:1999}). There are significant variations in solar
activity on time scales of one to three years (see
Section~\ref{sec:stvar}). These variations can produce double-peaked
maxima that are filtered out by a 24-month Gaussian filter. The
frequency responses of these filters are shown in Figure~\ref{fig:FilterResponse}.

\epubtkImage{FilterResponse.png}{%
\begin{figure}[htbp]
\centerline{\includegraphics[width=0.8\textwidth]{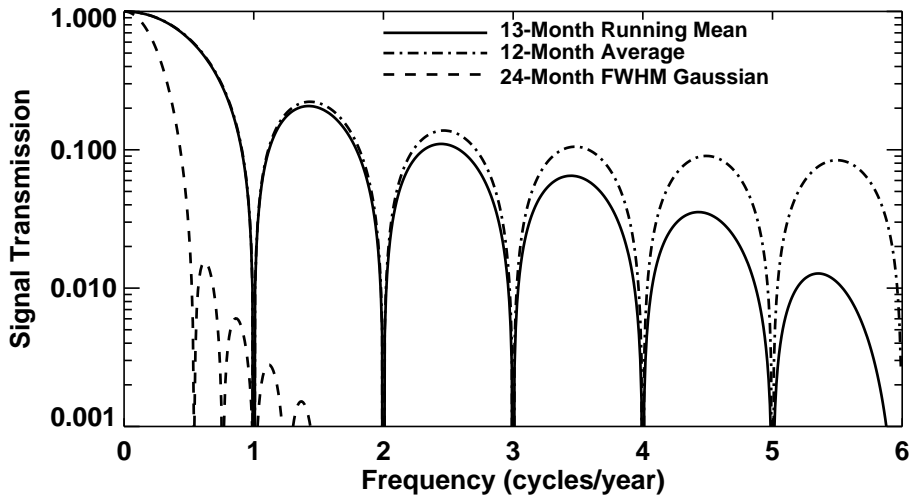}}
\caption{Signal transmission for filters used to smooth monthly sunspot 
numbers. The 13-month running mean and the 12-month average pass significant 
fractions (as much as 20\%) of signals with frequencies higher than 
one cycle per year. The 24-month FWHM Gaussian passes less than 0.3\% of those 
frequencies and passes less than about 1\% of the signal with frequencies 
of a half-cycle per year or higher.}
\label{fig:FilterResponse}
\end{figure}}

Using the 24-month FWHM Gaussian filter on the data used to create Table~\ref{tab3} 
gives far more consistent results for both maxima and minima. The results 
for maxima are shown in Table~\ref{tab4}. The ranges of dates for the last five 
maxima become: 1, 10, 13, 4, and 11 months---roughly half the ranges found 
using the 13-month running mean.

\begin{table}[htbp]
\caption{Dates and values of maxima using the 24-month FWHM Gaussian
  with sunspot number data, sunspot area data, and 10.7~cm radio flux
  data, as in Table~\ref{tab3}.}
\label{tab4}
\centering
\begin{tabular}{c| lr| lr| lr}
\toprule
Cycle & 
\multicolumn{2}{>{\centering\arraybackslash}p{1.1in}|}{24-month Gaussian \ifpdf\par\else\fi Maximum} &
\multicolumn{2}{>{\centering\arraybackslash}p{1in}|}{24-Month Gaussian \ifpdf\par\else\fi Sunspot Area} &
\multicolumn{2}{>{\centering\arraybackslash}p{1in}}{24-Month Gaussian \ifpdf\par\else\fi 10.7~cm Flux} \\
 ~ & Date    & Value & Date    & R-Value & Date  & R-Value \\
\midrule
 1 & 1761/05 &  72.9 &       ~ &     ~ &       ~ & ~ \\
 2 & 1770/01 & 100.5 &       ~ &     ~ &       ~ & ~ \\
 3 & 1778/09 & 137.4 &       ~ &     ~ &       ~ & ~ \\
 4 & 1788/03 & 130.6 &       ~ &     ~ &       ~ & ~ \\
 5 & 1804/06 &  45.7 &       ~ &     ~ &       ~ & ~ \\
\midrule
 6 & 1816/08 &  43.8 &       ~ &     ~ &       ~ & ~ \\
 7 & 1829/10 &  67.1 &       ~ &     ~ &       ~ & ~ \\
 8 & 1837/04 & 146.9 &       ~ &     ~ &       ~ & ~ \\
 9 & 1848/06 & 115.7 &       ~ &     ~ &       ~ & ~ \\
10 & 1860/03 &  92.1 &       ~ &     ~ &       ~ & ~ \\
\midrule
11 & 1870/11 & 138.5 &       ~ &     ~ &       ~ & ~ \\
12 & 1883/11 &  64.7 & 1883/10 &  70.8 &       ~ & ~ \\
13 & 1893/09 &  81.4 & 1893/09 &  84.7 &       ~ & ~ \\
14 & 1906/05 &  59.6 & 1906/04 &  62.4 &       ~ & ~ \\
15 & 1917/12 &  88.6 & 1918/01 &  79.6 &       ~ & ~ \\
\midrule
16 & 1927/12 &  71.6 & 1926/12 &  75.9 &       ~ & ~ \\
17 & 1937/11 & 108.2 & 1938/02 & 118.1 &       ~ & ~ \\
18 & 1948/03 & 141.7 & 1947/09 & 140.0 &       ~ & ~ \\
19 & 1958/02 & 188.0 & 1958/03 & 192.0 & 1958/03 & 188.1 \\
20 & 1969/03 & 106.6 & 1968/09 &  95.5 & 1969/07 & 104.6 \\
\midrule
21 & 1980/05 & 151.8 & 1981/06 & 140.2 & 1980/11 & 153.1 \\
22 & 1990/02 & 149.2 & 1990/06 & 141.7 & 1990/06 & 156.1 \\
23 & 2000/12 & 112.7 & 2001/11 & 106.2 & 2001/06 & 136.4 \\
\bottomrule
\end{tabular}
\end{table}

\subsection{Cycle periods}

The period of a sunspot cycle is defined as the elapsed time from the 
minimum preceding its maximum to the minimum following its maximum. This 
does not, of course, account for the fact that each cycle actually starts 
well before its preceding minimum and continues long after its following 
minimum. By this definition, a cycle's period is dependent upon the behavior 
of both the preceding and following cycles. The measured period of a cycle 
is also subject to uncertainties in determining the dates of minimum, as 
indicated in the previous subsections. Nonetheless, the length of a sunspot 
cycle is a key characteristic and variations in cycle periods have been well 
studied. The average cycle period can be fairly accurately determined by 
simply subtracting the date for the minimum preceding cycle~1 from the date 
for the minimum preceding cycle~23 and dividing by the 22 cycles those dates 
encompass. This gives an average period for cycles~1 to 22 of 131.7 months---almost exactly 11 years.

The distribution of cycle periods depends upon the cycles used and the 
methods used to determine minima. \cite{Eddy:1977} noted that the cycle periods 
did not appear to be distributed normally. \cite{Wilson:1987} included cycles~8 to 
20 and used the dates for minimum from the 13-month mean of the monthly 
sunspot numbers. He found that a bimodal distribution best fit the data with 
short-period (122 month) cycles and long-period (140 month) cycles separated 
by a gap (the Wilson Gap) surrounding the mean cycle length of 132.7 months. 
However, \cite{Hathaway:2002} used minima dates from the 
24-month Gaussian smoothing of the International Sunspot number for cycles~1 
to 23 and of the Group Sunspot Numbers for cycles --4 to 23 and found 
distributions that were consistent with a normal distributions about a mean 
of 131 months with a standard deviation of 14 months and no evidence of a 
gap. These cycle periods and their distributions are shown in Figure~\ref{fig:CyclePeriods}.

\epubtkImage{CyclePeriods.png}{%
\begin{figure}[htbp]
\centerline{\includegraphics[width=0.8\textwidth]{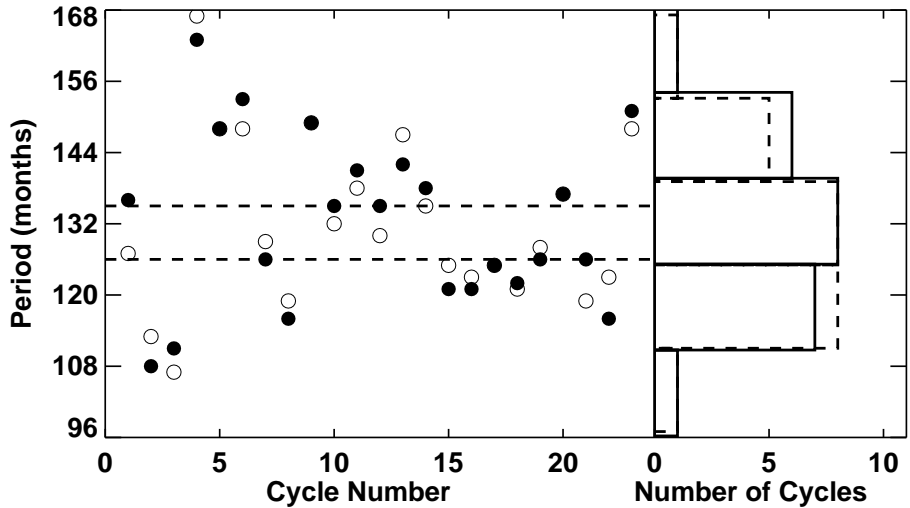}}
\caption{The left panel shows cycle periods as functions of Cycle Number. 
Filled circles give periods determined from minima in the 13-month mean, 
while open circles give periods determined from the 24-month Gaussian 
smoothing. Both measurements give a mean period of about 132 months with a 
standard deviation of about 14 months. The ``Wilson Gap'' in the periods between 
125 and 134 months from the 13-month mean is shown with dashed lines. 
The right panel shows histograms of cycle periods centered on the mean 
period with bin widths of one standard deviation. The solid lines show the 
distribution from the 13-month mean while the dashed lines show the 
distribution for the 24-month Gaussian. The periods appear normally 
distributed and the ``Wilson Gap'' is well populated with the 24-month 
Gaussian smoothed data.}
\label{fig:CyclePeriods}
\end{figure}}

\subsection{Cycle amplitudes}

The amplitude of a cycle is another key characteristic. As we have
seen in Tables~\ref{tab3} and \ref{tab4}, the actual value for the
amplitude of a cycle depends upon the activity index used and the type
of smoothing. These uncertainties can even change the relative
amplitudes of the cycles. In Table~\ref{tab3}, we see that the second
largest cycle is cycle~21 according to the 13-month mean of the
International Sunspot Numbers. But with the same smoothing, the second
largest cycle in sunspot area and 10.7~cm flux is cycle~22. Cycles~15 and 16
were very similar according to sunspot area but cycle~15 is
significantly larger than cycle~16 according to the International
Sunspot Number. The Group Sunspot Numbers do provide information on
earlier cycles but show systematic differences when compared to the
International Sunspot Numbers. The maxima determined by the 13-month
mean with the International Sunspot Numbers and the Group Sunspot
Numbers are given in Table~\ref{tab5}.

\begin{table}[htbp]
\caption{Cycle maxima determined by the 13-month mean with the
  International Sunspot Numbers and the Group Sunspot Numbers. The
  Group values are systematically lower than the International values
  prior to cycle~12.}
\label{tab5}
\centering
\begin{tabular}{c| lr| lr}
\toprule
Cycle& 
\multicolumn{2}{>{\centering\arraybackslash}p{1in}|}{International \ifpdf\par\else\fi Maximum} & 
\multicolumn{2}{>{\centering\arraybackslash}p{1in}}{Group SSN \ifpdf\par\else\fi Maximum} \\
~ & Date & Value & Date & Value \\
\midrule
--4 &       ~ &     ~ & 1705/05 &   5.5 \\
--3 &       ~ &     ~ & 1719/11 &  34.2 \\
--2 &       ~ &     ~ & 1730/02 &  82.6 \\
--1 &       ~ &     ~ & 1739/05 &  58.3 \\
0   &       ~ &     ~ & 1750/03 &  70.6 \\
\midrule
1   & 1761/06 &  86.5 & 1761/05 &  71.3 \\
2   & 1769/09 & 115.8 & 1769/09 & 106.5 \\
3   & 1778/05 & 158.5 & 1779/06 &  79.5 \\
4   & 1788/02 & 141.2 & 1787/10 &  90.5 \\
5   & 1805/02 &  49.2 & 1805/06 &  24.8 \\
\midrule
6   & 1816/05 &  48.7 & 1816/09 &  31.5 \\
7   & 1829/11 &  71.5 & 1829/12 &  64.4 \\
8   & 1837/03 & 146.9 & 1837/03 & 116.8 \\
9   & 1848/02 & 131.9 & 1848/11 &  93.2 \\
10  & 1860/02 &  98.0 & 1860/10 &  85.8 \\
\midrule
11  & 1870/08 & 140.3 & 1870/11 &  99.9 \\
12  & 1883/12 &  74.6 & 1884/03 &  68.2 \\
13  & 1894/01 &  87.9 & 1894/01 &  96.0 \\
14  & 1906/02 &  64.2 & 1906/02 &  64.6 \\
15  & 1917/08 & 105.4 & 1917/08 & 111.3 \\
\midrule
16  & 1928/04 &  78.1 & 1928/07 &  81.6 \\
17  & 1937/04 & 119.2 & 1937/04 & 125.1 \\
18  & 1947/05 & 151.8 & 1947/07 & 145.2 \\
19  & 1958/03 & 201.3 & 1958/03 & 186.1 \\
20  & 1968/11 & 110.6 & 1970/06 & 109.3 \\
\midrule
21  & 1979/12 & 164.5 & 1979/07 & 154.2 \\
22  & 1989/07 & 158.5 & 1991/02 & 153.5 \\
23  & 2000/04 & 120.7 & 2001/12 & 123.6 \\
24  & 2014/04 &  81.9 & 2014/04 & 82.0 \\
\bottomrule
\end{tabular}
\end{table}

These cycle maxima and their distributions are shown in
Figure~\ref{fig:CycleAmplitudes}. The mean amplitude of cycles~1 to 23 from the
International Sunspot Numbers is 114 with a standard deviation of
40. The mean amplitude of Cycles --4 to 23 from the Group Sunspot
Numbers is 90 with a standard deviation of 41.

\epubtkImage{CycleAmplitudes.png}{%
\begin{figure}[htbp]
\centerline{\includegraphics[width=0.8\textwidth]{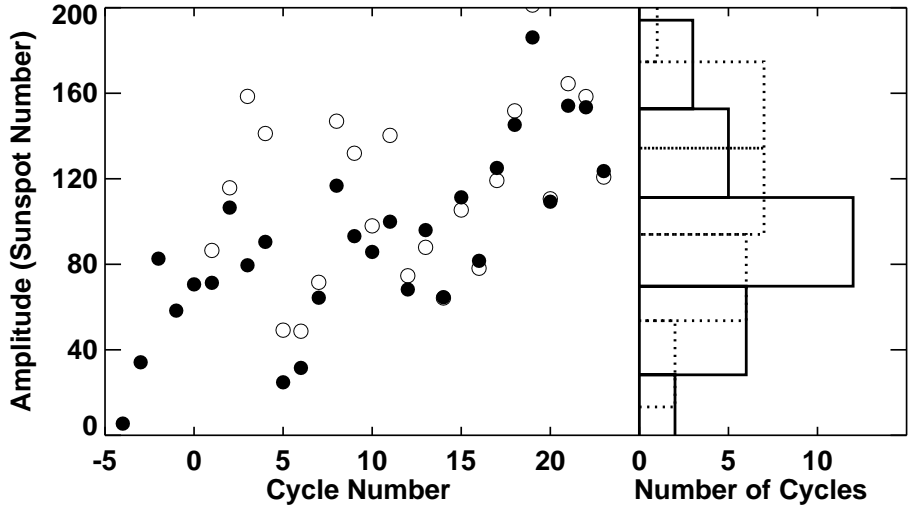}}
\caption{The left panel shows cycle amplitudes as functions of cycle 
number. The filled circles show the 13-month mean maxima with the Group 
Sunspot Numbers while the open circles show the maxima with the 
International Sunspot Numbers. The right panel shows the cycle amplitude 
distributions (solid lines for the Group values, dotted lines for the 
International values). The Group amplitudes are systematically lower than 
the International amplitudes for cycles prior to cycle~12 and have a nearly 
normal distribution. The amplitudes for the International Sunspot Number are 
skewed toward higher values.}
\label{fig:CycleAmplitudes}
\end{figure}}

\subsection{Cycle shape}

Sunspot cycles are asymmetric with respect to their maxima
\citep{Waldmeier:1935}. The elapsed time from minimum up to maximum is
almost always shorter than the elapsed time from maximum down to
minimum. An average cycle can be constructed by stretching and
contracting each cycle to the average length, normalizing each to the
average amplitude, and then taking the average at each month. This is
shown in Figure~\ref{fig:AverageCycle} for cycles~1 to 23. The average cycle
takes about 48 months to rise from minimum up to maximum and about 84
months to fall back to minimum again.

\epubtkImage{AverageCycle.png}{%
\begin{figure}[htbp]
\centerline{\includegraphics[width=0.8\textwidth]{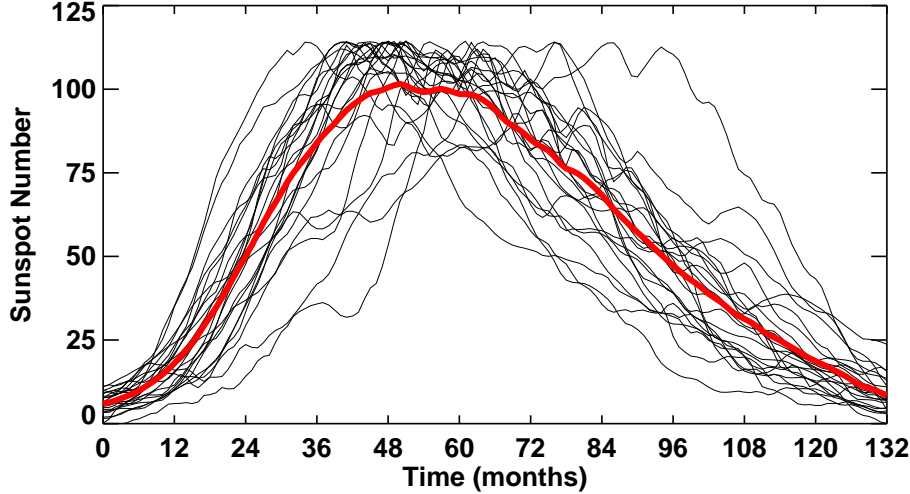}}
\caption{The average of cycles~1 to 23 (thick red line) normalized to the 
average amplitude and period. The average cycle is asymmetric in time with a 
rise to maximum over 4 years and a fall back to minimum over 7 years. The 23 
individual, normalized cycles are shown with thin black lines.}
\label{fig:AverageCycle}
\end{figure}}

Various functions have been used to fit the shape of the cycle and/or its 
various phases. \cite{Stewart:1938} proposed a single function for 
the full cycle that was the product of a power law for the initial rise and 
an exponential for the decline. They found the four parameters (starting 
time, amplitude, exponent for the rise, and time constant for the decline) 
that give the best fit for each cycle. \cite{Nordemann:1992} fit both the rise 
and the decay with exponentials that each required three parameters -- an 
amplitude, a time constant, and a starting time. \cite{Elling:1992} 
also fit the full cycle but with a modified $F$-distribution density function 
which requires five parameters. \cite{Hathaway:1994} suggested yet
another function -- similar to that of \cite{Stewart:1938} but with a
fixed (cubic) power law and a Gaussian for the decline. This function
of time
\begin{equation}
\label{eqn6}
F(t)=A\left( {\frac{t-t_0 }{b}} \right)^3\left[ {\exp \left( {\frac{t-t_0 
}{b}} \right)^2-c} \right]^{-1}
\end{equation}
has four parameters: an amplitude $A$, a starting time $t_{0}$, a rise time $b$, 
and an asymmetry parameter $c$. The average cycle is well fit with $A=195$, 
$b=56$, $c=0.8$, and $t_{0}= -4$ months (prior to minimum). This fit to the average 
cycle is shown in Figure~\ref{fig:AverageCycleFit}. \cite{Hathaway:1994} found 
that good fits to most cycles could be obtained with a fixed value for the 
parameter $c$ and a parameter $b$ that is allowed to vary with the amplitude -- 
leaving a function of just two parameters (amplitude and starting time) that
were well determine early in each cycle.

\epubtkImage{AverageCycleFit.png}{%
\begin{figure}[htbp]
\centerline{\includegraphics[width=0.8\textwidth]{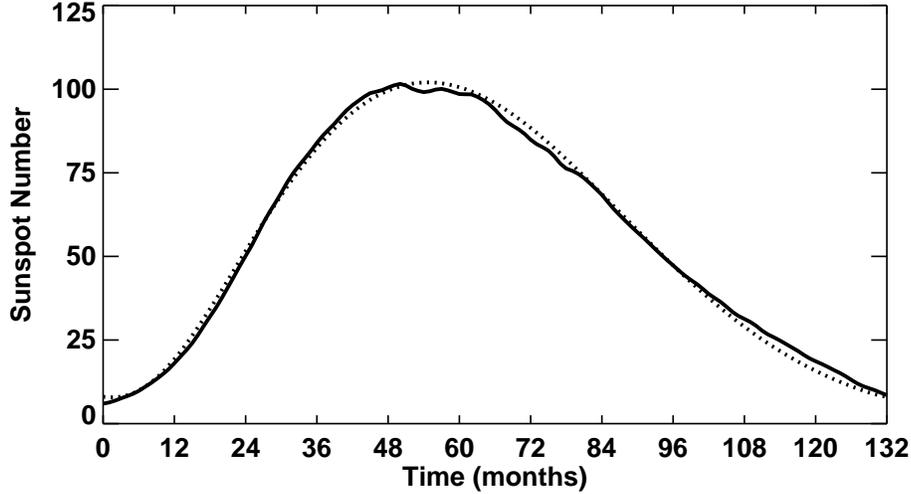}}
\caption{The average cycle (solid line) and the \cite{Hathaway:1994} 
functional fit to it (dotted line) from Equation~(\ref{eqn6}). This fit
has the average cycle starting 4 months prior to minimum, rising to
maximum over the next 54 months, and continuing about 18 months into
the next cycle.}
\label{fig:AverageCycleFit}
\end{figure}}

\cite{Li:1999} used a similar function to fit quarterly averages of the sunspot area
and also found that it could be reduced to a function of the same two parameters that were
well determine early in a cycle.
\cite{Volobuev:2009} introduced yet another (similar) function of four parameters for
sunspot numbers that could also be reduced to the same two parameters (note that
Volobuev refers to this as a one parameter fit by neglecting the need to fit or determine
the starting time).
Similar results have also been obtained by \cite{Du:2011}.

\subsection{Double peaks - the Gnevyshev Gap}
\label{sec:gnevyshev-gap}

These simple parametric functions all do a good job of fitting the average cycle shape
shown in Figures~\ref{fig:AverageCycle} and \ref{fig:AverageCycleFit}, but individual
cycles often have features that persistently deviate from these smooth profiles.
In particular, many cycles are observed to have double peaks.
\cite{Gnevyshev:1963} noted that cycle 19 had two maxima in solar activity as seen
in some activity indices (not so much in sunspot number but quite strong in coronal emission as seen in the
coronal green line at 5303 \AA) with a distinct 1-2 year gap (the Gnevyshev Gap).
He later \citep{Gnevyshev:1967, Gnevyshev:1977} suggested that the solar cycle is,
in general, characterized by two waves of activity and that these were responsible
for the double peaks.

This concept---two separate surges of solar activity---found further support
in the study of \cite{Feminella:1997} who noted that it is best seen in the occurrence
of large events (big flares but not small flares).

Another suggested source of double-peak behavior is north/south asymmetry in solar
activity (see Section~\ref{sec:ActiveHemispheres}).
Activity (e.g., sunspot number or area) can proceed in one hemisphere slightly out of
phase with activity in the other hemisphere.
This can result in an early peak associated with one hemisphere and a later peak associated
with the other hemisphere.
\cite{Norton:2010} examined this possibility and concluded that the Gnevyshev Gap
is a phenomena that occurs in both hemispheres and is not, in general, due to the
superposition of two hemispheres out of phase with each other.

\subsection{Rise time vs.\ amplitude - the Waldmeier Effect}
\label{sec:WaldmeierEffect}

A number of relationships have been found between various sunspot
cycle characteristics. Among the more significant relationships is the
Waldmeier Effect \citep{Waldmeier:1935, Waldmeier:1939} in which the
time it takes for the sunspot number to rise from minimum to maximum
is inversely proportional to the cycle amplitude. This is shown in
Figure~\ref{fig:WaldmeierEffect} for both the International Sunspot Number and the
10.7~cm radio flux data. Times and values for the maxima are taken
from the 24-month Gaussian given in Table~\ref{tab4}. Times for the
minima are taken from the average dates given in
Table~\ref{tab2}. Both of these indices exhibit the Waldmeier Effect,
but with the 10.7~cm flux maxima delayed by about 6 months. This is
larger than, but consistent with the delays seen by
\cite{Bachmann:1994}. The best fit through the Sunspot Number data
gives
\begin{equation}
\label{eq7}
\text{Rise Time (in months) } \approx 35 + 1800/\text{Amplitude (in Sunspot Number)}.
\end{equation}
While this effect is widely quoted and accepted it does face a number
of problems. \cite{Hathaway:2002} found that the effect was greatly
diminished when Group Sunspot Numbers were used (the anti-correlation
between rise time and amplitude dropped from --0.7 to
--0.34). Inspection of Figure~\ref{fig:WaldmeierEffect} clearly shows significant
scatter. \cite{Dikpati:2008-lett} noted that the effect is not seen for
sunspot area data. This is consistent with the data in
Tables~\ref{tab3} and \ref{tab4}, which show that significantly
different dates for maxima are found with sunspot area when compared
to sunspot number. The dates can differ by more than a year but
without any evidence of systematic differences (area sometimes leads
number and other time lags).

\epubtkImage{WaldmeierEffect.png}{%
\begin{figure}[htbp]
\centerline{\includegraphics[width=0.8\textwidth]{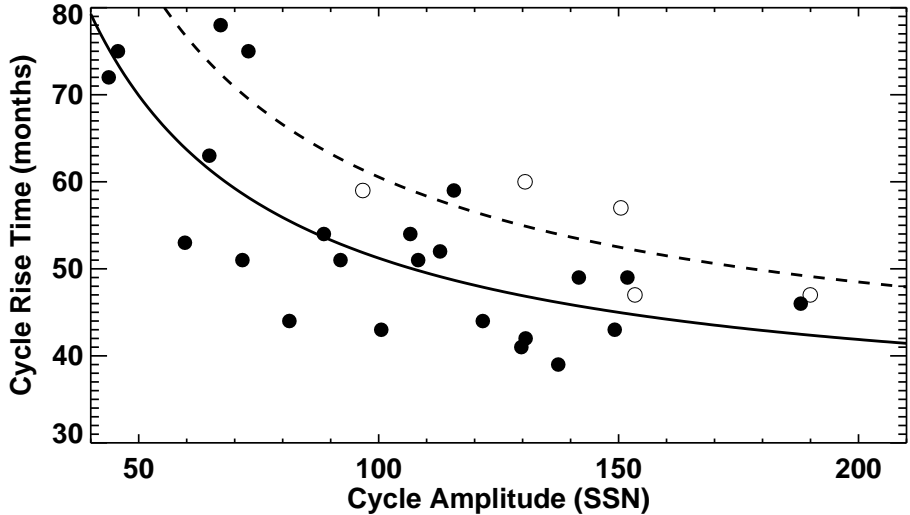}}
\caption{The Waldmeier Effect. The cycle rise time (from minimum to 
maximum) plotted versus cycle amplitude for International Sunspot Number 
data from cycles~1 to 23 (filled dots) and for 10.7~cm radio flux data from 
cycles~19 to 23 (open circles). This gives an inverse relationship between 
amplitude and rise time shown by the solid line for the Sunspot Number data 
and with the dashed line for the radio flux data. The radio flux maxima are 
systematically later than the Sunspot number data, as also seen in
Table~\ref{tab4}.}
\label{fig:WaldmeierEffect}
\end{figure}}

\subsection{Period vs.\ amplitude}
\label{sec:AmplitudePeriod}

Significant relationships are also found between cycle periods and 
amplitudes. The most significant relationship is between a cycle period and 
the amplitude of the following cycle \citep{Hathaway:1994,
Solanki:2002}. This is illustrated in Figure~\ref{fig:AmplitudePeriod}. The
correlation is fairly strong ($r = -0.68$, $r^{2} = 0.46$) and
significant at the 99\% level. While there is also a negative
correlation between a cycle period and its own amplitude, the
correlation is much weaker ($r = -0.37$, $r^{2} = 0.14$).

\epubtkImage{AmplitudePeriod.png}{%
\begin{figure}[htbp]
\centerline{\includegraphics[width=0.8\textwidth]{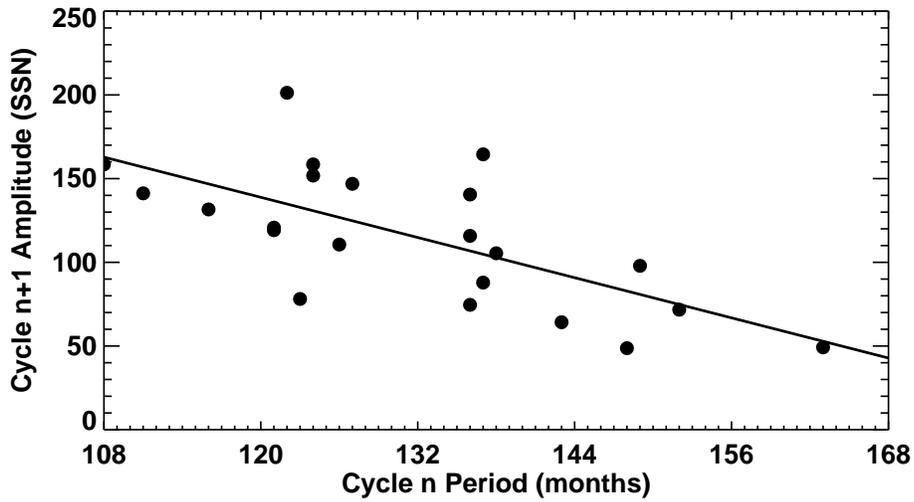}}
\caption{The Amplitude-Period Effect. The period of a cycle (from minimum 
to minimum) plotted versus following cycle amplitude for International 
Sunspot Number data from cycles~1 to 22. This gives an inverse relationship 
between amplitude and period shown by the solid line with Amplitude(n+1) = 380 -- 
2~\texttimes~Period(n).}
\label{fig:AmplitudePeriod}
\end{figure}}

\subsection{Maximum vs.\ minimum}
\label{sec:MaximumMinimum}

Although somewhat less significant, a relationship is also found between
cycle maxima and the previous minima \citep{Hathaway:1999}.
This is illustrated in Figure~\ref{fig:MaximumMinimum}. The
correlation is fair ($r = 0.56$, $r^{2} = 0.31$) and
significant at the 99\% level.

\epubtkImage{MaximumMinimum.png}{%
\begin{figure}[htbp]
\centerline{\includegraphics[width=0.8\textwidth]{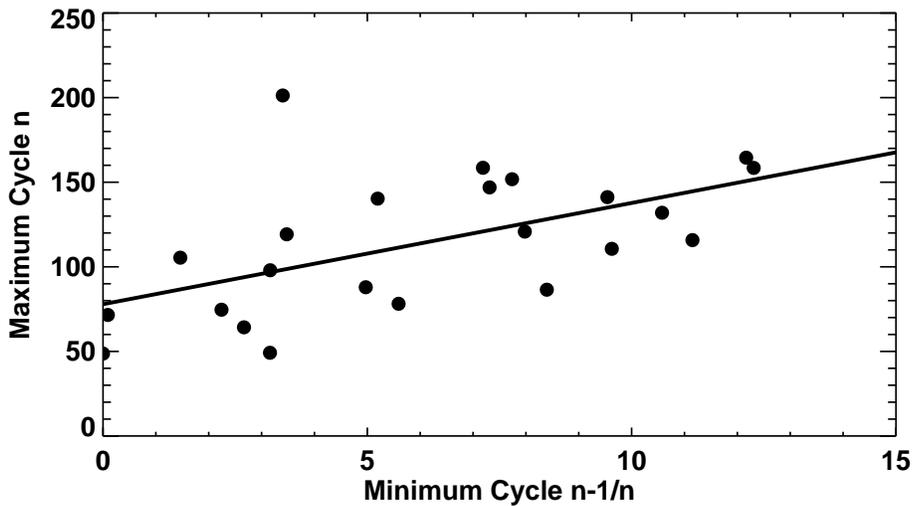}}
\caption{The Maximum-Minimum Effect. The maximum of a cycle plotted versus
minimum preceding the cycle given by the 13-month smoothed International 
Sunspot Number data from cycles~1 to 23. This gives a relationship 
between maximum and minimum shown by the solid line with Maximum(n) = 78 + 
6~\texttimes~Minimum(n).}
\label{fig:MaximumMinimum}
\end{figure}}

\subsection{Active latitudes - Sp\"{o}rer's Law}
\label{sec:ActiveLatitudes}

While Sp\"{o}rer's name is often attached to the concept of sunspot zones 
and their drift toward the equator, it appears that Carrington was the first 
to discover it. \cite{Carrington:1858} noted that the sunspots prior to the 
``minimum of frequency in February 1856'' were confined to an equatorial 
band well below 20\textdegree\ latitude. He went on to note that after that date 
two new belts of sunspots appeared at latitudes between 20\textdegree\ and 
40\textdegree\ latitude in each hemisphere.
This equatorward drift of the sunspot zones is a key characteristic of the
solar cycle---a characteristic that is often difficult to reproduce
in dynamo models with several possible different mechanisms proposed
\citep[see][]{Charbonneau:2010}.

Cycle-to-cycle variations in this equatorward drift have been reported and 
latitudes of the sunspot zones have been related to cycle amplitudes. 
\cite{Vitinskij:1976} used the latitudes of sunspot near minimum as a predictor 
for the amplitude of cycle~21. Separating the cycles according to size now 
suggests that this is a poor indicator of cycle amplitude. Regardless of 
amplitude class, all cycles start with sunspot zones centered at about
25\textdegree.

\cite{Li:2001} used the RGO/NOAA sunspot area and position data plotted
in Figure~\ref{fig:ButterflyDiagram} to quantify the latitudinal drift by
fitting the quarterly averaged sunspot group area centroid positions as
functions of time, with a quadratic in time relative to the time of minimum for each solar cycle.
The individual sunspot cycles can be separated near the time of minimum by the latitudes of the 
emerging sunspots (and more recently by magnetic polarity data as well).
They found that, on average from 1874 to 1999, both hemispheres had the same behavior
with faster equatorward drift early in the cycle and slower drift late in the cycle
and an average drift rate of $\sim1.6$\textdegree yr$^{-1}$. 

\cite{Hathaway:2003} used the same data to investigate the variation of the
equatorward drift with cycle period and amplitude.
They calculated the centroid positions of the sunspot group areas in each hemisphere
for each solar rotation in individual solar cycles, and fit those positions
to quadratics in time relative to the time of maximum for each cycle.
They found that cycles with higher drift rates at maximum tended to have shorter
periods and larger amplitudes but with a better correlation between drift rate and
the amplitude of the N+2 cycle \citep{Hathaway:2004}.

Recently, \cite{Hathaway:2011} found that the active latitudes follow a standard path for all
cycles (regardless of cycle amplitude) when time is measured relative to the cycle
starting times determined from fitting monthly sunspot numbers to the parametric
curves given by Equation~\ref{eqn6}.
The data show far less scatter when plotted  relative to this starting time, $t_0$, and are well
fit with an exponential function:

\begin{equation}
\bar{\lambda}(t) = 28^\circ \exp\left[-(t - t_0)/90\right]
\label{eqn8}
\end{equation}

\noindent
where $\bar{\lambda}$ is the active latitude and time, $t$, is measured in months.

These centroid positions are plotted as 
functions of time relative to $t_0$ in Figure~\ref{fig:ActiveLatitudeDrift}.
The area weighted averages of these positions in 6-month intervals are shown with the 
colored lines for different amplitude cycles. At the start of each cycle the centroid 
position of the sunspot areas is about 28\textdegree\ from the equator. The 
equatorward drift is more rapid early in the cycle and slows late in the 
cycle---eventually stopping at about 7\textdegree\ from the equator.

\epubtkImage{ActiveLatitudeDrift.png}{%
\begin{figure}[htbp]
\centerline{\includegraphics[width=0.8\textwidth]{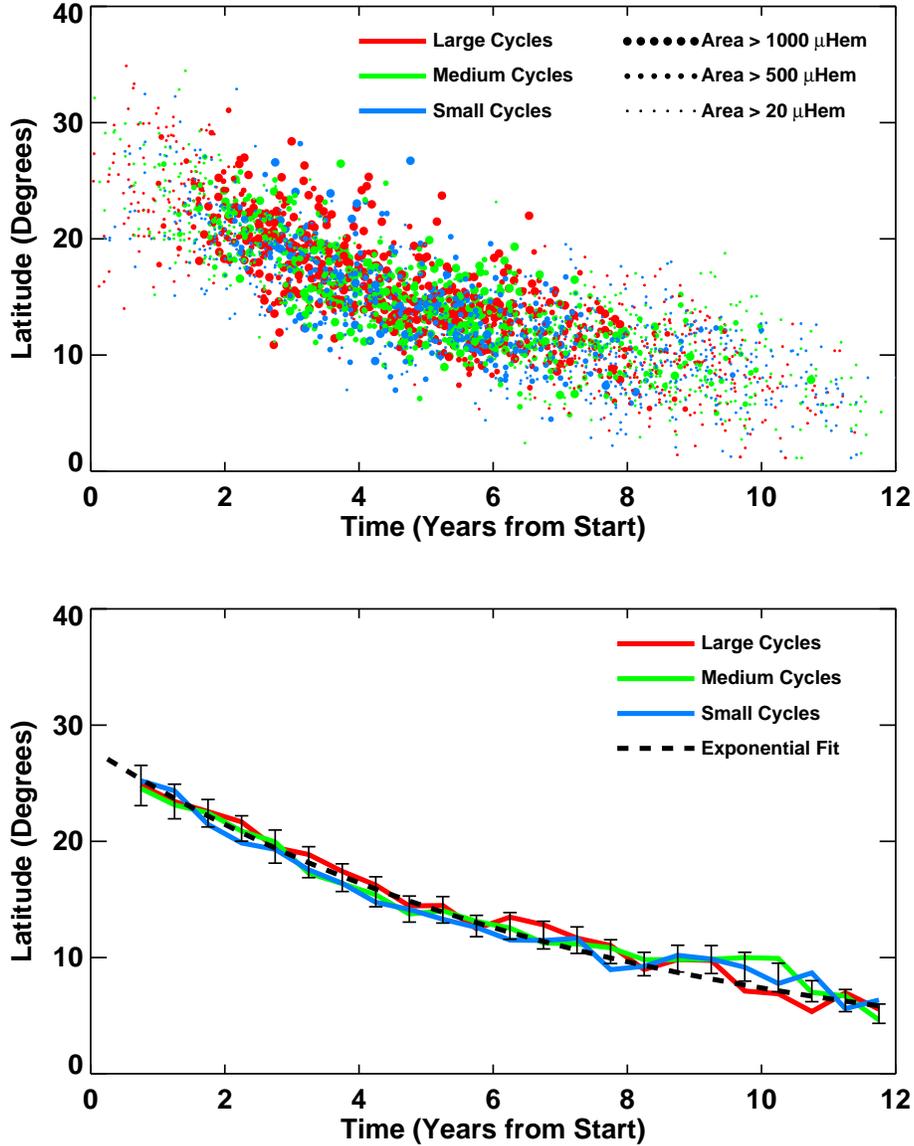}}
\caption{(Top) Latitude positions of the sunspot area centroid in each 
hemisphere for each Carrington Rotation as functions of time from cycle 
start. Three symbol sizes are used to differentiate data according to the 
average of the daily total sunspot area for each hemisphere and rotation.
(Bottom) The centroids of the centroids in 6-month intervals are shown 
for large amplitude cycles (red line), medium amplitude cycles (green line), 
and small amplitude cycles (blue line).
The exponential fit to the active latitude positions (Equation \ref{eqn8})
is shown with the black dashed line and $2\sigma$ error bars.}
\label{fig:ActiveLatitudeDrift}
\end{figure}}

The latitudinal width of the sunspot zones also varies over the cycle and as 
a function of cycle amplitude. This is illustrated in Figure~\ref{fig:ActiveLatitudeWidth},
where the latitudinal widths (standard deviation about the mean latitude)
of the sunspot zones are plotted for each hemisphere for each 
Carrington rotation as functions of time since the start of each cycle.
The active latitude bands are narrow at minimum, expand 
to a maximum width at the time of maximum, and then narrow again 
during the declining phase of the cycle. Larger cycles achieve greater 
widths than do smaller cycles.
\cite{Ivanov:2011} found a linear relationship between the width
(maximum latitude - minimum latitude, in their study) of the sunspot latitude bands and the
number of sunspot groups with no dependence on cycle amplitude.
Comparing the RMS width to the sunspot area confirms the lack of any
relational dependence on cycle strength but indicates a distinctly nonlinear
relationship with an asymptotic limit to the widths as the total sunspot area increases
(bottom panel of Figure~\ref{fig:ActiveLatitudeWidth}).
A satisfactory fit to the data (shown by the black line in the bottom panel
of Figure~\ref{fig:ActiveLatitudeWidth}) is given by

\begin{equation}
\sigma_\lambda(A) = 1.5^\circ + 3.8^\circ \left[1 - \exp(-A/400)\right]
\label{eqn9}
\end{equation}

\noindent
where $\sigma_\lambda$ is the RMS width of the sunspot zones and $A$ is the total sunspot
area in $\mu$Hem.

\epubtkImage{ActiveLatitudeWidth.png}{%
\begin{figure}[htbp]
\centerline{\includegraphics[width=0.8\textwidth]{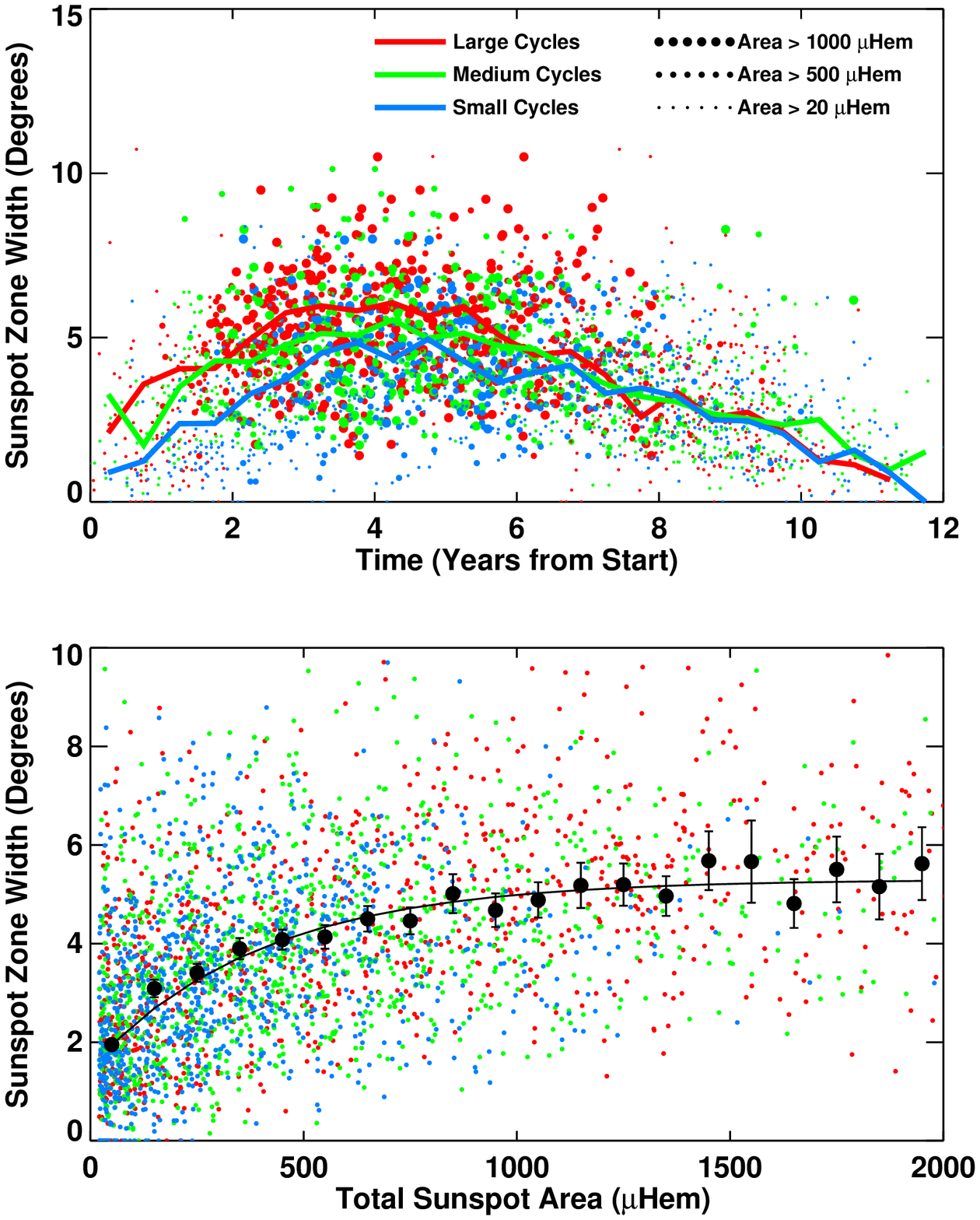}}
\caption{(Top) Latitudinal widths of the sunspot area centroid in each 
hemisphere for each Carrington Rotation as functions of time from cycle 
start. Three symbol sizes are used to differentiate data according to the 
daily average of the sunspot area for each hemisphere and rotation. The 
centroids of the centroids in 6-month intervals are shown
for large amplitude cycles (red line), medium amplitude cycles (green line), 
and small amplitude cycles (blue line).
(Bottom) Latitudinal widths as functions of total sunspot area with
color coded symbols for cycle strength. The black dots with $2\sigma$
error bars show the data binned in $100 \mu$Hem intervals.
The black line is given by Equation~\ref{eqn9}.}
\label{fig:ActiveLatitudeWidth}
\end{figure}}

\cite{Becker:1954} and \cite{Waldmeier:1955} had earlier noted that in large
cycles, the latitudes of the sunspot zones are higher at maximum than
in small cycles.
\cite{Li:2003} analyzed the latitudinal distribution of sunspot groups for each cycle
and found that the average latitudes were higher in bigger cycles and that there were
more sunspot groups at high latitudes ($35^\circ$ and higher) in bigger cycles.
This result was supported by the more extensive study of \cite{Solanki:2008}
who calculated the latitudinal moments of the sunspot group areas.
\cite{Solanki:2008} used the RGO data from 1874 to 1976, supplemented with Soviet data
from 1977 to 1985 and Mount Wilson data from 1986 to 2004 \citep[see][]{Balmaceda:2009}.
As with the earlier studies, they separated data from each cycle using diagonal
lines in the butterfly diagram.
They then calculated the latitudinal moments (total area, mean latitude, width, skew,
and kurtosis) of the sunspot area integrated over each individual solar cycle for
each hemisphere.
They found that bigger cycles had higher mean latitudes and greater sunspot zone widths
and that the distributions tended to be (weakly) skewed toward the equator but with no systematic
kurtosis.
(Somewhat surprisingly, they found slightly different behavior in the two hemispheres.
The range of variability from cycle to cycle in total area, mean latitude, and width was
less in the southern hemisphere and the correlations between total area and mean latitude
and total area and width were stronger in the southern hemisphere.)

These results are all consistent with the data shown in Figure~\ref{fig:ActiveLatitudeDrift}
and Figure~\ref{fig:ActiveLatitudeWidth}.
Large amplitude cycles reach their maxima sooner than do medium or small
amplitude cycles (the Waldmeier Effect -- Section~\ref{sec:WaldmeierEffect}).
Thus, the sunspot zone latitude at the maximum of a large cycle will be
higher simply because maximum occurs earlier and sunspot zones are still at
higher latitudes.
Likewise, the average latitude for a large cycle will be higher for the same reason.

\subsection{Active hemispheres}
\label{sec:ActiveHemispheres}

Comparisons of the activity in each solar hemisphere have long shown significant 
asymmetries. \cite{Spoerer:1889} and \cite{Maunder:1890, Maunder:1904} noted that there 
were often long periods of time when most of the sunspots were found 
preferentially in one hemisphere and not the other. \cite{Waldmeier:1971} found 
that this asymmetry extended to other measures of activity including 
faculae, prominences, and coronal brightness. \cite{Roy:1977} reported that major 
flares and magnetically complex sunspot groups also showed strong 
north--south asymmetry.

The nature of the asymmetry is often characterized in different ways that
can lead to different conclusions.
Simply quantifying the asymmetry itself is problematic.
Taking the difference between hemispheric measures of activity 
(absolute asymmetry) produces strong signals around the times of maxima simply
because the numbers are large.
Taking the ratio of the difference to the sum
(relative asymmetry) produces strong signals around the times of minima because
the differences are divided by small numbers.

One aspect of asymmetry might make activity in one hemisphere stronger
than in the other hemisphere, but without any shift in phase (cycle minima
and maxima occurring simultaneously in each hemisphere).
Another aspect of asymmetry might be reflected in a change in phase but without
a corresponding change in strength---one hemisphere rising to maximum before the other.
We find evidence for both of these aspects.
However, it is well worth noting that the two hemispheres never get very far out of phase
with each other (as seen in Figure~\ref{fig:ButterflyDiagram}) .
This is an indication of a fundamental linkage between the two hemispheres that must
be reproduced in dynamo models.

\cite{Carbonell:1993} examined the relative asymmetry
in sunspot areas with a variety of statistical tools and concluded
that the signal is dominated by a random (and intermittent) component,
but contains one component that varies over a cycle and a second component that
gives long-term trends. The variation in the strength of the asymmetry
over the course of an average cycle is strongly dependent upon how the
asymmetry is quantified (strong at minimum for relative asymmetry,
strong at maximum for absolute asymmetry).

Sunspot numbers (and most other solar activity indicators through their direct association
with the emergence of sunspot groups) follow Poisson statistics, which results in
variability proportional to the square-root of the number itself.
Taking the ratio of the hemispheric differences to the square-root of the sums
provides a measure of asymmetry that does not tend to favor either maximum or
minimum phases.

\epubtkImage{NormalizedAsymmetry.png}{%
\begin{figure}[htbp]
\centerline{\includegraphics[width=0.8\textwidth]{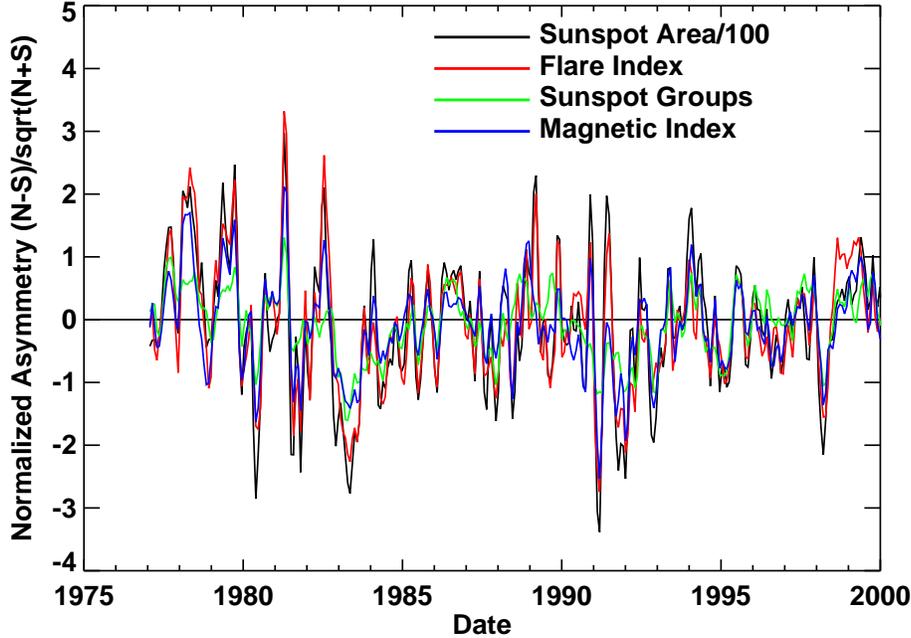}}
\caption{Normalized north-south asymmetry $(N-S)/\sqrt(N+S)$ in four different 
activity indicators for individual Carrington rotations.
Sunspot area is plotted in black. The Flare Index 
is shown in red. The number of sunspot groups is 
shown in green. The Magnetic Index is plotted in blue.}
\label{fig:NormalizedAsymmetry}
\end{figure}}

Figure~\ref{fig:NormalizedAsymmetry} shows this ``normalized'' asymmetry
for several key indicators. It is clear 
from this figure that hemispheric asymmetry is real (it consistently appears 
in all four indicators) and is often persistent---lasting for many years at 
a time.
Figure~\ref{fig:NormalizedSunspotAreaAsymmetry} uses the RGO/NOAA sunspot group
area data to extend the asymmetry record from 1874 to the present.
This longer time interval shows that the asymmetry usually switches from
north-dominant to south-dominant on time scales shorter than an 11-year sunspot
cycle.
Note that there are about 30 changes of sign in 130 years which gives a typical time
scale of about 4 years---on the order of half a solar cycle.

\epubtkImage{NormalizedSunspotAreaAsymmetry.png}{%
\begin{figure}[htbp]
\centerline{\includegraphics[width=0.8\textwidth]{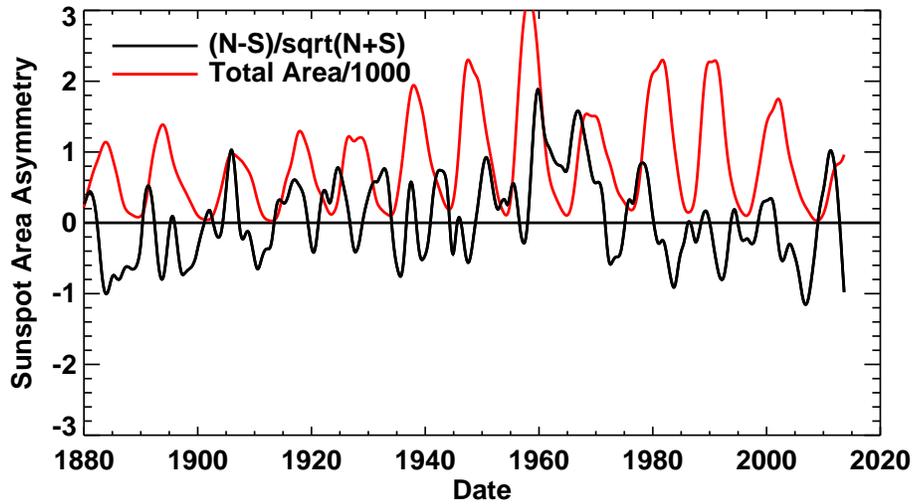}}
\caption{Smoothed, normalized north-south asymmetry in sunspot area.
The hemispheric asymmetry is shown by the black line while the total area scaled by 1/1000 
is shown by the red line for reference.}
\label{fig:NormalizedSunspotAreaAsymmetry}
\end{figure}}

Systematic variations over the course of a solar cycle or as a function of 
cycle amplitude have been suggested, but these variations have invariably 
been found to change from cycle to cycle.
For example, \cite{Newton:1955} showed that the northern hemisphere dominated in 
the early phases of cycles~12\,--\,15 with a switch to dominance in the south 
later in each cycle while the opposite was true for cycles~17\,--\,18.
\cite{Waldmeier:1957, Waldmeier:1971} noted that a significant
part of these variations can be accounted for by the fact that the two
hemispheres are not exactly in phase. When the northern hemisphere
activity leads that in the southern hemisphere, the north will
dominate early in the cycle while the south will dominate in the
declining phase.

\epubtkImage{NorthAndSouthSunspotArea.png}{%
\begin{figure}[htbp]
\centerline{\includegraphics[width=0.8\textwidth]{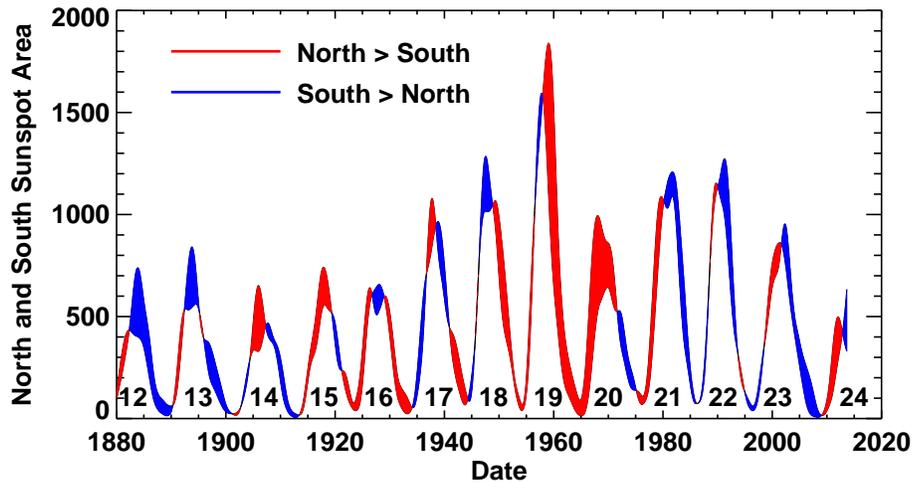}}
\caption{Smoothed monthly sunspot areas for northern and southern hemispheres separately.
The difference between the two curves is filled in red if the north dominates or
in blue if the south dominates.}
\label{fig:NorthAndSouthSunspotArea}
\end{figure}}

These changes in phase can be seen more clearly when data for each hemisphere is
plotted separately as was done by \cite{Temmer:2006} for hemispheric sunspot numbers
from 1945 to 2004.
Figure~\ref{fig:NorthAndSouthSunspotArea} uses the RGO/USAF sunspot area data
from 1874 to 2014 to illustrate this effect.
\cite{Zolotova:2006} employed cross-recurrence plots to explore the
phase relationship between northern and southern sunspot areas.
They found that the hemispheres drifted slightly out of phase over the course of
single or multiple cycles but could suddenly shift to the opposite phase.
This is seen in Figure~\ref{fig:NorthAndSouthSunspotArea} where cycles 17-19 have
the south leading the north, but then cycle 20 starts with the north leading the south.

\cite{Donner:2007} used a wavelet analysis on the sunspot area data and, as might be expected,
found near phase coherence only for periods near 11 years.
By examining the inter-hemispheric phase difference they found a similar pattern to that
found by \cite{Zolotova:2006} and concluded that, at a period of 10.75 years, the two
hemispheres never shifted out of phase by more than $\pm 10$ months or, equivalently,
10\% of the cycle period.
These small phase differences are consistent with the findings of \cite{Norton:2010} and
support their conclusion the the Gnevyshev gap is not due to the two hemispheres
getting out of phase (this would require phase shifts of 24 months or more).

\subsection{Active longitudes}
\label{sec:ActiveLongitudes}

Sunspots and solar activity also appear to cluster in ``active longitudes.''
\cite{Maunder:1905} noted that during cycle 13 (1891-1901) sunspots favored three specific
longitudes, with one longitude range in particular being more active than the others.
\cite{Bumba:1965} and \cite{Sawyer:1968} noted that new active regions grow in areas
previously occupied by old active regions and referred to these as ``complexes of activity''
while \cite{Castenmiller:1986} referred to similar structures as ``sunspot nests.''
\cite{Bogart:1982} found that this results in a periodic signal that is evident in
the sunspot number record.

Figure~\ref{fig:ActiveLongitudes} illustrates the active longitude phenomena.
In Figure~\ref{fig:ActiveLongitudes}a the sunspot area in 5\textdegree\ longitude bins,
averaged over 1805 solar rotations since 1878 and normalized to the
average value per bin, is plotted as a function of Carrington longitude.
The $2\,\sigma $ uncertainty in these values is represented by the dotted lines.
This $2\,\sigma $ limit is reached at several longitudes and significantly exceeded at two
(85\textdegree\,--\,90\textdegree\ and 90\textdegree\,--\,95\textdegree).
Figure~\ref{fig:ActiveLongitudes}b shows similar data for each individual cycle
with the normalized value offset in the vertical by the sunspot cycle number.
There are many peaks at twice the normal value and one, in cycle~18 at
85\textdegree\,--\,90\textdegree, at three times the normal value.
Some of these peaks persist from one cycle to the next,
a result that has been noted by many authors including \cite{Balthasar:1983},
\cite{Bumba:1991}, \cite{Miklailutsa:1994}, and \cite{Bai:2003}.

\epubtkImage{ActiveLongitudes.png}{%
\begin{figure}[htbp]
\centerline{\includegraphics[width=0.8\textwidth]{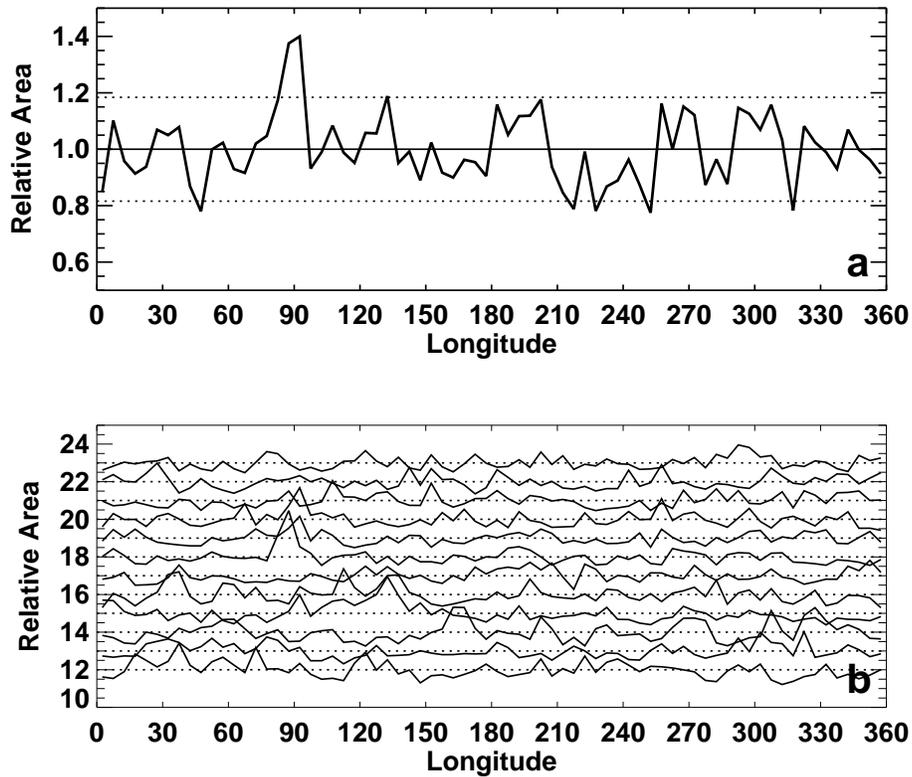}}
\caption{Active longitudes in sunspot area. The normalized sunspot area in 
5\textdegree\  longitude bins is plotted in the upper panel (a) for the years 
1878\,--\,2009. The dotted lines represent two standard errors in the normalized 
values. The sunspot area in several longitude bins meets or exceeds these 
limits. The individual cycles (12 through 23) are shown in the lower panel 
(b) with the normalized values offset in the vertical by the cycle number. 
Some active longitudes appear to persist from cycle to cycle.}
\label{fig:ActiveLongitudes}
\end{figure}}

\cite{Henney:2002} noted the persistence of magnetic structures in the
northern hemisphere at preferred longitudes (drifting slightly due to the latitude)
for two decades but also noted that the sunspot records suggests that two decades
is about the limit of such persistence (as seen in Figure~\ref{fig:ActiveLongitudes}b).
However, \cite{Berdyugina:2003} conclude that active longitudes can persist
for much longer if changes in rotation rate relative to the Carrington rate
are accounted for.
They also found that the active longitude in
the northern hemisphere tends to be shifted by 180\textdegree\ in
longitude from that in the southern hemisphere.

\subsection{Active region tilt - Joy's Law}
\label{sec:ActiveRegionTilt}

The tilt of active regions---Joy's Law---is another important characteristic
of the sunspot cycle.
First discovered by Joy as reported by \cite{Hale:1919}, this active region tilt
systematically places following-polarity magnetic flux at higher latitudes than
the leading-polarity magnetic flux.
Since the following-polarity is opposite in sign to the polar fields at the start of
each cycle, the poleward transport of this flux by diffusion and the meridional flow leads to
the polar field reversals at cycle maximum and the build up of new polar fields
during the declining phase of each cycle \citep[see][]{Sheeley:2005, Charbonneau:2010}.

\cite{Wang:1989} studied the tilt of some 2700 bipolar magnetic regions that erupted
during cycle 21 (1976-1986) by visually inspecting daily magnetograms from NSO/Kitt Peak.
They found that the average tilt angle increased with latitude at the rate of
about 4\textdegree\ for each 10\textdegree\ of latitude.
They noted that the scatter about this average was quite large with the standard
deviations about as large as the average value itself.
Furthermore, they noted that this RMS scatter was larger for smaller active regions.
They concluded that the latitudinal variation in the tilt
did not change systematically from 1977 to 1985 during the course of cycle 21.

\cite{Howard:1991a} analyzed daily Mount Wilson magnetograms acquired over two cycles
(cycles 20 and 21 from 1967 to 1990).
While he noted some differences between his results and those of \cite{Wang:1989}
(particularly at the highest and lowest latitudes where errors are large),
and suggested that these differences can be attributed to differences in data and
data analysis, it nonetheless appears that both studies find a similar dependence
of the tilt angles on latitude (about 4\textdegree\ for each 10\textdegree\ of latitude)
and no dependence on cycle or cycle phase.

In a companion study \cite{Howard:1991b} analyzed sunspot data derived from daily
white-light photographs taken at Mount Wilson from 1917 to 1985.
While this data lacks polarity information, the leading and following sunspots are
identified by their positions relative to the central meridian distance of the group.
He found a different relationship between tilt and latitude using this sunspot
data---about 2.5\textdegree\ for each 10\textdegree\ of latitude.
This result was also obtained by \cite{Sivaraman:1999} using a the same technique
on white-light photographs from both Mount Wilson and from Kodaikanal covering
the years 1906 to 1987.
\cite{Sivaraman:1999} also looked at the tilt angle residuals, the deviations of the
tilt angles for the sunspot groups from the average at the group's latitude, and found
no significant variation with average cycle phase for cycles 15 through 21.

In a later study \cite{Sivaraman:2007} noted that the sunspot group tilts relax
toward the average tilt at their emergent latitude after initial emergence.
This behavior, relaxation to the average tilt rather than zero tilt, was also
found by \cite{Kosovichev:2008} in their study of magnetograms from SOHO/MDI.

\epubtkImage{ActiveRegionTilt.png}{%
\begin{figure}[htbp]
\centerline{\includegraphics[width=0.8\textwidth]{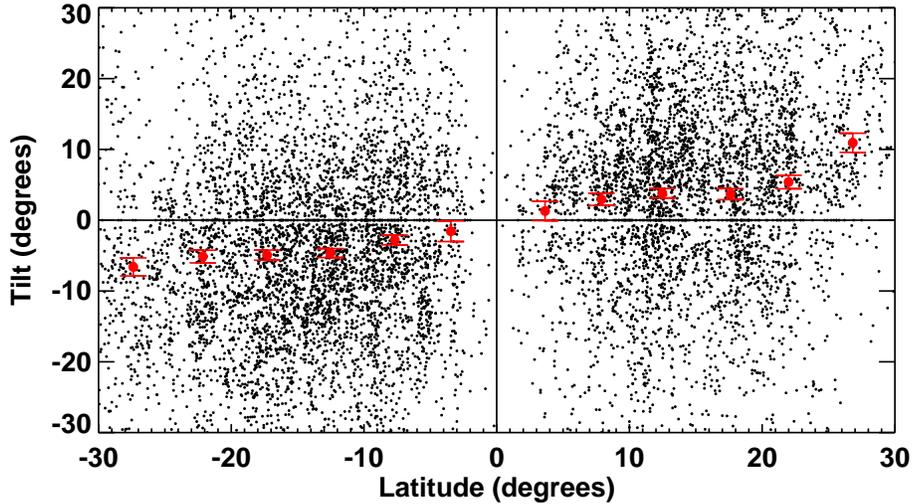}}
\caption{Active region tilt from SOHO/MDI magnetograms over cycle 23.
Each black data point gives the tilt (angle between leading-polarity center of gravity
and following-polarity center of gravity measured clockwise from the west to east
line) for a NOAA active region on a given day as a function of its latitude.
The red dots with $2\sigma$ error bars give the averages in 5\textdegree~ latitude
bins.}
\label{fig:ActiveRegionTilt}
\end{figure}}

The latitudinal variation in the active region tilt from an east-west orientation
seems decidedly different depending upon the source of the observations.
Measurements made from magnetograms \citep{Wang:1989, Howard:1991a, Tlatov:2010,
Li:2012, Stenflo:2012}
indicate higher values (4-5\textdegree\ for each 10\textdegree\ of latitude)
while measurements made from white-light images \citep{Howard:1991b, Sivaraman:1999}
indicate lower values (2.5\textdegree\ for each 10\textdegree\ of latitude).
Figure~\ref{fig:ActiveRegionTilt} illustrates the scatter in the measurements and
shows how the lower values can be obtained with magnetogram measurements when the data
is restricted to the sunspot groups themselves.

The first investigation of cycle-to-cycle variations in active region tilt was
undertaken by \cite{DasiEspuig:2010}.
They used the Mount Wilson and Kodaikanal white-light sunspot data and found
tilt angles as functions of latitude similar to those found from these data
by \cite{Howard:1991b} and by \cite{Sivaraman:2007}.
Although they were unable to find the Joy's Law relation for each cycle, they
did calculate a proxy---the average tilt angle divided by the average
latitude---and found that it varied inversely with cycle amplitude.
Part of this relationship could be due to the fact that the average latitude
is bigger in bigger cycles (see Section~\ref{sec:ActiveLatitudes}: big
cycles reach maximum early while the sunspot zones are at higher latitude)
but that should be offset by the larger tilt angles at the higher latitudes.
This relationship---less active region tilt in large cycles---could provide
an important feedback that regulates the amplitudes of the solar cycles
\citep[see][]{Cameron:2010}.

\subsection{The extended solar cycle}
\label{sec:ExtendedCycle}

The concept of ``extended'' solar cycles---solar cycles that extend
further back in time and to higher latitudes than indicated by the
sunspot zones---started with observations of ephemeral regions \citep{Martin:1979}
but gained support with observations of the torsional oscillations
\citep{Snodgrass:1987}.
The torsional oscillations \citep{Howard:1980, Howe:2009} are weak
($\sim 5$ m s$^{-1}$) perturbations
to the differential rotation profile in the form of a faster-than-average
stream on the equatorward side of a sunspot zone and a slower-than-average
stream on the poleward side (giving enhanced latitudinal shear in the sunspot
zones themselves).
These fast and slow zonal flow features move equatorward along with the sunspot
zones themselves over the course of each cycle.
However, \cite{Snodgrass:1987} noted that these features can be seen starting
at even higher latitudes well before the emergence of the first sunspots of a cycle.

\cite{Wilson_etal:1988} reported on additional observations that supported the
concept of an extended cycle of activity that begins near the poles at about the time
of the maximum of the previous cycle and drifts equatorward over the course
of 18---22 years.
These additional data include observations of coronal emissions and observations of
ephemeral active regions (see Section~\ref{sec:EMR}).

The coronal emission data are derived from scans around the limb obtained from
ground-based observatories in the green line of Fe XIV \citep[see][]{Altrock:1988}.
One component of this coronal emission emanates from coronal loops overlying
active regions.
This component follows the sunspot zones along their equatorward track.
A second component is associated with prominences in general and polar crown
filaments in particular.
This component moves poleward with the polar crown filaments as the polar
fields reverse at cycle maximum (these filaments lie over the neutral line
between the old polar fields and the following-polarity magnetic flux from the new cycle active regions).
Shortly after maximum (and the rush to the poles of the polar crown filaments)
a third component is seen at high latitudes to slowly move equatorward parallel
to the sunspot zones, eventually connecting to the sunspot zones of the next cycle.

This third component may be associated with the ephemeral regions.
Ephemeral regions tend to follow Hale's polarity rules but with an even larger
scatter in tilt angles than is seen with active regions.
Ephemeral regions can be associated with one cycle or the next by their latitude
distributions and their statistically dominant orientation \citep{Martin:1979}.
Ephemeral regions are found at higher latitudes than sunspots, with distributions
that suggest that they represent extensions of the sunspot zones back in time
and to higher latitudes --- starting at about the time of the previous cycle
maximum \citep{Harvey:1992, Tlatov:2010}. 
However, the interpretation of this coronal signal as an indication of
an extended cycle has been questioned by \cite{Robbrecht:2010}, who have
reproduced the signal with magnetic maps produced by the transport of
magnetic flux from active regions alone.

\newpage 

\section{Long-Term Variability}
\label{sec:ltvar}

Systematic variations from cycle to cycle and over many cycles could be 
significant discriminators in models of the solar cycle and might aid in 
predicting future cycles. Several key aspects of long-term variability have 
been noted: a 70-year period of extremely low activity from 1645 to 1715 
(the Maunder Minimum); a gradual increase in cycle amplitudes since the 
Maunder Minimum (a secular trend); an 80\,--\,90 year variation in cycle 
amplitudes (the Gleissberg Cycle); a two-cycle variation with odd-numbered 
cycles higher than the preceding even-numbered cycles (the Gnevyshev--Ohl 
Effect); a 210-year cycle in radio isotope proxies (the Suess Cycle); and 
other long term variations seen in radio isotopes. These aspects of 
long-term variability are examined in this section.

\subsection{The Maunder Minimum}

\cite{Maunder:1890}, reporting on the work of Sp\"{o}rer, noted that for a seventy-year
period from 1645\,--\,1715 the course of the sunspot cycle was interrupted. 
\cite{Eddy:1976} provided additional references to the lack of activity during 
this period and referred to it as the Maunder Minimum. He noted that many 
observers prior to 1890 had noticed this lack of activity and that both he 
and Maunder were simply pointing out an overlooked aspect of solar activity.

\cite{Hoyt:1998} compiled observations from numerous sources to
provide nearly complete coverage of sunspot observations during the
period of the Maunder Minimum. These observations (Figure~\ref{fig:MaunderMinimum})
clearly show the lack of activity and apparent cessation of the
sunspot cycle during the Maunder Minimum. Nonetheless,
\cite{Beer:1998} find evidence for a weak cyclic variation in
\super{10}Be during the Maunder Minimum, suggesting that the magnetic
cycle was still in progress but too weak to produce the intense
magnetic fields in sunspots. In addition, \cite{Ribes:1993} found that
the sunspots that were observed in the latter half of the Maunder
Minimum were at low latitudes and dominant in the southern hemisphere
-- another indication of weak/marginal magnetic fields.

\epubtkImage{MaunderMinimum.png}{%
\begin{figure}[htbp]
\centerline{\includegraphics[width=0.8\textwidth]{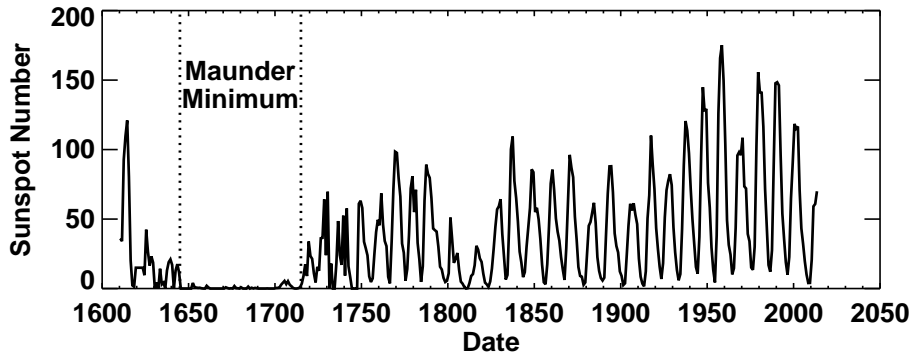}}
\caption{The Maunder Minimum. The yearly averages of the daily Group 
Sunspot Numbers are plotted as a function of time. The Maunder Minimum 
(1645\,--\,1715) is well-observed in this dataset.}
\label{fig:MaunderMinimum}
\end{figure}}

\subsection{The secular trend}

Since the Maunder Minimum there seems to have been a steady increase
in sunspot cycle amplitudes \citep{Wilson:1988}.
This is readily seen in the yearly Group Sunspot Numbers plotted in Figure~\ref{fig:MaunderMinimum} and in the cycle amplitudes for
Group Sunspot Numbers plotted in Figure~\ref{fig:CycleAmplitudes}.
\cite{Hathaway:2002} found a correlation coefficient of 0.7 between
cycle amplitude and cycle number.
Radioisotopes also show this recent trend \citep{Solanki:2004} and
indicate many upward and downward trends over the last 11,000 years.
It is well worth noting, however, that this linear trend is not so
apparent in the International Sunspot Numbers plotted in
Figure~\ref{fig:CycleAmplitudes}.
Furthermore, the recent reexaminations of the sunspot number
\citep{Svalgaard:2012}, indicate little if any secular increase in
cycle amplitudes since the Maunder Minimum.

\subsection{The Gleissberg Cycle}
\label{sec:gleissberg-cycle}

Numerous authors have noted multi-cycle periodicities in the sunspot cycle 
amplitudes.
\cite{Gleissberg:1939} noted a periodicity of seven or
eight cycles in cycle amplitudes from 1750 to 1928.
While \cite{Garcia:1998} suggest that a third period of this cycle can be
found in the sunspot data, others \citep{Hathaway:1999} suggest that
the period is changing or \citep{Rozelot:1994, Ogurtsov:2002} that it
consists of two different components (one with a 90\,--\,100 year
period and a second with a 50\,--\,60 year period).
A simple sinusoid fit to the residual cycle amplitudes when the secular trend
is removed now gives a 9.1 cycle periodicity.
This best fit is shown in Figure~\ref{fig:GleissbergCycle}. 

\epubtkImage{GleissbergCycle.png}{%
\begin{figure}[htbp]
\centerline{\includegraphics[width=0.8\textwidth]{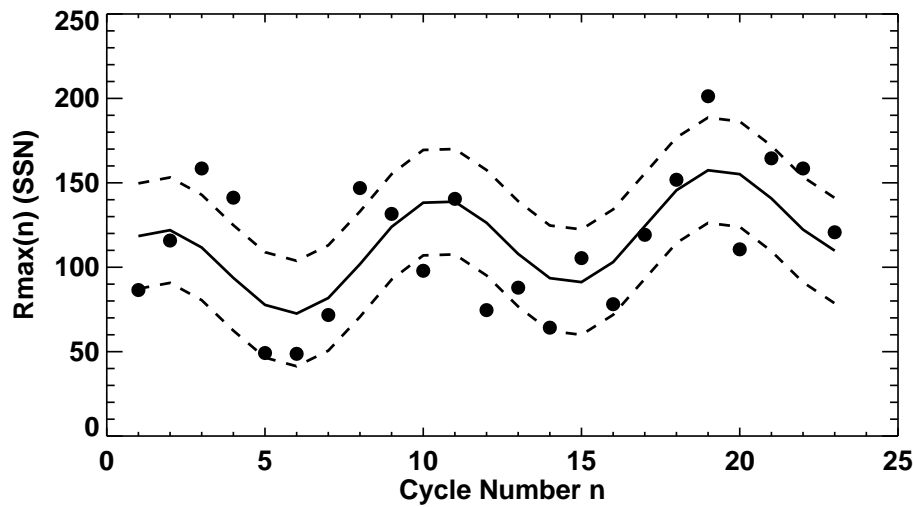}}
\caption{The Gleissberg Cycle. The best fit of cycle amplitudes to a 
simple sinusoidal function of cycle number is shown by the solid line (which 
includes the secular trend).}
\label{fig:GleissbergCycle}
\end{figure}}

In reference to the Gleissberg cycle, \cite{Waldmeier:1957} noted that the
phase shifts in the north-south asymmetry also seen to have a similar
period to that of the cycle amplitudes (Section~\ref{sec:ActiveHemispheres}).

\subsection{Gnevyshev--Ohl Rule (Even--Odd Effect)}
\label{sec:gnevyshev-ohl}

\cite{Gnevyshev:1948} found that if solar cycles are arranged in pairs 
with an even-numbered cycle and the following odd-numbered cycle then the 
sum of the sunspot numbers in the odd cycle is higher than in the even  cycle.
This is referred to as the Gnevyshev-Ohl Rule or Even-Odd Effect. 
This rule is illustrated in Figure~\ref{fig:EvenOddEffect}.
With the exception of the cycle 4/5 pair, this relationship held until
cycle~23 showed that the cycle~22/23 pair was also an exception.
If cycle amplitudes are compared then the cycle 8/9 pair is also an exception.
This rule also holds for other indicators of cycle amplitude such as sunspot area.
While much has been said about this rule relative to the 22-year Hale cycle,
it is difficult to understand why the order (even-odd vs.\ odd-even) of the pairing should make a difference. 
The observed effect does however impact flux transport models for the 
surface fields \citep[see][for a review]{Sheeley:2005}. Since the odd
cycles all have the same magnetic polarity, stronger odd cycles will
tend to build up polar fields of one polarity to the extent that the
transport during the even cycles cannot reverse the polar fields
without associated changes in transport. 

\epubtkImage{EvenOddEffect.png}{%
\begin{figure}[htbp]
\centerline{\includegraphics[width=0.8\textwidth]{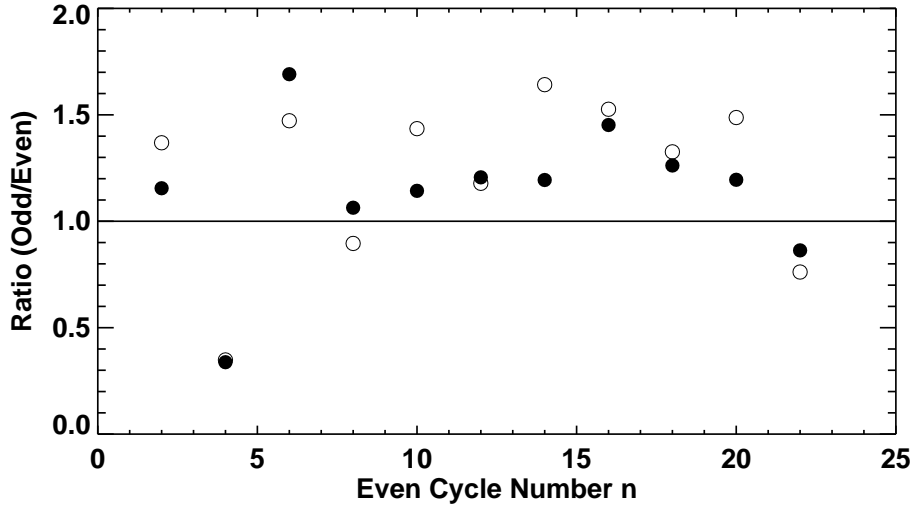}}
\caption{Gnevyshev-Ohl Rule. The ratio of the odd-cycle sunspot sum to the 
preceding even-cycle sunspot sum is shown with the filled circles. The ratio 
of the odd-cycle amplitude to the preceding even-cycle amplitude is shown 
with the open circles.}
\label{fig:EvenOddEffect}
\end{figure}}

\subsection{Long-term variations from radioisotope studies}

The solar cycle modulation of cosmic rays
(Section~\ref{sec:cosmic-rays}) leaves its imprint in the
concentration of the radioisotopes \super{14}C in tree rings and
\super{10}Be in ice cores (Section~\ref{sec:trees-and-ice}). The
connection between solar activity and radioisotope concentrations is
complicated by the transport and storage of these
radioisotopes. Nonetheless, estimates of solar activity levels over
time scales much longer than the 400-year sunspot record can be
obtained \citep[see][for a review]{Usoskin:2013}.

These reconstructions of solar activity reveal grand minima like the Maunder 
Minimum as well as grand maxima similar to the last half of the 20th 
century. The reconstructions suggest that the Sun spends about 1/6th of 
its current life in a grand minimum phase and about 1/10th in a grand 
maximum.

\subsection{The Suess cycle}

One periodicity that arises in many radiocarbon studies of solar activity 
has a well-defined period of about 210 years. This is often referred to as 
the Suess or de~Vries cycle \citep{Suess:1980}. Although the variations in the 
calculated production rates of \super{14}C and \super{10}Be are well correlated 
with each other \citep{Vonmoos:2006} and with the 400-year sunspot
record \citep{Berggren:2009}, there is little evidence of the Suess
cycle in the sunspot record itself \citep{Ma:2009}.

\newpage 

\section{Short-Term Variability}
\label{sec:stvar}

There are significant variations in solar activity on time scales shorter 
than the sunspot cycle. This is evident when the sunspot number record is 
filtered to remove both solar rotation effects (periods of about 27-days and 
less) and solar cycle effects. This signal is shown in
Figure~\ref{fig:ShortTermVariations} for the 
years 1850 to 2013. In this figure the daily sunspot numbers are filtered with 
a tapered Gaussian-shaped filter, Equation~(\ref{eq4}), with a FWHM of 54 days. This 
reduces all signals with periods shorter than 54-days to less than 2\% of 
their original amplitude. The resulting signal is sampled at 27-day 
intervals and then filtered again with a similar Gaussian with a FWHM of 24 
rotations. The lower panel of Figure~\ref{fig:ShortTermVariations} shows this final signal for the time 
period, while the upper panel shows the residual obtained when this smoothed 
sunspot number signal is subtracted from 54-day filtered data. This residual 
signal is quite chaotic but shows some interesting behavior and 
quasi-periodicities.

\epubtkImage{ShortTermVariations.png}{%
\begin{figure}[htbp]
\centerline{\includegraphics[width=0.8\textwidth]{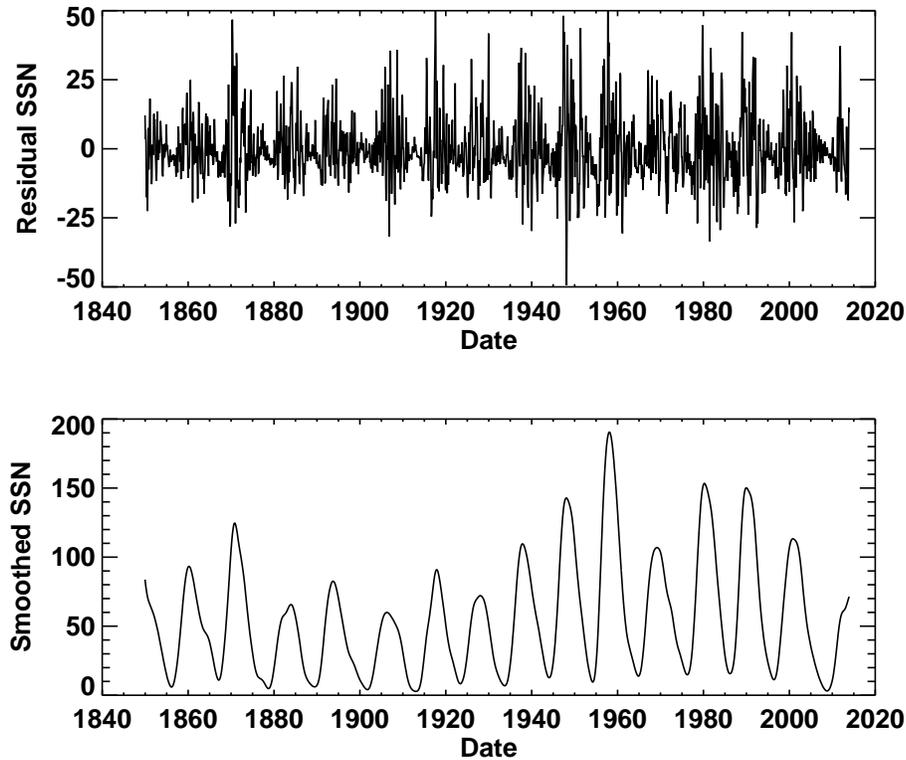}}
\caption{Short-term variations. The lower panel shows the daily 
International Sunspot Number (SSN) smoothed with a 24-rotation FWHM 
Gaussian. The upper panel shows the residual SSN signal smoothed with a 
54-day Gaussian and sampled at 27-day intervals.}
\label{fig:ShortTermVariations}
\end{figure}}

\subsection{154-day periodicity}

A 154-day periodic signal was noted in gamma-ray flare activity seen from 
SMM by \cite{Rieger:1984} for the time interval from 1980/02 to 1983/08. 
This signal was also found by \cite{Bai:1990} in proton flares for 
both this interval and an earlier interval from 1958/01 to 1971/12. 
\cite{Ballester:2002} found that this signal was also seen in the
Mt.~Wilson sunspot index for the 1980\,--\,1983 time
frame. \cite{Lean:1990} analyzed the signal in the sunspot area data
and found that it occurs in episodes around the epochs of sunspot
cycle maxima and that its frequency drifts as well. A wavelet
transform of the bandpass-limited (54 days $\lesssim$ period $\lesssim$
2~years) daily sunspot numbers shown in the upper panel of
Figure~\ref{fig:ShortTermVariations} is also shown in
Figure~\ref{fig:WaveletTransform}, with a horizontal
red line indicating periods of 154 days. The strong signal in the
early 1980s as well as other intermittent intervals is clearly evident
in this plot.

\epubtkImage{WaveletTransform.png}{%
\begin{figure}[htbp]
\centerline{\includegraphics[width=0.8\textwidth]{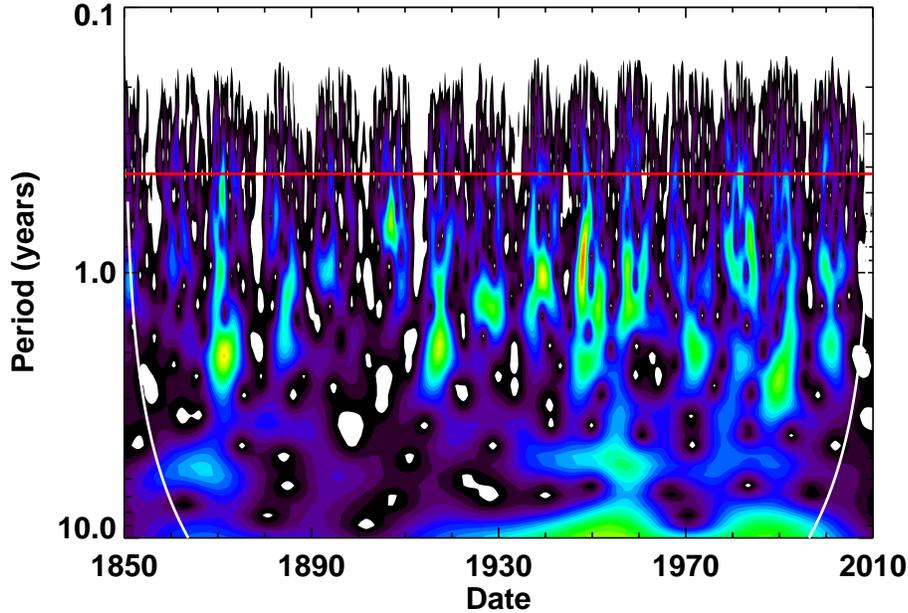}}
\caption{Morlet wavelet transform spectrum of the bandpass-limited daily 
International Sunspot Number. Increasing wavelet power is represented by 
colors from black through blue, green, yellow, and red. The 
Cone-Of-Influence (outside of which the data isn't long enough to give
good measurements of wavelet power) is outlined by the white curves. Periods of 154 days are 
indicated by the horizontal red line.}
\label{fig:WaveletTransform}
\end{figure}}

\subsection{Quasi-biennial variations and double-peaked maxima}

Another interesting periodicity is one found with a period of about two 
years \citep{Benevolenskaya:1995, Mursula:2003}.
This periodicity is particularly evident at cycle maximum in the form of the double peaks and the Gnevyshev Gap (see Section~\ref{sec:gnevyshev-gap}).
\cite{Wang:2003} found that the Sun's dipole magnetic moment and open
magnetic flux exhibits multiple peaks with quasi-periodicities of about
1.3 years, which they attributed to the stochastic processes of active region
emergence and a decay time of about 1 year set by the dynamical processes of
differential rotation, meridional flow, and supergranule diffusion.
These quasi-periodic variations are also evident in the
wavelet spectrum shown in Figure~\ref{fig:WaveletTransform}.
Multiple, significant peaks of power are seen intermittently at periods between
1 and 2 years and are most prevalent near the time of cycle maxima
\citep{Bazilevskaya:2000}.
A signal with a similar period was seen in the tachocline oscillations---periodic
variations in the shear at the base of the convection zone \citep{Howe:2000}.
These tachocline oscillations have also been found to be intermittent
\citep{Howe:2007}.

\newpage 

\section{Solar Cycle Predictions}
\label{sec:predict}

Predicting the solar cycle is indeed very difficult. A cursory examination 
of the sunspot record reveals a wide range of cycle amplitudes
(Figure~\ref{fig:MonthlySSN}). Over the last 24 cycles the average amplitude (in
terms of the 13-month smoothed monthly averages of the daily sunspot
number) was about 114. Over the last 400 years the cycle amplitudes
have varied widely---from basically zero through the Maunder Minimum,
to the two small cycles of the Dalton Minimum at the start of the 19th
century (amplitudes of 49.2 and 48.7), to the recent string of large
cycles (amplitudes of 151.8, 201.3, 110.6, 164.5, 158.5, and
120.8). In addition to the changes in the amplitude of the cycle,
there are changes in cycle length and cycle shape as discussed in
Section~\ref{sec:charact}.
A wide variety of methods have been used to predict the solar cycle.
For recent reviews see \cite{Petrovay:2010} and \cite{Pesnell:2012}.

\subsection{Predicting an ongoing cycle}

One popular and often-used method for predicting solar activity was first 
described by  \cite{McNish:1949}. As a cycle progresses, the smoothed 
monthly sunspot numbers are compared to the average cycle for the same 
number of months since minimum. The difference between the two is used to 
project future differences between predicted and mean cycle. The 
McNish-Lincoln regression technique originally used yearly values and only 
projected one year into the future. Later improvements to the technique use 
monthly values and use an auto-regression to predict the remainder of the 
cycle.

One problem with the modified McNish-Lincoln technique is that it does not 
account for systematic changes in the shape of the cycle with cycle
amplitude (i.e.\ the Waldmeier Effect, Section~\ref{sec:WaldmeierEffect}).
Another problem with the McNish-Lincoln method is its sensitivity to choices
for the date of cycle minimum.
Both the systematic changes in shape and the
sensitivity to cycle minimum choice can be accounted for with
techniques that fit the monthly data to parametric curves
\citep[e.g.][]{Stewart:1938, Elling:1992, Hathaway:1994}.
The two-parameter function of \cite{Hathaway:1994} given by Equation~(\ref{eqn6})
closely mimics the changing shape of the sunspot cycle.
Prediction requires fitting the data to the function with a best fit for an
initial starting time, $t_{0}$, and amplitude, $A$.

Both the modified McNish-Lincoln and the curve-fitting techniques
work nicely once a sunspot cycle is well under way. The critical point
seems to be 2 to 3 years after minimum  -- near the time of the
inflection point on the rise to maximum. Predictions for cycles~22 and
23 using the modified McNish-Lincoln and the Hathaway, Wilson, and
Reichmann curve-fitting techniques 24 months after minimum are shown
in Figure~\ref{fig:CyclePredictions}. Since cycle~23 had an amplitude very close to
the average of cycles~10\,--\,22, both of these predictions are very
similar. Distinct differences are seen for larger or smaller cycles
and when different dates are taken for minimum with the
McNish-Lincoln method.

\epubtkImage{CyclePredictions.png}{%
\begin{figure}[htbp]
\centerline{
  \includegraphics[width=0.48\textwidth]{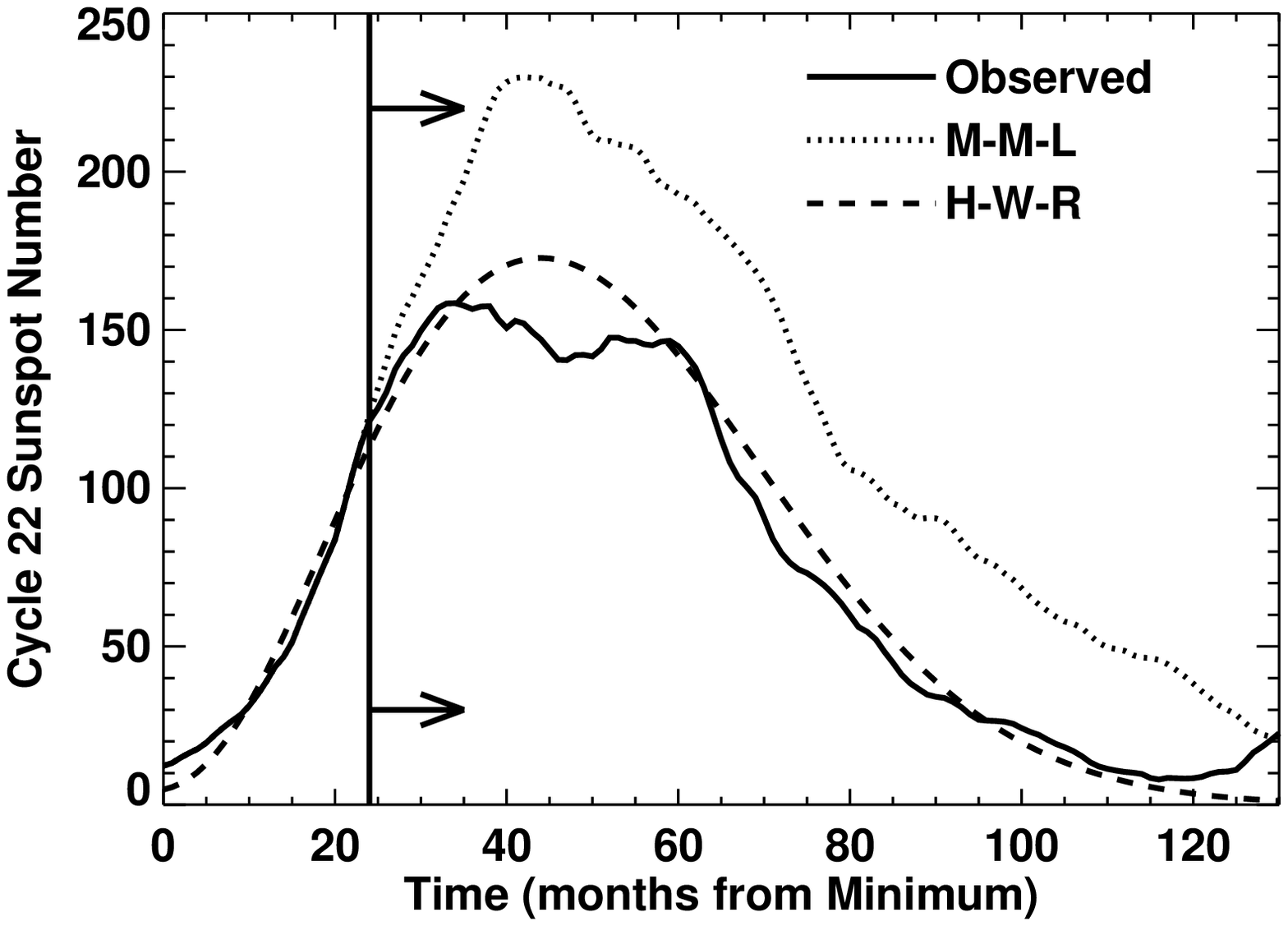}\hspace{4mm}
  \includegraphics[width=0.48\textwidth]{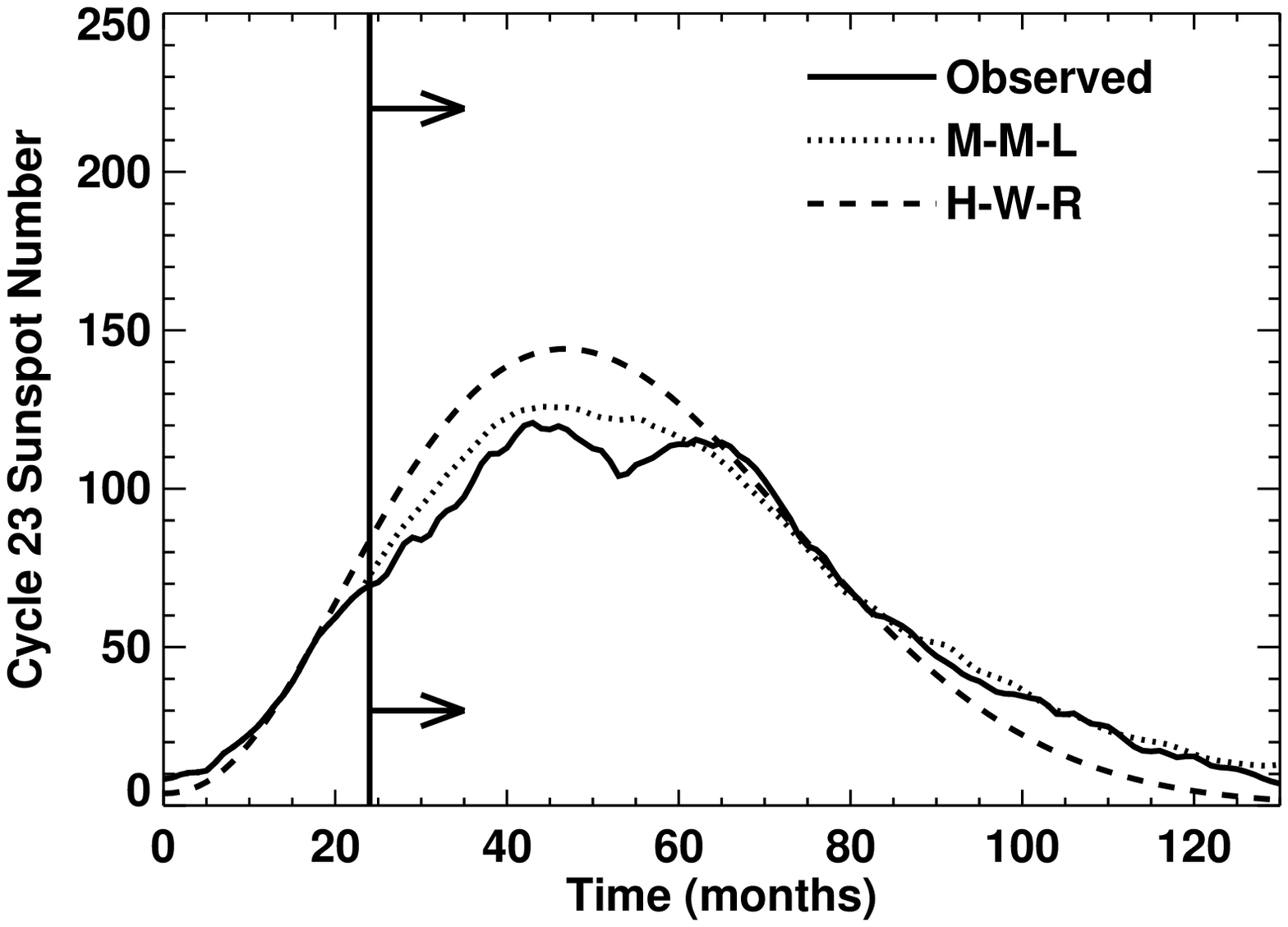}
}
\caption{Predictions for cycles~22 and 23 using the
  modified McNish-Lincoln (M-M-L) auto-regression technique and the
  Hathaway, Wilson, and Reichmann (H-W-R) curve-fitting technique 24
  months after the minima for each cycle.}
\label{fig:CyclePredictions}
\end{figure}}

Predicting the size and timing of a cycle prior to its start (or even during 
the first year or two of the cycle) requires methods other than 
auto-regression or curve-fitting. There is a long, and growing, list of 
measured quantities that can and have been used to predict future cycle 
amplitudes. Prediction methods range from simple climatological means to 
physics-based dynamos with assimilated data.

\subsection{Predicting future cycle amplitudes based on cycle
  statistics}

The mean amplitude of the last $n$ cycles gives the benchmark for other 
prediction techniques. The mean of the last 23 cycle amplitudes is
114.1~\textpm~40.4 where the error is the standard deviation about the
mean. This represents a prediction without any skill. If other methods
cannot predict with significantly better accuracy, they have little
use.

One class of prediction techniques is based on trends and periodicities in 
the cycle amplitudes. The Group Sunspot Number in particular indicates
an upward trend in cycle amplitudes since the Maunder Minimum.
Projecting this trend to the next cycle gives a prediction only slightly
better than the mean.
A number of periodicities have been noted in the cycle amplitude record.
\cite{Gleissberg:1939} noted a long period variation in cycle
amplitudes with a period of seven or eight cycles
(Section~\ref{sec:gleissberg-cycle} and Figure~\ref{fig:GleissbergCycle}).
\cite{Gnevyshev:1948} noted a two-cycle periodicity with the
odd-numbered cycle having larger amplitude than the preceding
even-numbered cycle (Section~\ref{sec:gnevyshev-ohl} and
Figure~\ref{fig:EvenOddEffect}).
\cite{Ahluwalia:1998} noted a three-cycle sawtooth-shaped periodicity
in the six-cycle record of the geomagnetic \textit{Ap} index.

Another class of prediction techniques uses the characteristics of the 
preceding cycle as indicators of the size of the next cycle.
\cite{Wilson:1998} found that the length (period) of the preceding
cycle is inversely correlated to the amplitude of the following
cycle. Another indicator of the size of the next cycle is the level of
activity at minimum---the amplitude of the following cycle is
correlated with the smoothed sunspot number at the preceding minimum
\citep{Brown:1976}. This type of technique has led to searches for
activity indicators that are correlated with future cycle amplitude.
\cite{Javaraiah:2007}, for example, has found sunspot areas from
intervals of time and latitude that correlate very well with future
cycle activity.

In spite of the statistical correlations, these methods based on
cycle statistics tend to be only marginally better than using the
average cycle \citep{Hathaway:1999}.

\subsection{Predicting future cycle amplitudes based on geomagnetic
  precursors}

One class of precursors for future cycle amplitudes that has worked well in 
the past uses geomagnetic activity either during the preceding cycle or near the 
time of minimum as an indicator of the amplitude for the next cycle. These 
``geomagnetic precursors'' use indices for geomagnetic activity (see
Section~\ref{sec:geomagnetic-activity}) that extend back to
1844. \cite{Ohl:1966} found that the minimum level of geomagnetic
activity seen in the \textit{aa} index near the time of sunspot cycle
minimum was a good predictor for the amplitude of the next cycle. This
is illustrated in Figure~\ref{fig:OhlsMethod}. The minima in \textit{aa} are
well-defined and are well-correlated with the following sunspot number
maxima ($r=0.93$). The ratio of max(\textit{R}) to min(\textit{aa})
gives
\begin{equation}
\label{eq8}
\mathrm{max}(R) = 7.95\,\mathrm{min}(aa) \pm 18
\end{equation}
This standard deviation from the relationship is significantly smaller than 
that associated with the average cycle prediction. The minimum 
level of the smoothed \textit{aa} index in 2009 indicated a small cycle~24 -- 
$R_{\mathrm{max}}(24)=70 \pm 18$ (open circles in Figure~\ref{fig:OhlsMethod}).
One problem with this method concerns the timing 
of the \textit{aa} index minima -- they often occur well after sunspot cycle minimum 
and therefore do not give a much advanced prediction.

\epubtkImage{OhlsMethod.png}{%
\begin{figure}[htbp]
\centerline{\includegraphics[width=0.8\textwidth]{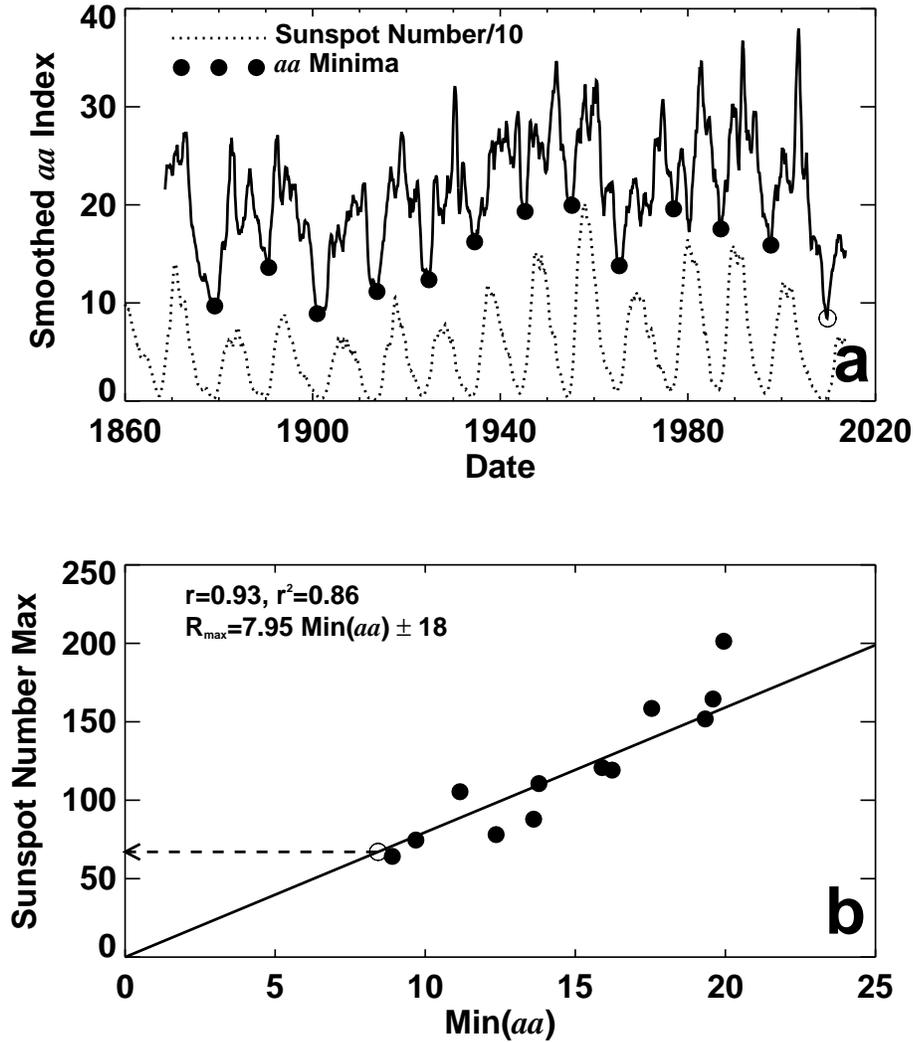}}
\caption{Ohl's method for predicting cycle amplitudes using the minima in 
the smoothed \textit{aa} index (panel a) as precursors for the maximum
sunspot numbers of the following sunspot number maxima (panel b).}
\label{fig:OhlsMethod}
\end{figure}}

Two variations on this method circumvent the timing
problem. \cite{Feynman:1982} noted that geomagnetic activity has two
different sources---one due to solar activity (flares,~CMEs, and
filament eruptions) that follows the sunspot cycle and another due to
recurrent high-speed solar wind streams that peak during the decline
of each cycle (see Section~\ref{sec:geomagnetic-activity} and
Figure~\ref{fig:aaVsTime}). She separated the two by finding
the sunspot number dependence of the base level of geomagnetic
activity and removing it to reveal the ``interplanetary'' component of
geomagnetic activity. The peaks in the interplanetary component prior
to sunspot cycle minimum have been very good indicators for the amplitude of
the following sunspot cycle, as shown in Figure~\ref{fig:FeynmanMethod}.

\epubtkImage{FeynmanMethod.png}{%
\begin{figure}[htbp]
\centerline{\includegraphics[width=0.8\textwidth]{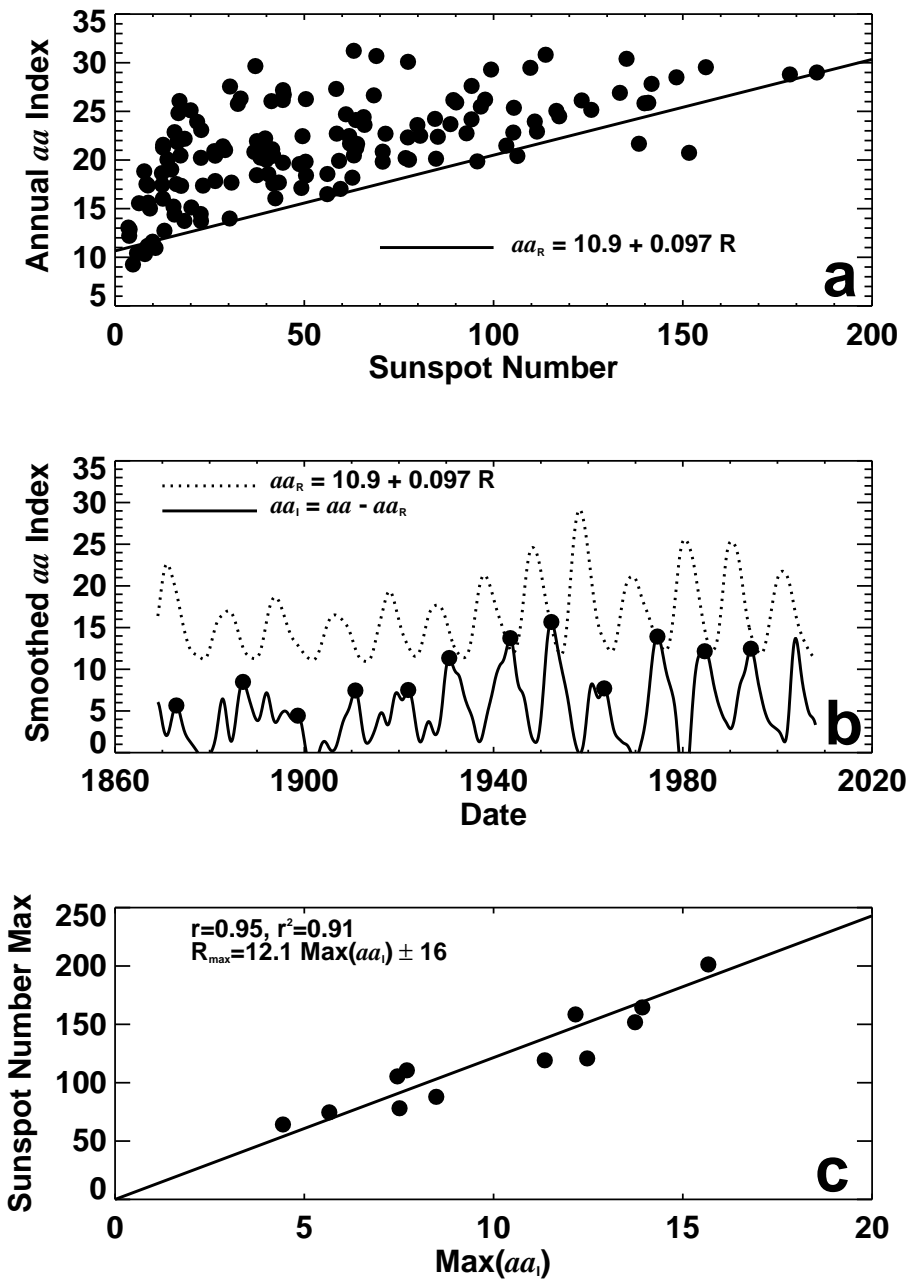}}
\caption{A modification of Feynman's method for separating geomagnetic 
activity into a sunspot number related component and an
``interplanetary'' component (panels a and b). The maxima in
$aa_{I}$ prior to minimum are well-correlated with the
following sunspot number maxima (panel c).}
\label{fig:FeynmanMethod}
\end{figure}}

\cite{Hathaway:2006} used a modification of this method to predict cycle~24.
At the time of that writing there was a distinct peak in
$aa_{\mathrm{I}}$ in late 2003 (in hind sight, clearly related to the
emergence of the large active regions in October/November of 2003).
This large peak led to a prediction of
$R_{\mathrm{max}}(24)=160 \pm 25$, which is clearly too high.
While this method does give predictions prior to sunspot number minimum
it is not without its problems.
Different smoothing of the data gives very different maxima
and different methods are used to extract the sunspot number component
for the data shown in Figure~\ref{fig:OhlsMethod}a. \cite{Feynman:1982} and
others chose to pass a sloping line through the two lowest points.
\cite{Hathaway:2006} fit a line through the 20 lowest points from 20
bins in sunspot number.
These variations add uncertainty in the actual predictions.

\cite{Thompson:1993} also noted that some geomagnetic activity during
the previous cycle served as a predictor for the amplitude of the
following cycle but, instead of trying to separate the two, he simply
related the geomagnetic activity (as represented by the number of days
with the geomagnetic \textit{Ap} index $\ge 25$) during one cycle to
the sum of the amplitudes of that cycle and the following cycle
(Figure~\ref{fig:ThompsonMethod}).
Predictions for the amplitude of a sunspot cycle
are available well before minimum with this method.
The number of geomagnetically disturbed days during cycle~23 gives a prediction of
$R_{\mathrm{max}}(24)=131 \pm 28$, which is also too high.
Disadvantages with this method include the fact that two cycle amplitudes
are involved, the longer cycle will have more disturbed days simply due
to its length.
In addition, outbursts like those in the fall of 2003 can also impact this method.

\epubtkImage{ThompsonMethod.png}{%
\begin{figure}[htbp]
\centerline{\includegraphics[width=0.8\textwidth]{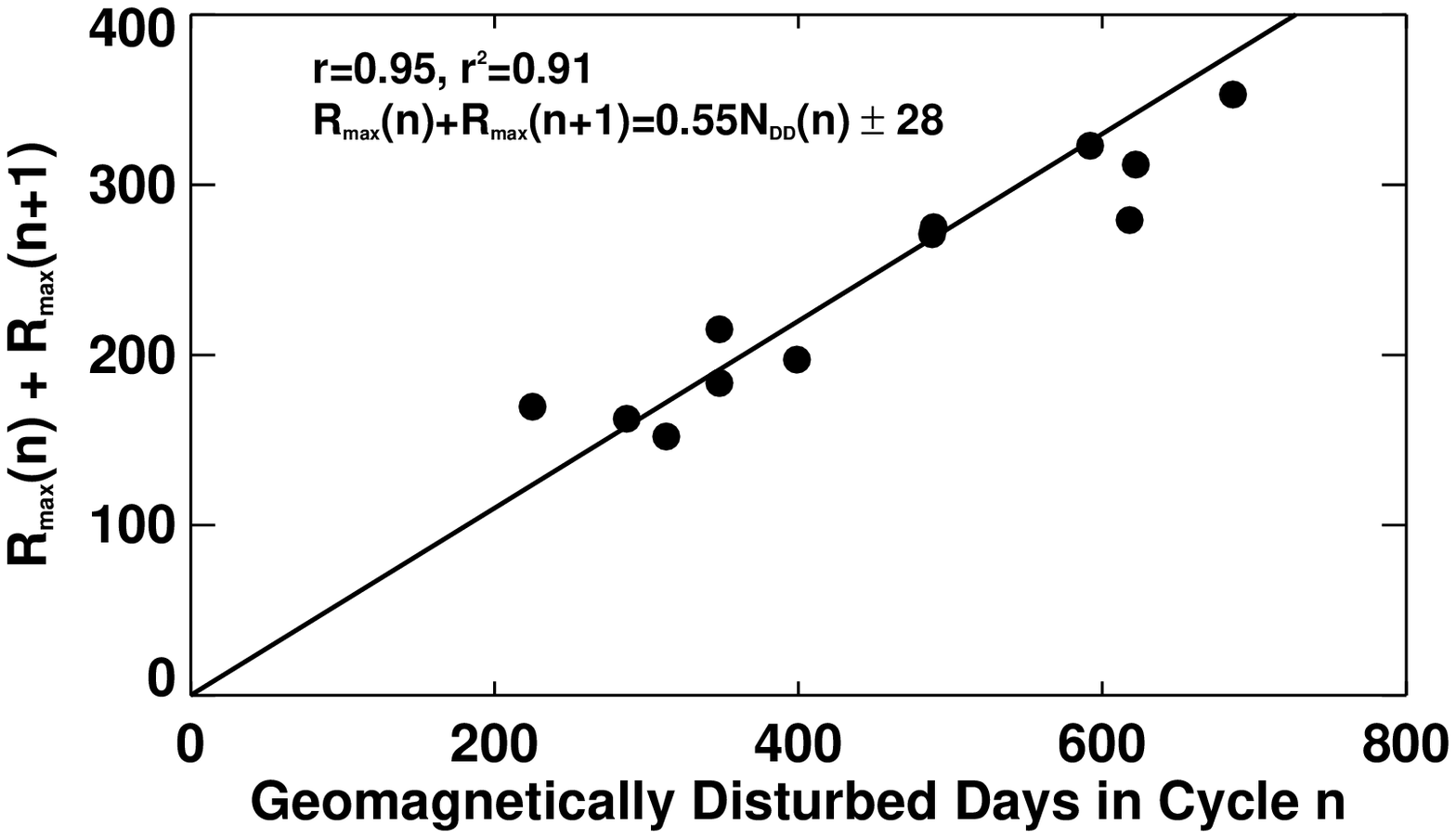}}
\caption{Thompson method for predicting sunspot number maxima. The number 
of geomagnetically disturbed days in a cycle is proportional to the sum of 
the maxima of that cycle and the next.}
\label{fig:ThompsonMethod}
\end{figure}}

\cite{Hathaway:1999} tested these precursor methods by backing up in
time to 1950, calibrating each precursor method using only data prior
to that time, and then using each method to predict cycles~19\,--\,22,
updating the data and recalibrating each method for each remaining
cycle. The results of this test were examined for both accuracy and
stability (i.e., whether the relationships used in the method vary
significantly from one cycle to the next). An updated version of their
Table~\ref{tab3} (including cycle~23 and corrections to the data)
is given here as Table~\ref{tab6}. The RMS errors in
the predictions show that the geomagnetic precursor methods (Ohl's
method, Feynman's method, and Thompson's method) consistently
outperform the other tested methods. Furthermore, these geomagnetic
precursor methods are also more stable. For example, as time
progressed from cycle~19 to cycle~23 the Gleissberg cycle period
changed from 7.5-cycles to 8.5-cycles and the mean cycle amplitude
changed from 103.9 to 114.1, while the relationships between
geomagnetic indicators and sunspot cycle amplitude were relatively
unchanged. 

\begin{table}[htbp]
\caption{Prediction method errors for cycle~19\,--\,23. The three
  geomagnetic precursor methods (Ohl's, Feynman's, and Thompson's)
  give the smallest errors.}
\label{tab6}
\centering
\begin{tabular}{lrrrrrr}
\toprule
Prediction Method & Cycle~19 & Cycle~20 & Cycle~21 & Cycle~22 &
Cycle~23 & RMS \\
\midrule
Mean Cycle           &  --97.4  &  --1.6 & --55.4 & --46.7 & --6.9 & 54.4 \\
Even--Odd            &  --60.1  &      ? & --26.7 & ? & 61.4 & 52.0 \\
Maximum--Minimum     & --109.7  &   24.9 & --18.6 & --8.1 & 5.2 & 51.2 \\
Amplitude--Period    &  --75.3  &   18.4 & --73.5 & --25.6 & 15.0 & 49.6 \\
Secular Trend        &  --96.4  &   14.6 & --40.6 & --25.4 & 18.9 & 49.3 \\
Three Cycle Sawtooth &  --96.5  &   14.6 & --38.5 & --25.4 & 18.8 & 49.0 \\
Gleissberg Cycle     &  --64.8  &   48.0 & --36.9 & --31.8 & --0.9 & 42.1 \\
Ohl's Method         &  --55.4  &  --5.9 & 2.3 & --9.1 & 10.5 & 28.7 \\
Feynman's Method     &  --43.3  & --22.4 & --1.0 & --14.8 & 25.9 & 28.6 \\
Thompson's Method    &  --17.8  &    8.7 & --26.5 & --13.6 & 40.5 & 27.0 \\
\bottomrule
\end{tabular}
\end{table}

The physical mechanisms behind the geomagnetic precursors are uncertain. The geomagnetic 
disturbances that produce the precursor signal are primarily due to high-speed
solar wind streams from low-latitude coronal holes late in a cycle. 
\cite{Schatten:1987} suggested that this geomagnetic activity near the 
time of sunspot cycle minimum is related to the strength of the Sun's polar 
magnetic field which is, in turn, related to the strength of the following 
maximum (see next Section~\ref{sec:PolarFieldPredictions} on dynamo based
predictions). \cite{Cameron:2007} suggest that it is simply the
overlap of the sunspot cycles and the Waldmeier Effect that leads to
these precursor relationships with the next cycle's
amplitude. \cite{Wang:2009} argue that Ohl's method has closer
connections to the Sun's magnetic dipole strength and should therefore
provide better predictions.

\subsection{Predicting future cycle amplitudes based on polar fields}
\label{sec:PolarFieldPredictions}

Dynamo models for the Sun's magnetic field and its evolution have led to 
predictions based on aspects of those models. \cite{Schatten:1978} suggested 
using the strength of the Sun's polar field as a predictor for the amplitude 
of the following cycle, based on the \cite{Babcock:1961} dynamo model. In the 
Babcock model, the polar field at minimum is representative of the poloidal 
field that is sheared out by differential rotation to produce the toroidal 
field that erupts as active regions during the following cycle. Diffusion of 
the erupting active-region magnetic field and transport by the meridional 
flow (along with the Joy's Law tilt of these active regions) then leads to 
the accumulation of opposite polarity fields at the poles and the ultimate 
reversal of the polar fields, as seen in Figure~\ref{fig:MagBfly}.

Good measurements of the Sun's polar field are difficult to obtain. The 
field is weak and predominantly radially directed, and thus nearly transverse 
to our line of sight. This makes the Zeeman signature weak and prone to the 
detrimental effects of scattered light. Nevertheless, systematic 
measurements of the polar fields have been made at the Wilcox Solar 
Observatory since 1976 and have been used by Schatten and his colleagues to 
predict cycles~21\,--\,24. These polar field measurements are shown in
Figure~\ref{fig:PolarFields} along with smoothed sunspot numbers.

While the physical basis for these predictions is appealing, the fact that
direct measurements were only available for the last three cycles was
a distinct problem.
The number of polar faculae (available from Mount Wilson photographs
since 1906) was recognized as an indicator of polar fields by
\cite{Sheeley:1966}, but an examination \citep{Layden:1991} of the use of
the polar faculae counts \citep{Sheeley:1964, Sheeley:1976} as a predictor
suggested they have little predictive power.
Recently, however, \cite{MunozJaramillo:2012} recalibrated these polar faculae
counts and found that the revised counts and relationship to polar fields
do, in fact, provide very good predictions of cycle amplitudes \citep{MunozJaramillo:2013A}.

It was unclear when these polar field measurements should be taken.
Predictions based on the polar fields for previous cycles have given
different values at different times.
Using the polar faculae measurements, \cite{MunozJaramillo:2013B} found that
predictions of cycle amplitudes within $2\sigma$ of the observed amplitude
were successfully made for more than 80\% of the cycles for measurements within 2 years
of cycle minimum. The success rate dropped rapidly for predictions made earlier. 

The polar fields were obviously much weaker during the cycle 23/24 minimum.
This has led to a prediction of $R_{\mathrm{max}}(24)
= 75 \pm 8$ by \cite{Svalgaard:2005}---about half the size of the
previous three cycles, based on the polar fields being about half as
strong.
(As of this writing, the maximum of cycle~24 appears to be the peak
of 81.9 realized in April 2014.)
While in previous minima the strength of the polar fields (as
represented by the average of the absolute field strength in the north
and in the south) varied as minimum approached, this did not happen on
the approach to cycle~24 minimum in late 2008. This suggests that the
early prediction made in 2005 was not compromised.

\subsection{Predicting future cycle amplitudes based on flux transport dynamos}
\label{sec:DynamoPredictions}

The use of polar fields at minimum as a predictor for the amplitude of the next cycle
is loosely based on dynamo theories like that of \cite{Babcock:1961}.
Predictions have also been made using more fully developed dynamo
models along with the assimilation of data from previous years and cycles.
\cite{Kitiashvili:2008} used a 1D (time) dynamo model in which the alpha-effect
(the lifting and twisting that converts toroidal field into poloidal field)
depended non-linearly on the magnetic field itself.
After assimilating data from previous cycles, they showed that their model
had good predictive powers and predicted a sunspot number maximum of about
80 for cycle~24.

Over the last two decades, 3D (latitude, depth, and time) flux-transport
dynamo models have been
developed to include the kinematic effects of the 
Sun's meridional circulation, finding that it can play a significant role 
in the magnetic dynamo \citep{Dikpati:1999}. In these models 
the speed of the meridional circulation sets the cycle period and influences 
both the strength of the polar fields and the amplitudes of following 
cycles. Two predictions were made based on flux transport 
dynamos with assimilated data, with very different results.

\cite{Dikpati:2006} predicted an amplitude for cycle~24 of 
150\,--\,180 using a flux transport dynamo that included a rotation profile and a 
near surface meridional flow based on helioseismic observations. They 
modeled the axisymmetric poloidal and toroidal magnetic field using a 
meridional flow that returns to the equator at the base of the convection 
zone and used two source terms for the poloidal field---one at the surface 
due to the Joy's Law tilt of the emerging active regions, and one in the 
tachocline due to hydrodynamic and MHD instabilities. 
They drive the model with a surface source of poloidal
field that depends upon the sunspot areas observed since
1874. Measurements of the meridional flow speed prior to 1996 are
highly uncertain \citep{Hathaway:1996}, so they maintained a
constant flow speed prior to 1996 and forced each of those earlier
cycles to have a constant period as a consequence. The surface
poloidal source term drifted linearly from 30\textdegree\ to
5\textdegree\ over each cycle with an amplitude that depended on the
observed sunspot areas. The prediction was based on the strength of
the toroidal field produced in the tachocline. They found excellent
agreement between this toroidal field strength and the amplitude of
each of the last eight cycles (the four earlier cycles---during the
initialization phase---were also well-fit, but not with the degree of
agreement of the later cycles). The correlation they found between the
predicted toroidal field and the cycle amplitudes is similar to that
found with the geomagnetic precursors and polar field strength
indicators. When they kept the meridional flow speed at the same
constant level during cycle~23, they found $R_{\mathrm{max}}(24) \sim
180$. When they allowed the meridional flow speed to drop by 40\%, as
was seen from 1996\,--\,2002, they found $R_{\mathrm{max}}(24) \sim
150$, and further predicted that cycle~24 would start late due to the slower
meridional flow.

\cite{Choudhuri:2007} predicted an amplitude for cycle~24 
of 80 using a similar flux-transport dynamo, but using the surface poloidal 
field at minimum as the assimilated data. They used a similar axisymmetric 
model for the poloidal and toroidal fields, but with a meridional flow that 
extends below the base of the convection zone and a diffusivity that remains 
high throughout the convection zone. In their model, the toroidal field in 
the tachocline produces flux eruptions when its strength exceeds a given 
limit. The number of eruptions is proportional to the sunspot number 
and was used as the predicted quantity. They assimilate data by 
instantaneously changing the poloidal field at minimum throughout most of 
the convection zone to make it match the dipole moment obtained from
the Wilcox Solar Observatory observations (Figure~\ref{fig:PolarFields}). They
found an excellent fit to the last three cycles (the full extent of
the data) and found $R_{\mathrm{max}}(24) \sim 80$, in agreement with
the polar field prediction of \cite{Svalgaard:2005}.

Criticism has been leveled against all of these flux-transport-dynamo-based
predictions. \cite{Dikpati:2006} criticized the use of polar field
strengths to predict the sunspot cycle peak (that follows four years later),
by questioning how those fields could be carried down to the low-latitude
tachocline in such a short time. \cite{Cameron:2007} produced
a simplified 1D flux transport model and showed that with similar
parameters to those used by \cite{Dikpati:2006}, the flux transport
across the equator was an excellent predictor for the amplitude of the
next cycle, but the predictive skill was lost when more realistic
parameterizations of the active region emergence were
used. \cite{Yeates:2008} compared an advection-dominated model like
that of \cite{Dikpati:2006} to a diffusion-dominated models like that
of \cite{Choudhuri:2007}, and concluded that the diffusion-dominated
model was better because it gave a better fit to the relationship
between meridional flow speed and cycle amplitude. \cite{Dikpati:2008}
returned with a study of the use of polar fields and cross equatorial
flux as predictors of cycle amplitudes, and concluded that their
tachocline toroidal flux was the best indicator. Furthermore, they
found that the polar fields followed the current cycle so that the
weak polar fields at this minimum were due to the weakened meridional
flow. The strongest criticism of these dynamo-based predictions was
given by \cite{Tobias:2006} and \cite{Bushby:2007}. They conclude that
the solar dynamo is deterministically chaotic and thus inherently
unpredictable.

A much more significant problem with these models and their predictions
arises from recent measurements of the meridional flow itself.
The flux-transport dynamo modelers all assume that the poleward meridional
flow observed at the surface sinks inward in the polar regions and returns
equatorward at the base of the convection zone, turning equatorward
starting at the mid-point depth of 100 Mm.
Measurements of this equatorward return flow using both the motions of supergranules
\citep{Hathaway:2012} and acoustic waves \citep{Zhao:2013}, indicate
that the equatorward return flow starts at a much shallower depth (50-60 Mm).
The deeper probing with the acoustic waves indicates that the flow
turns poleward again deeper down (below 120 Mm).
This double cell structure for the Sun's meridional flow cannot accommodate
any of these flux-transport dynamo models.

An additional problem for these flux transport dynamo models arises from
the time dependence of the meridional flow.
\cite{Nandy:2011} used a flux transport dynamo model to explain both the weak
polar fields at the end of cycle~23 and the long period of cycle~23, by
assuming that the meridional flow was fast at the beginning of the cycle
and slow at the end. 
In general, the meridional flow is observed to be fast at cycle minima and slow
at cycle maxima due to inflows toward the active latitudes
\citep{Komm:1993, Gizon:2004, Zhao:2004, GonzalezHernandez:2008, Hathaway:2010}.
But in particular, the meridional flow was slower at the cycle~22/23 minimum in 1996
and faster at the cycle~23/24 minimum in 2008
\citep{Hathaway:2010}---exactly the opposite to the variations proposed
by \cite{Nandy:2011}.

While we ultimately expect dynamo theory to provide us with a better understanding
of the solar cycle and to provide us with better predictions, much progress
is still needed.


\newpage 
\section{Cycle 23/24 minimum and cycle 24}
\label{sec:Cycle23/24}
 
The long, unexpected delay in the start of cycle~24 left behind a solar
cycle minimum unlike any seen in living memory.
In December of 2008, the 13-month smoothed sunspot number dropped to
1.7---its lowest value since July of 1913, and
the smoothed number of spotless days in a month reached its highest
value since August of 1913.
In September of 2009, the geomagnetic {\em aa}-index dropped to its
lowest value on record (since 1868, see Figure~\ref{fig:aaVsInternational}),
while the galactic cosmic ray flux reached record highs (since 1953).

Since that minimum, we have seen cycle~24 rise slowly through one peak
and then another to a maximum smoothed sunspot number of 81.9 in April 2014.
While this behavior is exceptional in view of living memory,
it is well within the bounds of normal behavior for solar cycles.

We have seen that small cycles start late and thereby leave behind
a long cycle (the Amplitude-Period Effect, Section~\ref{sec:AmplitudePeriod})
and a low minimum (the Maximum-Minimum Effect, Section~\ref{sec:MaximumMinimum}).
Small cycles also take a longer time to rise to maximum
(the Waldmeier Effect, Section~\ref{sec:WaldmeierEffect}).

Relative to previous cycle behavior, an amplitude of about 82 for cycle~24
suggests (Figure~\ref{fig:AmplitudePeriod}) a period of 150 months for
cycle~23 (147 months observed); (Figure~\ref{fig:MaximumMinimum}) a
minimum of 0.3 for cycle~23/24 minimum (1.7 observed); and
(Figure~\ref{fig:WaldmeierEffect}) a rise time from minimum to maximum
of 56 months (63 observed).

While this behavior is not exceptional in terms of the historical record,
it is exceptional when considering that the last time this was seen was
100 years ago.
Furthermore, we are blessed with wide-ranging and detailed observations
that were not available 100 years ago to help us understand the origin
of this behavior.
Many papers have been written concerning the chain of events that
led to this deep minimum and weak cycle.
While some have led us down dead-end paths, others may lead us to a
better understanding of how solar cycle amplitudes are modulated
and how and why deeper minima, like the Maunder Minimum, occur.

\subsection{Deviations from previous behavior in 10.7 cm flux}
\label{sec:DeviousBehavior}

\cite{Tapping:2011} noted that the relationship between the 10.7 cm radio flux
and the International Sunspot Number changed during cycle 23.
This change is evident in Figure~\ref{fig:F107VsInternational} where the
data for cycles 23 and 24 indicates higher values for the 10.7 cm radio
flux relative to the values of the International Sunspot Number,
particularly after the first (and higher) peak in sunspot number in early 2000.

\subsection{The Livingston-Penn Effect}
\label{sec:LivingstonPenn}

\cite{Penn:2006} and \cite{Livingston:2009} reported that their measurements
of the magnetic field strength and emergent intensity at the darkest points
in sunspot umbrae indicated a linear trend in sunspot properties that would lead
to the total loss of sunspots by 2015.
Their measurements were made spectroscopically using the highly magnetically
sensitive spectral line of neutral iron at 1564.8 nm.
Measurements from individual sunspot umbrae were made by moving the
entrance aperture of the spectrograph from sunspot to sunspot.
Their measurements indicated that, on average, the field strength in sunspot
umbrae was getting weaker at the same time that the emergent intensity
was getting brighter.

\cite{Schad:2010} examined this effect using the full disc daily
line-of-sight magnetograms produced by the NASA/NSO spectromagnetograph
at the Kitt Peak Vacuum Telescope between 1993 and 2003.
They used an automated algorithm to detect sunspot umbrae and measured
the field strength, emergent intensity, and size of nearly 13,000 sunspot umbrae.
Their results recover historical relationships between the
size of sunspot umbrae and the field strength and emergent intensity,
but did not show any systematic secular trend.

\cite{Livingston:2012} extended the measurements in the infrared line through
cycle~23/24 minimum and into the rise of cycle~24.
Although the trends seemed to be leveling off, their data still showed
a continued decline in umbral magnetic fields and a rise in umbral intensities. 
They also noted a minimum  of about 1500 G for the field strength in the smallest
(and brightest) sunspots.
While this relationship may be critical for understanding connections between
dynamo theory and the sunspot cycle (the production or non-production
of sunspots depends upon a threshold strength of the generated fields),
their data showed that few, if any, of these small spots were measured earlier
in the program (from 1998 to about 2005).
They concluded that this lack of small sunspots was a characteristic of 
cycle~23 and was related to the changes in the 10.7 cm radio flux.

However, \cite{deToma:2013B}, using photometric data from the San Fernando Observatory,
found no systematic trends in the emergent intensities from sunspot umbrae
over the 27 years from 1986 to 2012.

Finally, \cite{Watson:2014} compared sunspot umbral field strengths and emergent intensities
measured by \cite{Livingston:2012} and compared them with measurements made from
space by SOHO/MDI and SDO/HMI using an automated detection algorithm.
They noted that the number of umbral measurements per day, obtained with the
space-based instruments, was well-correlated with the sunspot number.
However, the ground-based measurements obtained at Kitt Peak did not show this
correlation until sometime after 2003.
One must conclude that large sunspot umbrae were selectively measured prior
to 2003 and that measurements representative of the full sunspot distribution
were not made until after that time---the Livingston-Penn effect is largely
due to selection effects, rather than a systematic trend in sunspot properties. 

\subsection{Sunspot Size Distributions}
\label{sec:SunspotDistribution}

Sunspot umbral areas were found to be distributed log-normally.
\cite{Bogdan:1988} measured umbral areas on Mount Wilson
white-light plates collected from 1917 to 1982 and found that the
same log-normal distribution is obtained for all phases of the solar cycle,
and for the individual cycles, as well.

However, \cite{Kilcik:2011} found a difference in the relative numbers of small and
large sunspot groups in cycle~23 when compared to cycle~22, with a deficit of small
sunspots in cycle~23.
This was based not so much on actual size measurements, but rather on the Zurich sunspot
classifications reported by the USAF observers.
\cite{Lefevre:2011} performed a detailed analysis of the size distributions
of sunspot groups in the USAF database, and of individual sunspots in the Debrecen
database, and concluded that cycle~23 did have significantly fewer small
($\lesssim 17$ $\mu$Hem) sunspots and sunspot groups.
This is shown by the distributions plotted in
Figure~\ref{fig:RegionSizeDistributionCycle22vs23}.

\epubtkImage{RegionSizeDistributionCycle22vs23.png}{%
\begin{figure}[htbp]
\centerline{\includegraphics[width=0.8\textwidth]{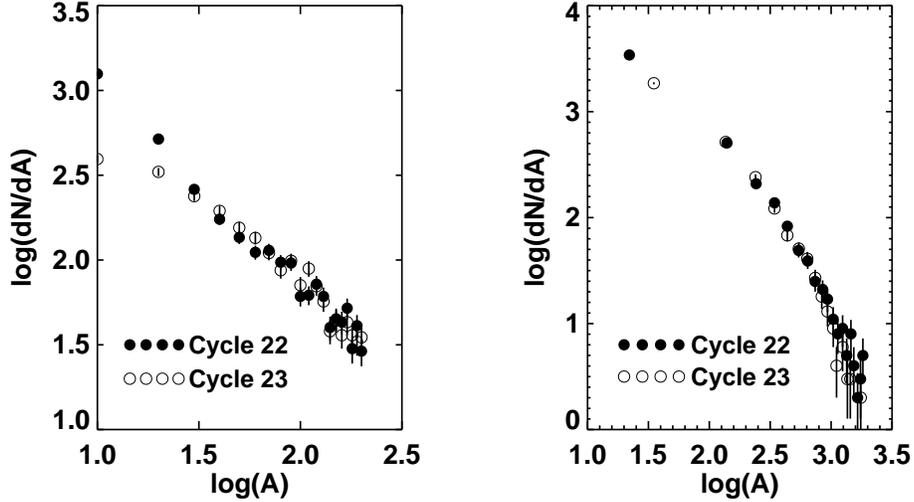}}
\caption{Sunspot group size (area) distributions for cycles 22 and 23.
The maximum sizes attained by active regions in the USAF database are used
to count each active region in a size bin for its respective cycle.
On the left, the bins extend from 10 to 200 $\mu$Hem in 10 $\mu$Hem increments.
On the right, the bins extend from 100 to 2000 $\mu$Hem in 100 $\mu$Hem increments.
The deficit of small spots (with areas of 10 and 20 $\mu$Hem) in cycle 23 appears to be significant.
Cycle~23 also appears to have fewer large active regions.
While those deviations are well within the $1\sigma$ errors, there are consistently
fewer numbers in all area bins above $\sim$700 $\mu$Hem.}
\label{fig:RegionSizeDistributionCycle22vs23}
\end{figure}}

However, these small sunspots are difficult to measure.
Furthermore, the measurements acquired by the USAF involved several observing
sites and dozens of observers, all of which changed over the course of the two cycles.
\cite{deToma:2013A} made measurements of sunspot areas from photometric images
obtained over cycle~22 and 23 at the San Fernando Observatory, and found that
the important difference was in the numbers of large ($\gtrsim$ 700 $\mu$Hem)
sunspots---cycle~23 had fewer of them.
This is seen to some extent in the USAF data plotted in 
Figure~\ref{fig:RegionSizeDistributionCycle22vs23}.
The cycle~23 data points fall consistently below the cycle~22 data points
for sunspot group areas $\gtrsim$ 700 $\mu$Hem.

\subsection{Flow variations}
\label{sec:FlowVariations}

The axisymmetric flows -- differential rotation and meridional flow -- also
change with the solar cycle.

If a long-term average rotation profile is subtracted from the instantaneous
rotation profile, it reveals the torsional oscillations
\citep{Howard:1980}---faster and slower than average zonal flows
that drift equatorward with the sunspot zones.
The faster zone is found on the equatorward side of the sunspot zones,
while the slower zone is found on the poleward side of the sunspot
zones---thereby enhancing the latitudinal shear in the rotation rate at the
latitudes where sunspots emerge.
The instantaneous deviations from the average rotation profile
also include changes in the high-latitude (polar) zonal flows, with a speed-up
starting at about the time of cycle maximum that is balanced by
a slow-down starting at about the time of cycle minimum.

Likewise, if a long-term-average meridional flow profile is subtracted
from the instantaneous meridional flow profile it reveals in-flows
toward the active latitudes \citep{Gizon:2004, Zhao:2004}.
These in-flows have the effect of making the meridional flow slower
at cycle maxima and faster at cycle minima as was observed
earlier by \cite{Komm:1993}.

Both the torsional oscillations and the in-flows toward the active latitudes
were seen in cycle 23 and the rise of cycle 24, but with interesting differences
from previous cycles.
\cite{Howe_etal:2009} noted that the delayed start of cycle 24 could be seen
in the slower equatorward progression of the torsional oscillations during
the decline of cycle 23.
\cite{Howe:2013} noted that the high-latitude branch of the torsional
oscillations (the polar spin-up) was not yet evident (unless
a shorter term average was removed from the data).
\cite{Hathaway:2010} noted that the meridional flow was slower at cycle 22/23
minimum in 1996 than it was at cycle 23/24 minimum in 2008.

The relationships between these flow variations and solar cycle variability
are being explored.
Changes in the meridional flow in particular can lead to changes in the polar fields
through surface flux transport \citep{Sheeley:2005}.
If these changes are directly related to solar activity, then they
may help to modulate cycle amplitudes \citep[see][]{Cameron:2012}.


\newpage 
\section{Conclusions}
\label{sec:Conclude}

Understanding the solar cycle remains one of the biggest problems in 
solar physics. It is also one of the oldest. Several key features of the 
solar cycle have been reviewed here and must be explained by any viable 
dynamo theory or model. They include:

\begin{itemize}

\item The solar cycle has a period of about 11 years but varies in
  length with a standard deviation of about 14 months.

\item Each cycle appears as an outburst of activity that overlaps with
  both the preceding and following cycles.
  This overlap is only about 18 months when measured by the occurrence of sunspots
  but stretches to years when measured by ephemeral regions, torsional oscillations,
  and coronal emissions.

\item Solar cycles are asymmetric with respect to their maxima---the
  rise to maximum is shorter than the decline to minimum, and the rise
  time is shorter for larger amplitude cycles.

\item Big cycles usually start early and leave behind a short
  preceding cycle and a high minimum of activity.

\item Sunspots erupt in low-latitude bands on either side of the
  equator, and these bands drift toward the equator as each cycle
  progresses with little variation from cycle to cycle.

\item The activity bands widen during the rise to maximum and narrow
  during the decline to minimum.
  This width is primarily a direct function of the sunspot number or area
  with little, if any, further dependence on cycle size or phase. 

\item At any time, one hemisphere may dominate over the other, but the
  northern and southern hemispheres never get out of phase by more than
  about 10 months.
  
\item Sunspot groups tend to emerge at longitudes where previous
  groups had emerged (active longitudes/activity nests).

\item Sunspots erupt in groups extended in longitude but more
  constrained in latitude, with one magnetic polarity associated with
  the leading spots (leading in the direction of rotation) and the opposite
  polarity associated with the following spots.

\item The magnetic polarities of active regions reverse from northern
  to southern hemispheres and from one cycle to the next, but exceptions occur.

\item The polar fields reverse polarity during each cycle at about the
  time of cycle maximum.

\item The leading spots in a group are positioned slightly equatorward
  of the following spots, and this tilt increases with latitude.
  This tilt may vary inversely with the amplitude of a cycle.
  There is a wide scatter of tilt angles about the mean, and this scatter
  is even larger for the smaller and weaker ephemeral regions.

\item Cycle amplitudes exhibit weak quasi-periodicities like the 7 to
  8-cycle Gleissberg Cycle (100 years).

\item Cycle amplitudes exhibit extended periods of inactivity, like the
  Maunder Minimum, where sunspots are not observed but low level magnetic
  activity continues.

\item Solar activity exhibits quasi-periodicities at time scales
  shorter than 11 years (quasi-biennial).

\item Predicting the level of solar activity for the remainder of a
  cycle is reliable 2\,--\,3 years after cycle minimum.

\item Predictions for the amplitude of a cycle based on the Sun's
  polar field strength or on geomagnetic activity near cycle minimum
  are significantly better than using the climatological mean.

\end{itemize}

These characteristics provide strong constraints on dynamo theory.
If we consider the \cite{Babcock:1961} dynamo as a straw-man for dynamo theory
then the toroidal-to-poloidal process is fairly well understood in terms of the
surface flux transport of the magnetic field (which emerges
in the form of tilted active regions), by the observed meridional flow, differential rotation,
and convective motions \citep{Sheeley:2005}.
We also understand the poloidal-to-toroidal process, but are faced with three different
shear flows (the surface shear layer, the latitudinal shear in the bulk of the convection zone,
and the tachocline shear layer) that may each participate in the process.

While the origin of the tilt in active regions is thought to be due to the Coriolis force acting
on flux tubes rising from the tachocline \citep{Fan:2009}, those models have difficulty
explaining why active regions have such limited longitudinal extent, and why active regions
tend to relax to the tilt associated with their latitude of emergence---rather than no tilt at all.

One of the biggest difficulties remains in explaining why the active regions emerge
in low-latitude bands that drift equatorward.
The earliest dynamo theories explained this in terms of a dynamo wave in the bulk of the convection zone.
Observations from helioseismology, and the theory of buoyant flux tubes, pushed the dynamo
wave to the tachocline where it then had a problematic poleward branch.
Flux-transport dynamos solved this problem by  invoking a deep equatorward meridional flow to drive the process,
but now, even that seems unlikely.

All of these problems are further compounded by the extensive overlap between cycles.
Indeed, to some extent, each of the solar cycle characteristics itemized above need to be
better understood with improved dynamo models.

Future observations of sub-surface flows and (perhaps) magnetic fields,
observations of activity cycles on other stars, and continued
efforts at modeling the solar dynamo will undoubtably lead us closer to better understanding
the solar cycle.

\newpage 


\end{document}